\documentclass[twocolumn]{aastex62}

\usepackage{xspace,nicefrac,multirow,amsmath,upgreek}

\usepackage{color,ulem,soul}
\definecolor{dark-red}{rgb}{0.75, 0.00, 0.00}
\definecolor{dark-orange}{rgb}{0.85, 0.45, 0.00}
\definecolor{magenta}{rgb}{0.75, 0.25, 0.75}

\definecolor{hlcolor}{rgb}{1.00, 0.94, 0.92}\sethlcolor{hlcolor}

\newcommand{\Sec}[1]{Sect.~\ref{sec:#1}}
\newcommand{\Section}[1]{Section~\ref{sec:#1}}

\newcommand{\Fig}[1]{Fig.~\ref{fig:#1}}
\newcommand{\Figure}[1]{Figure~\ref{fig:#1}}

\newcommand{\Tab}[1]{Tab.~\ref{tab:#1}}
\newcommand{\Table}[1]{Table~\ref{tab:#1}}

\newcommand{\Eqn}[1]{Eqn.~(\ref{eq:#1})}

\newcommand{\beq}{\begin{equation}}
\newcommand{\eeq}{\end{equation}}

\newcommand\m{\,\rm m}
\newcommand\cm{\,\rm cm}
\newcommand\mm{\,\rm mm}
\newcommand\um{\,\upmu\rm m}
\newcommand\s{\,\rm s}
\newcommand\g{\,\rm g}

\newcommand\K{\,\rm K}
\newcommand\yr{\,\rm yr}

\newcommand\opa{\,\cm^2\g^{-1}}

\newcommand\au{\,\rm au}
\newcommand\mG{\,\rm mG}

\newcommand\Rsun{\,R_\odot}
\newcommand\Msun{\,M_\odot}
\newcommand\Mdot{\,M_\odot\,{\rm yr}^{-1}}
\newcommand\mJy{\,{\rm mJy}}
\newcommand\GHz{\,{\rm GHz}}

\newcommand\tms{\!\times\!}
\newcommand\cdt{\!\cdot\!}

\newcommand\eq{\!=\!}
\newcommand\mi{\text{-}}

\newcommand{\ee}[1]{\tms 10^{#1}}

\newcommand\zz{\hat{{\mathbf z}}}
\newcommand\bb{\hat{{\mathbf e}}_b}
\newcommand\nn{\hat{{\mathbf n}}}

\newcommand{\dv}{\nabla\cdt}
\newcommand{\cl}{\nabla\tms}

\newcommand{\dd}{{\rm d}}

\newcommand\rr{\mathbf r}

\newcommand\V{\mathbf v}
\newcommand\U{\mathbf u}
\newcommand\A{\mathbf A}
\newcommand\B{\mathbf B}

\newcommand\Vp{v_{\rm p}}
\newcommand\Bp{B_{\rm p}}

\newcommand\Rc{s}
\newcommand\cS{c_{\rm s}}
\newcommand\zb{z_{\rm b}}
\newcommand\vK{v_{\rm K}}

\newcommand\nir{\textsc{nirvana-iii}\xspace}
\newcommand\rad{\textsc{radmc-3d}\xspace}
\newcommand\lime{\textsc{lime}\xspace}
\newcommand\krome{\textsc{krome}\xspace}

\newcommand\etao{\eta_{\rm \Omega}}
\newcommand\etad{\eta_{\rm A}}

\newcommand\EO{\mathbf E_{\Omega}}
\newcommand\EAD{\mathbf E_{\rm A}}
\newcommand\Ediss{\mathbf E_{\rm d}}

\newcommand\bp{\beta_{\rm p}}

\newcommand\Mx{T_{z\!\phi}^{\rm Mx}|_{\zb}}

\newcommand\eR{\mathcal{E}}
\newcommand\Fr{\mbox{\boldmath{${\cal F}$}}_{\!\rm r}}
\newcommand\pR{\mathcal{P}_{\rm r}}

\newcommand\Qirr{\mathcal{Q}^+_{\rm irr}}
\newcommand\Qpdr{\mathcal{Q}^{\nicefrac{+}{-}}_{\rm pdr}}
\newcommand\Qd{\mathcal{Q}^+_{\rm d}}
\newcommand\Qad{\mathcal{Q}^+_{\rm ad}}

\newcommand\Fm{\mathcal{F}_{\rm m}}
\newcommand\Fe{\mbox{\boldmath{${\cal F}$}}_{\!\rm e}}

\newcommand\Sm{\mbox{\boldmath{${\cal S}$}}_{\rm m}}
\newcommand\Se{\mathcal{S}_{\rm e}}

\newcommand\kP{\kappa_{\rm P}}
\newcommand\kR{\kappa_{\rm R}}
\newcommand\aR{a_{\rm R}}

\newcommand{\HH}{{\rm H}}

\newcommand{\simgt}%
           {\,\hbox{\lower0.35ex\hbox{$\sim$}\llap{\raise0.35ex\hbox{$>$}}}\,}
\newcommand{\simlt}%
           {\,\hbox{\lower0.35ex\hbox{$\sim$}\llap{\raise0.35ex\hbox{$<$}}}\,}

\newcommand\yes{\textbullet}
\newcommand\no{$\circ$}

\newcommand\pit{\!\nicefrac{\pi}{2}\!}

\newcommand\gt{GTNM15\xspace}
\newcommand\bai{B17\xspace}
\newcommand\blf{BLF17\xspace}

\newcommand\aip{Leibniz-Institut f{\"u}r Astrophysik Potsdam (AIP),
               An der Sternwarte 16, 14482, Potsdam, Germany}

\newcommand\nbia{Niels Bohr International Academy, The Niels Bohr Institute, Blegdamsvej 17, DK-2100, Copenhagen \O, Denmark}

\newcommand\kitp{Kavli Institute for Theoretical Physics, University of California, Santa Barbara 93106, USA}

\newcommand\starplan{Centre for Star \& Planet Formation, Natural History
  Museum of Denmark, University of Copenhagen, \O ster Voldgade 5-7, DK-1350 Copenhagen,
  Denmark}

\newcommand\aop{Astrophysics \& Planetary Science, The Niels Bohr
  Institute, Blegdamsvej 17, DK-2100, Copenhagen \O, Denmark}

\newcommand\qmul{Astronomy Unit, Queen Mary University of London, Mile End
  Road, London E1 4NS, UK}

\newcommand\jpl{Jet Propulsion Laboratory, California Institute of
  Technology, Pasadena, CA 91109, USA}

\newcommand\mpe{Max-Planck-Institut f{\"u}r extraterrestrische Physik, Giessenbachstra{\ss}e 1, 85748 Garching, Germany}

\newcommand\uva{Department of Astronomy, University of Virginia, Charlottesville, VA 22904, USA}

\received{March 31, 2020}\revised{April 30, 2020}\accepted{\today}

\shorttitle{Hydromagnetic simulations of irradiated PPDs}
\shortauthors{Gressel et al.}

\begin{document}

\title{\bf\large Global Hydromagnetic Simulations of Protoplanetary Disks with\\
  Stellar Irradiation and Simplified Thermochemistry}


\author[0000-0002-5398-9225]{Oliver Gressel} 
\affiliation{\aip}\affiliation{\nbia}\affiliation{\kitp}

\author[0000-0002-3835-3990]{Jon P. Ramsey} 
\affiliation{\uva}\affiliation{\starplan}

\author[0000-0002-5074-7183]{Christian Brinch} 
\affiliation{\nbia}\affiliation{\aop}

\author{Richard P. Nelson} 
\affiliation{\kitp}\affiliation{\qmul}

\author[0000-0001-8292-1943]{Neal J. Turner} 
\affiliation{\kitp}\affiliation{\jpl}

\author{Simon Bruderer} 
\affiliation{\mpe}


\begin{abstract}
  Outflows driven by large-scale magnetic fields likely play an important role in the evolution and dispersal of protoplanetary disks, and in setting the conditions for planet formation.
  We extend our 2-D axisymmetric non-ideal MHD model of these outflows by incorporating radiative transfer and simplified thermochemistry, with the twin aims of exploring how heating influences wind launching, and illustrating how such models can be tested through observations of diagnostic spectral lines.
  Our model disks launch magnetocentrifugal outflows primarily through magnetic tension forces, so the mass-loss rate increases only moderately when thermochemical effects are switched on.
  For typical field strengths, thermochemical and irradiation heating are more important than magnetic dissipation.
  We furthermore find that the entrained vertical magnetic flux diffuses out of the disk on secular timescales as a result of non-ideal MHD.
  Through post-processing line radiative transfer, we demonstrate that spectral line intensities and moment-1 maps of atomic oxygen, the HCN molecule, and other species show potentially observable differences between a model with a magnetically driven outflow and one with a weaker, photoevaporative outflow.
  In particular, the line shapes and velocity asymmetries in the moment-1 maps could enable the identification of outflows emanating from the disk surface.
\end{abstract}

\keywords{accretion, accretion disks -- magnetohydrodynamics --
  radiative transfer -- astrochemistry -- methods: numerical --
  planetary systems: protoplanetary disks}


\section{Introduction} \label{sec:intro}

The planet-forming regions of circumstellar disks represent a class of astonishingly complex ecosystems where gas and dust microphysics intricately couple to large-scale dynamics \citep{2011ARA&A..49...67W}.
Only when considered together, can the dynamical state and evolution of the system be determined successfully.
In this paper, we aim to produce the most realistic computer models to-date of the inner regions of protoplanetary disks (PPDs), taking into account radiative, thermodynamic and non-ideal magnetohydrodynamic aspects simultaneously.

Stars accrete a sizable fraction of their final mass during their pre-main-sequence evolution \citep[see][for a comprehensive review]{2016ARA&A..54..135H}, and the surrounding PPD acts as a conduit for this mass.
Given the occurrence rate of T~Tauri stars with infrared excess in young stellar clusters \citep{2001ApJ...553L.153H,2009AIPC.1158....3M}, one can estimate a median disk lifetime of about three million years.
There are two fundamental drivers of disk evolution, which will ultimately lead to the dispersal of the PPD on timescales comparable to this observational constraint.
First, radiation energy deposited by the star can drive photoevaporation \citep[see][]{2014prpl.conf..475A}, that is, the dispersal of the PPD via thermochemical heating (from X-rays, EUV/FUV photons) driven winds that are primarily associated with mass loss.
Second, the redistribution of angular momentum enables accretion of disk material onto the central star with observations inferring typical rates of $\dot{M}\simeq 10^{-8\pm1}\Mdot$ \citep{1998ApJ...492..323G}.\footnote{Note that, for sufficient magnetic lever arms, direct loss of material via magnetically-powered winds is theoretically expected to remain sub-dominant compared with the accreting mass flux.}

For typical class-II objects, the disk surface density has decreased to levels where the gravitational instability cannot drive spiral features in the disk.
In this regime, angular momentum transport can therefore be sub-divided into three distinct paradigms:
\begin{enumerate}\itemsep0pt
\item enhanced viscous torques between adjacent disk annuli, e.g., caused by hydrodynamic and/or magnetorotational turbulence,
\item macroscopic Maxwell stresses due to large-scale coherent horizontal magnetic fields, \item vertical extraction of disk angular momentum by magnetocentrifugal winds \end{enumerate} \citep[see, e.g.,][for recent reviews]{2011ppcd.book..283K,2014prpl.conf..411T,2014prpl.conf..451F}.

In sufficiently ionized parts of the disk \citep{2009ApJ...703.2152T} magnetorotational instability \citep[MRI,][]{1991ApJ...376..214B,1991ApJ...376..223H} will produce turbulent Maxwell and Reynolds stresses that can account for the typically observed accretion rates \citep{1998ApJ...495..385H,2007MNRAS.376.1740K}.
This includes the warm inner regions of the PPD, i.e., inside of $\sim 0.3\au$, where thermal ionization is likely to occur \citep{2015ApJ...811..156D}, as well as the dilute outer reaches, where cosmic rays can penetrate down to the midplane \citep{2015MNRAS.454.1117S}. The latter scenario may, for instance, explain the observed relatively weak velocity dispersion in TW Hya \citep{2018ApJ...856..117F}.

However, when considering the detailed ionization structure in the inner regions of PPDs, large stretches of the disk are expected to remain laminar owing to the combined dissipative effect of ambipolar diffusion (AD) and Ohmic resistivity \citep{2013ApJ...769...76B,2013ApJ...772...96B}.
Moreover, in regions where the Hall effect is dominant, the Hall-shear instability (HSI) can nevertheless amplify large-scale horizontal magnetic field and provide a means for non-turbulent accretion.
When accounting for all three non-ideal MHD effects, a plethora of field morphologies has been obtained
\citep{2014A&A...566A..56L} --- including self-organization \citep{2013MNRAS.434.2295K,2016A&A...589A..87B,2018ApJ...865..105K}, bursts of turbulence \citep{2015MNRAS.454.1117S}, and hemispherical asymmetries.

In the presence of even weak vertical flux --- as is likely inherited from the parent molecular cloud during the initial protostellar collapse \citep[see, e.g.,][and references therein]{2014prpl.conf..173L} --- both of the aforementioned scenarios can be complemented with a picture where angular momentum is (additionally) transported in the vertical direction by means of a magnetocentrifugal disk wind. Laminar AD-dominated winds were first studied in limited local box models \citep{2013ApJ...769...76B}, and later confirmed by means of semi-global \citep{2017MNRAS.471.1488N} and global models with a prescribed temperature structure \citep{2015ApJ...801...84G} or near-isothermal equation of state \citep{2017MNRAS.468.3850S,2018MNRAS.477.1239S,2019MNRAS.484..107S}.  The most physically-complete global simulations to date have been performed by \citet[\bai]{2017ApJ...845...75B} and \citet[\blf]{2017A&A...600A..75B} --- both with simplified thermodynamics --- as well as \citet{2019ApJ...874...90W}, and we aim to compare to these works wherever possible.

Theoretical work in the context of star formation \citep[as, e.g., reviewed in][]{2007ARA&A..45..565M} often invokes an hour glass-shaped field topology with magnetic torques mediating the angular momentum during the collapse of the pre-stellar core.
An important consideration is whether all systems will eventually evolve into a self-regulated final state, where the amount of flux is limited by the microphysics, or whether there instead is a distribution of systems with a broad range of mass-to-flux ratios \citep[see discussion in][]{2017ApJ...836...46B}.

Theoretically, it is expected that, for disk formation to proceed, the molecular cloud has to shed at least part of its magnetic flux to avoid excessive magnetic braking \citep[see the discussion in section~3.1 of][]{2014prpl.conf..173L} during its collapse. On empirical grounds, the vertical magnetic flux in PPDs is likely weak.  Paleomagnetic measurements for the solar system derived from the Semarkona meteorite \citep{2014Sci...346.1089F} predict values in a range of 50 to $500\mG$ for the background magnetic field strength.  Quantitative determination of the field strength in nearby PPDs, however, remains challenging \citep{2017A&A...607A.104B,2019A&A...624L...7V}.

Since vertical fields are sub-dominant with respect to the gas pressure at the disk midplane by many orders of magnitude, thermal processes are therefore important in setting the mass loading of the disk wind
\citep{2016ApJ...818..152B}.
This implies that the thermal structure of the disk plays an important
role in setting the timescale on which the system evolves \citep{2016ApJ...821...80B,2016A&A...596A..74S,2019MNRAS.tmp.1959S}.

Consequently, we aim to improve our previous simulations presented in \citet*[hereafter \gt]{2015ApJ...801...84G} by complementing them with i) stellar irradiation, ii) diffuse reprocessing of radiation, and, iii) a simplified, tabulated thermochemical model --- the latter providing heating and cooling mechanisms driven by the FUV flux from the central star.
Fully dynamical simulations of photoevaporative hydrodynamic winds are relatively few --- notable recent exceptions being the work of \citet{2017ApJ...847...11W} which includes a detailed thermochemical treatment, as well as \citet{2018ApJ...857...57N}, who combine a chemical reaction network with radiative transfer. \citet*{2019ApJ...874...90W}, for the first time, have combined consistent dynamical thermochemistry with magnetohydrodynamics and found that outflow rates are moderately affected by thermal effects (\citet{2019arXiv191104510R} follow a similar approach but neglect AD).

We aim to complement \citeauthor*{2019ApJ...874...90W}'s work, which assumed a prescription for the ambient heating from diffuse reprocessed infrared radiation, by including more complete radiation physics. Moreover, in contrast to their ``magneto-thermal'' winds\footnote{A term coined by \bai based on the observation that his winds are launched by vertical gradients in the azimuthal magnetic field, and rotational velocities generally remain sub-Keplerian.}, we robustly find truly magneto-centrifugal (but not necessarily cold) winds, which we will henceforth call ``thermally-assisted centrifugal outflows'' (TACOs).

Beyond the determination of accretion rates, and mass-loss rates in the outflow, a side effect of conducting self-consistent non-ideal MHD simulations is to establish the importance of Joule and AD dissipation heating on the temperature structure of the PPD. This may help to constrain, e.g., the precise location of the water-ice line \citep[see][]{2017SSRv..212..813H}, which has important implications for planet formation theory \citep[e.g.,][]{2017A&A...608A..92D} and should be detectable in nearby PPDs \citep{2015ApJ...815L..15B,2018RNAAS...2c.169C} with the Atacama Large Millimeter/sub-millimeter Array (ALMA).

By means of post-processing our simulation data with chemistry and radiative transfer, we hope to derive observational signatures \citep[similar to the ones for photoevaporative winds reviewed in ][sect.~2.4]{2014prpl.conf..475A} that might help to distinguish between photoevaporative \citep[e.g.,][]{2019MNRAS.487..691P}, wind-driven and turbulent accretion mechanisms \citep[see][for recent studies of this type]{2016ApJ...831..169S,2017ApJ...847....6N,2019ApJ...870...76B}.
While we are generally interested in the overall (steady-state) structure of the forming outflows, recent observations of clumpy winds, for instance, in EX Lup \citep{2018ApJ...859..111H} also warrant a closer inspection of potential mechanisms that could produce time-variability in the mass-loading of the launching mechanism and their subsequent observational signatures.

This paper is organized as follows: In Section~\ref{sec:methods} we introduce the numerical methods, and the specific disk model is then introduced in Section~\ref{sec:model}. We present our simulation results in Section~\ref{sec:results}, and showcase the derived synthetic observables in Section~\ref{sec:synthetic}. We summarize our results in Section~\ref{sec:summary}.


\section{Numerical methods} \label{sec:methods}
The results presented in this paper are based on a set of 2D-axisymmetric radiation magnetohydrodynamic (RMHD) simulations of protoplanetary disks. Snapshots from selected simulations were post-processed with the \krome astrochemistry package \citep{2014MNRAS.439.2386G}, followed by \rad \citep{2012ascl.soft02015D} and \lime \citep{2010A&A...523A..25B} radiative transfer tools.  In the following, we outline the underlying equations and applied numerical techniques; the post-processing pipeline, and the assumptions adopted therein, are described in Section~\ref{sec:lime}.

\subsection{Simulation code} 

For the dynamical RMHD simulations, we employ the \nir fluid code, which is built around a standard second-order accurate finite-volume scheme \citep{2004JCoPh.196..393Z,2016A&A...586A..82Z}.
The version of the \nir code that we use here is derived from the curvilinear generalization of the method described in \citet{2011JCoPh.230.1035Z}, and moreover features the second-order accurate Runge-Kutta-Legendre (RKL2) super-time-stepping scheme of \citet{2012MNRAS.422.2102M} for various parabolic source terms, that are separated from the main update via Strang splitting.
For our purposes here, the code operates on uniformly-spaced meshes in spherical-polar coordinates $(r,\theta,\phi)$ representing spherical radius, co-latitude and azimuth (not used here), respectively.\footnote{The same 2D mesh was used by \rad, while the data was interpolated onto a 3D Delaunay graph by \lime, where the interpolation is set to concentrate resolution elements near gradients in optical depth for improved performance and accuracy.}

For obtaining interface states entering the magnetohydrodynamic fluxes, we use the HLLD approximate Riemann solver introduced by \citet{2005JCoPh.208..315M}. In \nir, all electromotive forces are staggered according to the constrained transport paradigm \citep{1988ApJ...332..659E}, with the result that the divergence-free constraint of the magnetic field is maintained to within machine precision. Our implementation deviates from the public distribution of \nir in that we adopt an upwind reconstruction \citep{2008JCoPh.227.4123G} to interpolate the edge-centered electric field on the curvilinear mesh \citep{2010ApJS..188..290S}. This allows us to directly use the quantities obtained via the HLLD solver, without having to formulate a two-dimensional Riemann problem at the cell edges.

When considering a non-ideal plasma, situations may arise where the collisional coupling frequencies between the various charged and uncharged species are similar to relevant dynamical timescales.
In general, such cases may demand a treatment where each class of particles possesses its own inertia.
Multi-fluid MHD simulations by \citet{2016MNRAS.463..134R}, however, suggest that for typical number densities found in disks around T~Tauri stars, it is fair to assume the strong-coupling limit. Accordingly, the equations reduce to those of a single-fluid, which tracks the motion of the neutral component.

\subsection{Equations of motion} \label{sec:eom} 

The equations implemented in \nir are formulated in conserved variables $\rho$ (density), $\rho\V$ (momentum), $e\equiv \epsilon + \rho\V^2\!/2 + \B^2\!/2$ (total energy), and $\B$ (magnetic flux density). Moreover, we have equations for  $\epsilon$ (internal energy density), and $\eR$ (radiation energy density). Suppressing explicit factors of the vacuum permeability, $\mu_0$, the system of equations reads
\begin{eqnarray}
                                                            \label{eq:rho}
  \partial_t\rho +\dv(\rho \V) & ~=~ & 0                    \,,       \\[6pt]
  \partial_t(\rho\V) +\dv\Fm & ~=~ & \Sm                    \,,       \\[6pt]
  \partial_t \B -\!\cl\big(\,                               \label{eq:ind}
     \V\tms\B + \Ediss \,\big) & ~=~ & \mathbf{0}           \,,       \\[6pt]
  \partial_t e + \dv\Fe                                     \label{eq:ene}
       & ~=~ & \Se + \Sm\cdt\V                              \nonumber \\
       & & +\,\nabla\cdot\big(\,\Ediss\tms\B\,\big)\,,                \\[6pt]
  \partial_t \epsilon + \dv (\epsilon\V) + p\dv\V           \label{eq:dual}
      & ~=~ & \Se + \etao\,\big|\cl\B\big|^2                \nonumber \\
      & & +\,\etad\,\big|\,\bb\tms\cl\B\,\big|^2,\qquad               \\[6pt]
  \frac{c}{\hat{c}}\,
  \partial_t \eR + \dv (\eR\V) & ~=~ &                      \label{eq:erad}
             c\rho\,\kP\! \left(\aR T^4\!-\eR\right)        \nonumber \\
             & & - \dv\Fr \,- \pR\!:\!\nabla\V              \,,
\end{eqnarray}
where $\Ediss$ is the dissipative electromotive force (see \Section{thermo} for details), $\Fr$ denotes the radiation flux, $\pR$ is the radiation pressure tensor, $\aR\equiv 4\, \sigma/c$ is the radiation constant, $\sigma$ is the Stefan-Boltzmann constant, $c$ is the speed of light, and $\kR$ and $\kP$ are the Rosseland and Planck mean opacities, respectively. In the above equations, the fluxes, $\mathcal{F}$, and source terms, $\mathcal{S}$, for the momentum and energy equations are given by
\begin{eqnarray}
  \Fm & ~=~ & \rho\mathbf{vv}+p^{\star} \mathbb{I}-\!\mathbf{BB}\,, \\[6pt]
  \Fe & ~=~ & (e\!+\!p^{\star})\V - (\V\cdt\B)\,\B\,,       \\[6pt]
  \Sm & ~=~ & - \rho \nabla\Phi + \rho\kR\,c^{-1}\,\Fr\,,   \\[6pt]
  \Se & ~=~ & - c\rho\,\kP\! \left(\aR T^4\!-\eR\right) + \Qirr + \Qpdr\,,
\end{eqnarray}
respectively, where $p^{\star}\equiv p+B^2/2$ is the total pressure, and where the gravitational potential $\Phi(r) \equiv -G M_\star/r$ is that of a simple point-mass.

\subsubsection{Dual vector potential} \label{sec:vp} 

The advective and dissipative electromotive forces computed for advancing \Eqn{ind} can conveniently be used to advance a magnetic vector potential, $\A(\rr,t)$, in a gauge-agnostic fashion via
\beq
  \partial_t \A - \big(\,\V\tms\B + \Ediss \,\big) ~=~ \mathbf{0} \,.
  \label{eq:vp}
\eeq
This is possible because $\A$ itself is not used anywhere in the system of equations (\ref{eq:rho})--(\ref{eq:erad}), but simply integrated alongside the other variables, solely using source terms computed from $\B$ \citep[also see appendix~B of][]{2012ApJS..199...13R}. We periodically evaluate the diagnostic quantity $\B^{\star}\equiv \cl\A$, and find that $|\B^{\star}-\B|\,/\,|\B|$ evolves in a similar manner as the $|\dv(\B)|\,/\,|\B|$ error.

In axisymmetry, a useful property of the azimuthal component $A_\phi(r,\theta)$ is that isocontours of the flux function $\Psi\equiv r\sin(\theta)\,A_\phi$ trace poloidal field lines. Compared to integrating field lines \emph{a posteriori} using $\B_{\rm p}(r,\theta)$, this happens in an ``absolute'' sense. Hence, by selecting a specific potential surface $\Psi=\Psi_0$, one can follow the motion of a given field line in time --- informing us about the overall global evolution of the magnetic flux.

\subsubsection{Dual energy formalism} 

Note that Equations~(\ref{eq:ene}) and (\ref{eq:dual}) are strictly speaking redundant. We exploit this redundancy in the form of a dual energy formalism. While the conservative formulation of (\ref{eq:ene}) is generally preferable, recovering the thermal energy density as $\tilde{\epsilon} = e - \rho\V^2\!/2 - \B^2\!/2$ can become inaccurate in regions of high Mach number and/or low plasma parameter -- that is, where the kinetic and/or magnetic energy dominate the total energy. In extreme cases, the derived internal energy $\tilde{\epsilon}$ can even become negative. To avoid this issue, which is intrinsic to floating point arithmetic, we override the internal energy whenever $\tilde{\epsilon}\le 0.05\, e$, and instead use the explicit value of $\epsilon$ evolved via Equation~(\ref{eq:dual}). While this sacrifices exact energy conservation, it greatly benefits the numerical robustness of the integration scheme.
To maintain consistency, $\epsilon$ is re-initialized (i.e., during each cycle) with $\tilde{\epsilon}$ in cells where the above criterion is not met.

\subsubsection{Thermodynamics \& dissipative MHD terms} \label{sec:thermo} 

As for thermodynamic relations, we assume an ideal gas, that is, $T\equiv \bar{\mu}m_{\rm H}\,k_{\rm B}^{-1}\;p/\rho$ is the gas temperature, with the gas pressure $p = (\gamma-1)\epsilon$. We assume constant values $\bar{\mu}=2.34$ and $\gamma=7/5$ representing a diatomic mixture of molecular hydrogen and helium gas. This specifically means that we ignore the effect that, for instance, hydrogen dissociation has on the adiabatic index, $\gamma$, and mean molecular weight, $\bar{\mu}$.

For the sake of building-up a firm understanding of the underlying physics, we ignore the complications associated with the Hall effect in the current paper.
The dissipative electromotive force, $\Ediss\equiv \EO\!+\!\EAD$, appearing in the induction \Eqn{ind} is due to Ohmic resistivity, $\EO$, and ambipolar diffusion, $\EAD$, given by
\begin{eqnarray}
  \EO  & \equiv & -~ \etao \,(\cl\B)\,,\quad\textrm{and}
  \label{eq:emf_o} \\
  \EAD & \equiv & +~ \etad \,\big[\,(\cl\B)\times\bb\,\big]\times\bb\,,
  \label{eq:emf_ad}
\end{eqnarray}
respectively, where $\bb$ denotes the unit vector along the direction of the magnetic field. Note the presence of the divergence of the Poynting flux, $\Ediss\tms\B$, in the total energy equation~(\ref{eq:ene}), which accounts for the ``non-locality'' of the Joule heating (as well as the dissipation via ambipolar collisions) in the \emph{total} energy formalism. Note also that these two effects appear explicitly as $\etao\,|\cl\B|^2$ and $\etad\,|\bb\tms\cl\B|^2$ in the \emph{internal} energy equation~(\ref{eq:dual}).

It is instructive to study the impact of various heating mechanisms on the thermal structure of the disk. To this end, we can choose to disable the non-ideal MHD heating (see $\Qd$ in \Tab{models}). This is achieved by omitting the $\etao\,|\cl\B|^2$ and $\etad\,|\bb\tms\cl\B|^2$ terms in \Eqn{dual}, and \emph{subtracting} them from \Eqn{ene}. We stress that it is \emph{not} sufficient (and, in fact, inconsistent) to omit $\nabla\cdot(\Ediss\tms\B)$ in \Eqn{ene}, since dissipation of magnetic fields in \Eqn{ind} is intrinsically converted into thermal energy by virtue of the total energy formulation.

In both the internal and total energy equations, external heat sources from frequency-integrated stellar irradiation (in the visible), and from a simplified FUV thermochemical model, appear as source terms, denoted by $\Qirr$ and $\Qpdr$, respectively. These are discussed in the following sections.

\subsection{Radiative transfer and stellar irradiation} \label{sec:radi} 

The flux-limited-diffusion (FLD) approach taken here treats the radiation field's angular variation less precisely than characteristics-based methods \citep[e.g.][]{2012ApJS..199...14J}.
However, in combination with ray-tracing for a central radiation source, its merit has been demonstrated for providing a reasonable approximation to the thermodynamics of passively irradiated disks \citep[see, e.g.,][in the context of star formation and PPDs]{2000ApJ...539..258R,2010A&A...511A..81K,2013A&A...549A.124B,2013A&A...555A...7K,2015A&A...574A..81R}.
We here only briefly outline this hybrid ray-tracing + FLD scheme, that we employ for radiative transfer within the dynamical MHD simulations. A more detailed account of the method can be found in \citet{2017JPhCS.837a2008G}.
Extending this previous work, we complement the diffusive fluxes with a simple two-stream approximation in the vertical direction \citep{2007ApJ...665.1254B}, which improves the temperature structure in the optical transition region.

\subsubsection{Radiation diffusion} 

In the set of equations (\ref{eq:rho})--(\ref{eq:erad}), listed in \Sec{eom}, the radiation energy flux and radiation pressure tensor need to be specified in the form of a closure relation.
An elegant but effortful way to do this is the variable Eddington tensor method \citep{1992ApJS...80..819S,2012ApJS..199...14J}, where one solves for the time-independent angular distribution of the radiation intensity and integrates the moments, at each grid location, to compute the Eddington factor.
Another popular method is to assume that the radiation flux can be approximated by the gradient of the radiation energy density, that is,
\beq
  \Fr = -\lambda(R)\,\frac{c}{\rho\kR}\nabla\,\eR\,,
  \label{eq:flux}
\eeq
leading to a purely diffusive redistribution of energy. In \Eqn{flux}, the diffusion coefficient can be identified as $D\equiv \lambda(R) c /\rho\kR$, where we have introduced a dimensionless quantity $R\equiv |\nabla\eR| / (\rho\kR \eR )$, which traces how sudden or gradual $\eR(\rr)$ varies in relation to the photon mean-free-path expressed in terms of the extinction coefficient $\alpha\equiv \rho\kR$.

Going back to the work of \cite{1981ApJ...248..321L}, a limiter function $\lambda(R)$ is introduced to guarantee that $|\Fr|$ remains smaller than $c\,\eR$ (as mandated by causality) in regions of vanishing optical depth.
It is easy to show that $\lambda(R)$ needs to go as $1/R$ for $R\rightarrow \infty$, and $\lambda(R)\rightarrow 1/3$ for $R\rightarrow 0$, corresponding to the classical Eddington approximation.
Here we use the limiter function suggested by \citet{1989A&A...208...98K}. In this framework, the radiation pressure is given by
\beq
  \pR \equiv \Big[\; \frac{1}{2}\,(1-f_{\rm edd})\,\mathbb{I}
    \,+\, \frac{1}{2}\,(3f_{\rm edd}-1)\,\nn\nn \;\Big]\, \eR\,,
\eeq
with a scalar Eddington factor $f_{\rm edd} \equiv \lambda + \lambda^2 R^2$, and with $\nn\equiv \nabla\eR/|\nabla\eR|$ the normal vector along the energy density gradient \citep{2001ApJS..135...95T}.

\subsubsection{Method of discrete ordinates} 

We have found that FLD alone is not sufficient to capture the vertical temperature structure of the disk near the optical depth transition.
As a remedy, we have implemented a first-order variant of the short characteristic method by \citet{1987JQSRT..38..325O} --- see their eqns.~(15) and (16).
This type of method has been shown by \citet{2007ApJ...665.1254B} to provide a good first approximation to the vertical cooling in the context of PPDs \citep[also see][]{2006A&A...448..731H}.
For simplicity, we integrate along the $\theta$~direction of our computational grid (rather than the vertical direction), which greatly reduces complexity and communication overhead, and which is justifiable as long as we only apply the solution near the midplane.

Tracing downward (upward) rays with intensity $I_+$ ($I_-$), and inclination $\mu\equiv \cos(\theta_{\rm r}) = 1/\sqrt{3}$, and assuming a spatially uniform source function $S_0\equiv c\,\eR/4\pi$, and extinction coefficient $\alpha_0\equiv\rho\kappa$ within a cell, we obtain
\begin{eqnarray}
  I_+^{\,i,j} & = & I_+^{\,i,j-1}\;{\rm e}^{-\Delta\tau/\mu}
                + (1-{\rm e}^{-\Delta\tau/\mu})\;S_0^{\,i,j-1/2},    \\[2pt]
  I_-^{\,i,j} & = & I_-^{\,i,j+1}\;{\rm e}^{-\Delta\tau/\mu}
                + (1-{\rm e}^{-\Delta\tau/\mu})\;S_0^{\,i,j+1/2},\quad
\end{eqnarray}
with $\Delta\tau/\mu=\alpha_0\,r\Delta\theta/\mu$ being the slanted optical depth (i.e., along the ray direction) across the grid cell.

Domain decomposition is handled as in \citet{2006A&A...448..731H}, that is, first evaluating the integrals within each domain, and then augmenting with the attenuated intensity from blocks above (below), before sending the corrected intensity to the next adjacent block.
This interleaves computation and communication for somewhat improved scaling efficiency.
For models with reflective symmetry at the midplane, the initial condition for the upward ray is simply taken as the (flipped) final value of the downward ray.

With $I_+$ and $I_-$ computed over the entire grid, we then evaluate the flux at vertical cell interfaces as
\beq
  F_{\theta\,{\rm ray}}(r,\theta) \,\equiv\, 2\pi\mu\,(I_+ - I_-)\,,
\eeq
and we override $F_{\theta\,{\rm fld}}(r,\theta)$, obtained via FLD, with this solution at the vertical optical depth-transition region in our disk. The tapering is based on the FLD equivalent of the Eddington factor, $f_{\rm edd}(R)$, which characterizes the optical regime.
More specifically, we chose
\beq
  F_{\theta}(r,\theta) =
  f_{\rm op}\;F_{\theta\,{\rm fld}} \,+\,
  (1-f_{\rm op})\;F_{\theta\,{\rm ray}}\,,
\eeq
with $f_{\rm op}\equiv 0.5 + 0.5\; {\rm erf}(\,16 (f_{\rm edd}-\mu)\,)$, providing a gradual transition between the two fluxes. In Appendix~\ref{sec:relax}, we check our implementation against the relaxation benchmark presented in Sect.~4.1 of \citet{2007ApJ...665.1254B}.

\subsubsection{Reduced speed of light approximation} 

Note that we integrate the diffusion term in a time-explicit fashion. To render this feasible, we make use of the reduced-speed-of-light approximation \citep[RSLA, ][]{2001NewA....6..437G}.
This method has been applied, in the context of an M1 closure, in simulations of the interstellar medium \citep{2013ApJS..206...21S}.

The RSLA is valid as long as the radiation-diffusion timescale resulting from the adopted artificial value of $\hat{c}$ in \Eqn{erad} is short compared to any other relevant dynamical timescale.
During the T~Tauri phase of low-mass protostars, radiative effects are important for setting a consistent temperature structure within the disk, while true radiation hydrodynamic effects (such as driving of winds by radiation pressure) are likely less important \citep{1998apsf.book.....H}.
Because the factor $c/\hat{c}$ only enters the partial time derivative on the LHS of \Eqn{erad}, it is implied that the steady-state solution, where $\partial_t\rightarrow 0$, is recovered in an exact manner. However, the RSLA changes the timescale on which this solution is achieved.
To integrate the radiation diffusion term, which can become stiff, we employ the second-order accurate RKL2 scheme of \citet{2012MNRAS.422.2102M} mentioned earlier, which is also used for updating Ohmic diffusion and AD.

As part of the individual RKL2 sub-steps, and deviating from the treatment adopted in \citet{2017JPhCS.837a2008G}, we here integrate the radiation-matter coupling term, $c\rho\,\kP\! \left(\aR T^4\!-\eR\right)$, using the fully-implicit update described in \citet{2013ApJS..206...21S}.
By combining initial bisection with subsequent Newton-Raphson iteration, the solver typically converges within a few steps.

\subsubsection{Irradiation heating} 

It has recently become clear that the thermal disk structure plays an important role in setting the mass loading of the disk wind \citep[e.g.][]{2016ApJ...818..152B}, and ultimately the evolution timescale of the system \citep{2016ApJ...821...80B,2016A&A...596A..74S}.
Yet our previous global disk models \citep{2013MNRAS.435.2610N,2013ApJ...779...59G,2015ApJ...801...84G} have either assumed an isothermal temperature $T=T(\Rc)$, constant on cylinders, or have used  adiabatic treatment combined with a so-called ``$\beta$ cooling'' --- reinstating the nominal $T(\Rc)$ on a timescale comparable to $2\pi\Omega^{-1}$.

In contrast, even classic models of dust absorption and re-radiation of star
light in the disk surface \citep{1997ApJ...490..368C}, arrive at a more complex temperature distribution within the PPD.

To account for this, we include a frequency-integrated irradiation flux of the form
\beq
  \Fr\,\!_{,\rm irr}(r) \equiv F(r_\star)\;(r_\star/r)^2\; \exp\,\left(-\tau_\star(r)\right)\,\hat{\mathbf{r}}
\eeq
\citep[e.g.][]{2015A&A...574A..81R}, which is attenuated by the optical depth $\tau_\star(r)$ towards the central star.
Following \citet{2013A&A...555A...7K}, we obtain the energy source term as
\beq
  \Qirr(r,\theta) \equiv - \nabla\cdt \Fr\,\!_{,\rm irr}(r,\theta)\,.
\eeq
In grid cells across which the optical depth is low (that is, $\Delta\tau_\star\le 10^{-3}$), we instead use the integral formulation
\beq
  \Qirr(r_i) \equiv \frac{3\,\rho\kP}{r^3_{i+\frac{1}{2}} \!-r^3_{i-\frac{1}{2}}}
  \int_{r_{i-\frac{1}{2}}}^{r_{i+\frac{1}{2}}}
  \hat{\mathbf{r}}\cdt\Fr\,\!_{,\rm irr}(r')\,r'^2{\rm d}r'\,.
\eeq
This formulation has been found to provide a more accurate solution on the discretized mesh when differences across cells are small \citep{1999A&A...348..233B}.
To keep matters tractable in terms of complexity, we restrict ourselves to using a fixed, wavelength-independent opacity coefficient $\kappa_\star = \kP = \kR$ within the MHD simulations.

\subsection{Simplified thermo- and photo-chemistry} \label{sec:pdr} 

Direct heating due to stellar irradiation is not the only means by which radiation affects the thermal structure of the disk.
Thermo- and photo-chemical effects, such as photoelectric heating, cooling due to line radiation or H$_2$ formation and dissociation can dominate the temperature structure of the disk under certain conditions \citep[see, e.g.,][]{2009A&A...501..383W}.
The surface layers of a PPD irradiated by FUV photons is, in fact, similar to a photon-dominated region (PDR) within the interstellar medium: radiation impinging on a substantial matter density gradient results in a temperature gradient and a layered chemical structure where species are sequentially dissociated or ionized as the attenuation decreases, that is, as the intensity of the radiation field increases \citep[e.g.,][and references therein]{2007A&A...467..187R}. To include a fully self-consistent treatment of photo- and thermo-chemistry in multi-dimensional, dynamic, RMHD simulations is, however, very restrictive and can indeed dominate the cost of the simulation.
Here, we instead take a different approach and develop a simplified and efficient thermo- and photo-chemical model based on, and calibrated against, comprehensive PDR models \citep{2007A&A...467..187R}.
The simplified treatment provides reasonable heating and cooling rates and therefore a more realistic temperature structure in the disk, while taking into account the dynamically changing environment.

As described in detail in Appendix \ref{app:simple_pdr}, this PDR module solves for the net heating/cooling rate given the local FUV radiation field, the visual extinction $A_{\rm V}$, the gas density and temperature, plus the vertical and radial column densities of certain chemical species (see Sect.~\ref{app:chemistry}).

The PDR module not only returns the net heating/cooling rate, but also the updated abundances of the chemical species (used in turn for calculating the shielding column densities).
In the version of the module used here, the included species are H, H$_2$, C$^+$, C, CO, O and e$^-$.
The inclusion of line cooling rates from these species, plus several other heating/cooling processes (see Sect.~\ref{app:heatcool}) is sufficient to recover the temperature structure of a standard PDR model but at a small fraction of the cost of a typical fully-fledged PDR code (see Sect.~\ref{app:pdr_benchmark}).

In the models presented here, the PDR module is called once every five computational RMHD steps. The resulting heating/cooling rate, $\Qpdr$, is stored (advection with the flow can be enabled, but is currently ignored) and applied as a source term as part of the implicit radiation-matter coupling update during each cycle. The effect of $\Qpdr$ is naturally also included in the calculation of the permissible simulation time step.

\subsection{Improvements to the ionization model} \label{sec:ion} 

Our approach to modeling the ionization state is described in detail in section~2.1 of \gt. Here, we only briefly recapitulate its main features and highlight modifications.

The diffusion coefficients $\etao(\rr,t)$ and $\etad(\rr,t)$ are updated every five computational cycles.
As in our previous work, we use a look-up table derived from a simplified equilibrium ionization chemistry with a minimal set of gas-phase reactions but accounting for grain charging.
The network is based on \texttt{model4} of \citet{2006A&A...445..205I},  with one representative metal and one molecular ion --- also see section~2.2. of \citet{2013ApJ...771...80L} and section~4.2 of \citet{2013ApJ...764...65M}.

As in our previous simulations, we do not apply any cap on the coefficient $\etao$ \citep[see the discussion in][appendix B1]{2012MNRAS.422.1140G}.
To mitigate against severe timestep constraints, we limit $\etad$ such that
$\Lambda_{\rm AD}\,\beta_{\rm p} \ge 0.001$, where $\beta_{\rm p}\equiv 2p/B^2$ is the plasma parameter, and $\Lambda_{\rm AD} \equiv v_{\rm A}^2/(\Omega\,\eta_{\rm AD})$.
We have performed a reference run without such a limiter, and have found that the results are not affected by the procedure.

\subsubsection{Attenuation of X-rays/FUV and self-shielding} \label{sec:XR} 

The chemical equilibrium state critically depends on the amount of ionizing radiation permeating the disk.
To this end, shielding columns are integrated along radial, and vertical\footnote{Recall that ``vertical'' rays follow the $\theta$ coordinate.} rays.
Note that the radial gas column, $\Sigma_\star(\rr,t)$, contains a contribution from a ``virtual'' inner disk (i.e. with $r\le r_{\rm in}$) not covered by our computational grid.
We instead use $\rho(\rr,0)$ given in \Eqn{rho_init} down to an inner truncation radius of $r_{\rm tr}=5 R_\odot$.
When integrating shielding columns for the species entering the PDR thermochemistry (see \Sec{pdr}), we assume spatially uniform fractional abundances (along the ray) beyond the active domain.

Vertical shielding involves two sets of rays originating from above and below the disk, respectively.
In the case where we only evolve \emph{one} hemisphere, we first compute rays from the disk surface to the midplane and then augment this column to the second set of rays originating from below the disk.

For the absorption of the direct X-ray component,
we use a simple fit to the results of \citet{1999ApJ...518..848I} based on the mass column \citep[as previously, all coefficients are adopted from][]{2009ApJ...701..737B}.
For consistency with the recent work of \citet{2017ApJ...845...75B} --- see their equation (15) -- we now also introduce a geometric factor of five in $\Sigma_{\rm X,abs}$ when attenuating the direct X-ray component to account for the fact that the original work had considered a vertical shielding column.

\subsubsection{FUV ionization layer} \label{sec:fuv} 

In \gt, the effect of the FUV ionization layer \citep[see][for details]{2011ApJ...735....8P} was based on the assumption that all gas-phase carbon and sulfur atoms --- which are most susceptible to losing electrons when struck by energetic photons --- are ionized.
The prescription had a constant ionization fraction of $f=2\times10^{-5}$ below an assumed $\Sigma_{\rm FUV}=0.03\g\,\cm^{-2}$, and a collision rate of $2\ee{-9} \m^3\s^{-1}$.
For the models presented here, we supersede this parametric prescription and we instead compute $\etao$ and $\etad$ directly from the number density of $e^{-}$ and $C^{+}$, obtained self-consistently from the PDR module (see \Sec{pdr}), which assumes a relative gas-phase abundance of $\chi_{\rm C}=10^{-4}$ for carbon.

As discussed in section 2.4.3 of \citet{2017ApJ...845...75B}, we boost the ionization fraction $x_e(\rr,t)$ by a factor
\beq
  \propto \exp\left( \frac{0.3\, \Sigma_{\rm FUV}}
                  {\Sigma_\star(\rr,t) + 0.03 \Sigma_{\rm FUV}} \right)\,,
\eeq
to mimic ideal-MHD behavior at very low densities, and $\etao,\etad \propto x_e^{-1}$ are attenuated accordingly.

While the flux from scattered FUV photons may be significant \citep{2011ApJ...739...78B}, it is currently not feasible to include this process in our dynamical treatment.
Meanwhile, due to its lower amplitude compared to FUV photons from the star, we neglect the ambient illumination of the disk by any nearby massive stars.

\subsection{Post-processing: Astrochemistry, continuum, and line radiative transfer} \label{sec:lime} 

In order to derive observational predictions from our models, we post-process the simulation outputs using a combination of dust radiative transfer, astrochemistry and line radiative transfer modelling. This is accomplished using a Python-based pipeline to seamlessly translate the \nir simulation output into the formats required by the various other codes that we employ.

The first step is to calculate the temperature of the dust using an accurate radiative transfer method. For this task we use the well-proven \rad code \citep{2012ascl.soft02015D}.
Dust densities are taken from \nir, where we co-evolve a set of dust fluids (with individual mass density $\varrho_i$ and velocity $\U_i$) via
\begin{eqnarray}
  \partial_t \varrho_i + \dv(\varrho_i \U_i) & \,=\, & 0\,, \\
  \partial_t ( \varrho_i \U_i ) + \dv( \varrho_i \U_i \U_i )  & \,=\, &
  - \varrho_i \nabla\Phi + \mathbf{F}_{\rm d} ( \V - \U_i )\,,\quad
\end{eqnarray}
using second-order upwind Godunov fluxes for the transport step. Neglecting the back-reaction onto the gas, the aerodynamic drag force, $\mathbf{F}_{\rm d}(\V-\U_i)$, is implemented analytically \citep[as described in section~2.4 of][for particles] {2010MNRAS.409..639N} and covers both the Epstein and Stokes regimes, assuming a solid density of $\rho_{\rm s}\eq2\g\cm^{-3}$.
We arbitrarily chose $n_{\rm d}\eq 4$ species, representing particles of sizes $20\um$, $80\um$, $0.32\mm$, and $1.28\mm$, respectively.
For simplicity, all species are initialized with the same density $\varrho_i=0.01\,\rho/n_{\rm d}$, i.e., amounting to a total dust-to-gas mass ratio of one percent.

Dust opacities for \rad are obtained from the adopted grain sizes and the application of a modified version of the Mie-scattering code by B.~Draine \citep[see][]{1983asls.book.....B}. The optical data used corresponds to pyroxene ($\mathrm{Mg_{0.7}Fe_{0.3}SiO_3}$; \citealt{1995A&A...300..503D}) and was obtained from the Jena Dust Database\footnote{\url{https://www.astro.uni-jena.de/Laboratory/OCDB/}}.

Once we obtain the dust temperatures from \rad, we apply the \krome \citep{2014MNRAS.439.2386G,2017MNRAS.466.1259G} astrochemistry package as a consistency check of the abundances obtained from the simplified thermochemistry scheme described in \Sec{pdr}.
This step is not strictly required for post-processing, but with \krome we can easily extend the chemical network to include many more molecular species than would be possible with the simplified thermochemistry module --- albeit at the expense of losing the direct link with the dynamics. Here, we use a chemical network derived from the KIDA database\footnote{\url{http://kida.obs.u-bordeaux1.fr/}}, with ``not recommended'' reactions and all species containing more than three carbon atoms removed; the network used corresponds to the state of the KIDA database in January 2017. The network also includes simplified photochemistry for all species in the network that have cross-sections available \citep[i.e., in][]{2017A&A...602A.105H}. We integrate the cross-sections over frequency to get photodissociation and photoionization rates, assuming a two-component blackbody comprised of a normal stellar contribution (see \S \ref{sec:model}) and an accretion hotspot with a given FUV luminosity, $L_{\rm FUV}$ (see \S \ref{sec:uv}). The shielding against/attenuation of the FUV radiation at any given location is calculated using a simple exponential and shielding function of the form
\beq
  \propto \exp(\-\gamma_{\rm exp}A_{\rm V}),
  \label{eq:fuv_shielding}
\eeq
where $A_{\rm V}$ is the visual extinction, $\gamma_{\rm exp}$ is the shielding function, and values of the shielding function are from \citet{2017A&A...602A.105H} for a 4000K blackbody. The visual extinction is calculated by integrating the total hydrogen density along the column to a given cell and applying the typical conversion factor ($A_{\rm V} = N_{\rm tot} \cdot 6.3\times10^{-22}\cm^2$).
We run \krome for $10^3\yr$ on every cell from an RMHD simulation snapshot, but with the temperature given by the \rad calculation. The initial abundances for the chemical models are taken from \citet[][Tab.~6]{2009ApJS..183..179B} and otherwise set to their solar values \citep[e.g.,][]{2009ARA&A..47..481A}.

With the chemical abundances in hand, we then run the molecular excitation and line radiative transfer code \lime. In order to calculated the expected emergent spectral lines, \lime needs the gas and dust densities and temperatures, as well as the chemical abundances and the velocity field, all of which are provided by either a previous step in the post-processing pipeline or by the RMHD simulation itself. \lime calculates spectra at arbitrary spatial and spectral resolution, and the final step of the post-processing is to run these spectra through \textsc{casa} \citep{2007ASPC..376..127M} to convolve the images with a beam size and subtract the continuum, before finally arriving at synthetic observations of spectral lines. The resulting spectral data cubes can either compared with existing observational data or be used to make predictions for future targeted observations.


\section{Model description} \label{sec:model}

We adopt a system of code units where we measure length in astronomical units, mass in solar masses, and where we choose time units such that $G\equiv1$, implying that velocities are in terms of the orbital velocity at radius $\Rc_0$. We here limit ourselves to the case of a solar-mass star (that is, $M_\star \eq \Msun$), which we furthermore assume to have an effective surface temperature of $T_\star=4400\K$, and a surface radius of $R_\star=2\Rsun$, representing a typical solar-like star in its T~Tauri phase.

\subsection{Equilibrium disk model} \label{sec:ic} 

Our fiducial model assumes the same equilibrium disk structure as used in the previous simulations of \gt. With the square of the isothermal sound speed, $\cS^2(\!\Rc)\equiv c_{\rm s0}^2\, (\Rc/\Rc_0)^{\bar{q}}$, constant on cylinders, and $\Rc_0=1$, the initial disk structure is given by
\begin{eqnarray}
  \rho(\rr,t_0) &\,=\,& \rho_0\, \Big( \frac{\Rc}{\Rc_0} \Big)^{\!\bar{p}}
                        \exp{\left(\,\frac{G M_\star}{\cS^2(\!\Rc)}
                        \left[ \frac{1}{r} -\!\frac{1}{\Rc}
                          \right]\,\right)}\,,
  \label{eq:rho_init} \\[4pt]
 v_\phi^2(\rr,t_0) &\,=\,&  \Big( 1+\bar{q}
                                 + (\bar{p} + \bar{q}) \frac{H^2}{\Rc^2}
                                 - \frac{\bar{q} \Rc}{r}\,\Big)
                                 \frac{GM_\star\!}{\Rc}\,,\quad
  \label{eq:vphi_init} \\[4pt]
  \epsilon(\rr,t_0) &\,=\,& \cS^2(\!\Rc)\,\rho(\rr,t_0)/(\gamma-1)\,,
  \label{eq:ene_init} \\[4pt]
  \eR(\rr,t_0) &\,=\,& \aR\Big( \bar{\mu}\,m_{\rm H}\,k_{\rm B}^{-1}\,
                                 \cS^2(\!\Rc) \Big)^4\,,
  \label{eq:erad_init} \\[4pt]
  \B(\rr,t_0) &\,=\,& B_0\, \Big( \frac{\Rc}{\Rc_0} \Big)^{\!\bar{n}}\;\zz
  \label{eq:mag_init}\,,
\end{eqnarray}
with $B_0^2 \equiv 2\rho_0\,c_{\rm s0}^2 / \bp(t_0)$ and $\bar{n}\equiv (\bar{p}+\bar{q})/2$.
Our fiducial model has a radial power-law index for the midplane gas density of $\,\bar{p}=-1.5$, and is slightly flared with a power-law index of $\,\bar{q}=-0.75$ for the gas temperature as a function of cylindrical radius.
Note that, in order to avoid divergent behavior towards the coordinate axis, all power laws switch to constants for $\Rc \le \Rc_1 = r_{\rm in}$, that is, they are truncated to their value at that location.

For the initial gas density, $\rho(\rr,t_0)$, we choose $\rho_0$, such that $\Sigma_{\rm g}\simeq340{\rm \,g\,cm^{-2}}$ at $1\au$, and $\Sigma_{\rm g}\simeq140{\rm \,g\,cm^{-2}}$ at $10\au$, respectively. Note that we limit the dynamic range in density such that $\rho(\rr)> 10^{-8} \rho_0\,(\Rc/\Rc_0)^{\bar{p}}$ at any given radius, $\Rc$, which is moreover enforced during the simulation. This is required to avoid unreasonably low densities and the associated low plasma parameter that severely restricts the computational time step, mostly because of the increasing Alfv{\'e}n speed.

With the convention $H\equiv \cS\Omega^{-1}$, the initial disk aspect ratio (specified via $c_{\rm s0}$) is $H/\Rc=0.05$ at $1\au$, and increases to $\simeq 0.07$ at $20\au$. The radiation energy, $\eR(\rr,t_0)$, is initially set such that the radiation temperature equals the gas temperature. The initial magnetic field, $\B(\rr,t_0)$, is purely vertical, and specified in terms of its midplane plasma parameter, $\bp(t_0)$, which is $10^4$ for our fiducial model.  The initial condition (\ref{eq:mag_init}) for $\B(\rr,t_0)$ is implemented in terms of the vector potential (omitting powers of $\Rc_0=1$)
\beq
  A_\phi(\Rc) = B_0\,\Bigg\{
  \begin{array}{lr}
    \frac{1}{2} \Rc\, \Rc_1^{\bar{n}}
    \hfill\textrm{if~} \Rc \le \Rc_1 \\[6pt]
    \frac{1}{\bar{n}+2}\,\Rc^{\bar{n}+1} +\; C(s)
    \quad\textrm{if~} \Rc > \Rc_1
  \end{array}\Bigg. \,,
\eeq
with an integration ``constant'' $C(s) \equiv \frac{\bar{n}}{2(\bar{n}+2)}\, \Rc^{-1}\, \Rc_1^{(\bar{n}+2)}$ demanded by continuity at $\Rc=\Rc_1=r_{\rm in}$.

Preliminary simulations showed that the developing wind creates a considerable bow shock, when expanding into a disk atmosphere at rest, that is, barring its rotational velocity. While this did not prompt any difficulties in our previous near-isothermal simulations, it leads to undesirable behavior when including radiative and thermochemical physics as we do here. We therefore prescribe an initial poloidal velocity $v_r(\rr,t_0) = v_0\, H\,r/\Rc^2$, with $v_0 = 1/2$ in code units. This is, however, only set in regions where $\rho(\rr)\le 10^{-8}$ times the midplane density. The initial velocity field is erased by the expanding bow shock, and (as desired) only a moderate shock front arises ahead of the wind during its expansion into the unperturbed disk atmosphere. The initial shock is typically flushed out of the simulation domain within a few dynamical times at the inner radius.

\subsection{Simulation domain \& boundaries} \label{sec:bc} 

Our simulation domain extends from an inner radius of $r_{\rm in}=0.75\au$, out to $r_{\rm out}=22.5\au$, spanning a factor of 30 in radial dynamic range. This puts us in-between recent simulations by \citet{2017A&A...600A..75B} and \citet{2017ApJ...845...75B}, whose respective models had a dynamic range of 10 and 100. The vertical domain in our simulations covers $\theta \in [0,\pi/2]$ for cases that impose a mirror symmetry at the midplane, and $\theta \in [0,\pi]$ for models that allow for a hemispheric asymmetry to evolve. Nominally, we resolve the domain with $N_r\tms N_\theta = 480 \tms 180$ equidistant cells (per hemisphere), yielding a vertical resolution of six to eight grid cells per pressure scale height.
We have moreover performed a model (OA-b4-hr) with twice the nominal resolution, i.e., $960\tms 360$, and generally find agreement to within about ten percent, with the exception of the inferred mass accretion rate, which is lower in the higher resolved simulation (see \Tab{results}).

\subsubsection{Boundary conditions \& radial buffer zones} 

Our radial boundary conditions are of the standard ``outflow'' type.
For consistency with the equilibrium disk model, we extrapolate the mass density $\rho(\rr)$ according to the power law with $\bar{p}$.
At the inner boundary, we moreover apply a diffusive/damping radial buffer zone \citep{2006A&A...457..343F}, with a characteristic damping timescale of $0.1\times 2\pi\Omega^{-1}$ at $r_{\rm in}$.
The buffer is tapered-off with a  functional profile $\,1-{\rm erf}(4\;|\,r_{\rm in}-\Rc\,|)$, resulting in diminished wave reflections compared with a linear ramp.
Parallel magnetic field components have vanishing gradients (amplitudes) at the inner (outer) radial boundary, while the normal component is reconstructed from the $\nabla\cdot\B=0$ condition.
This is with the exception of $B_\theta(r_{\rm out})$, which we compute such that the toroidal current, $J_\phi$, is zero \citep[see appendix A in][]{2010ApJ...709.1100P}.
With $J_\phi=0$, the poloidal component of the Lorentz force also has to vanish, which avoids spurious collimation of the outflow \citep[also see discussion in ][]{1999ApJ...526..631K}.
We, moreover, force $B_\phi$ to zero at $r=r_{\rm out}$, creating a boundary layer with a slight magnetic pressure gradient that gives material leaving the domain an additional boost.
We have found that this reduces issues with the boundary influencing the flow upstream in the sub-Alfv{\'e}nic region of the outflow.
However, it will be prudent in future work to more systematically search for the best magnetic boundary conditions \citep[see][]{1999ApJ...516..221U} or, alternatively, explore boundary conditions based on a self-consistent characteristic wave analysis \citep[e.g.][]{1989JCoPh..84..343V,2001A&A...367..705D}.

As demanded by the azimuthal symmetry, at axial boundaries we apply standard ``reflecting'' boundary conditions, that is, the latitudinal and azimuthal components of vector fields must vanish at the axis, that is $(v_\theta,\,v_\phi,\,B_\theta,\,B_\phi)=0$, while the radial (i.e., vertical) component is free to evolve, i.e., $\partial_\theta (v_r,\,B_r)=0$.
For hemispherical models, boundary conditions at the disk midplane apply $v_\theta=0$ and $\partial_\theta(v_r,v_\phi)=0$, as well as, $(B_r,B_\phi)=0$ and $\partial_\theta B_\theta=0$, which precludes the buildup of a disk azimuthal field.

The radiation energy density is treated as a regular scalar variable at the axial and hemispherical boundaries.
To account for the inner disk that is not part of the computational domain, we impose a power-law with index $4\bar{q}$ onto $\eR(r_{\rm in})$, unless this would result in a radiation temperature higher than that of the initial model.
This has the merit of avoiding a runaway situation with catastrophic heating.
At the outer radial boundaries, in regions with $\Delta\tau_{\rm cell} \le 0.001$, we solve for the free streaming limit, that is, $\Fr(r_{\rm out})=c\,\eR(r_{\rm out})$, with $\Fr$ taken from \Eqn{flux}.
This is done via Newton-Raphson iteration, which typically converges within a few steps.
In regions of high optical depth near the midplane, we apply a zero-gradient boundary condition instead.
We currently do not use a tapering between the two regimes, which results in a noticeable (but inconsequential) artifact at the location of the transition.


\section{Simulation results} \label{sec:results}

The central motivation behind this paper is to extend the previous wind models of \gt into a more global domain, and improve the physical treatment in terms of the thermodynamics, adding radiative energy transport.

\begin{table}\begin{center}
    \caption{Summary of simulation sets.
      \label{tab:models}}
\begin{tabular}{p{1.67cm}ccccp{10pt}p{10pt}p{10pt}c}\hline\hline
  \\[-12pt]
  Label & $\bp$ & $L_\theta$ & $N_\theta$ & $\,\kR\,$ &
          \multicolumn{3}{c}{$\!\!\Qd\ \Qirr\ \Qpdr\!\!$} & $\dot{M}$
  \\[4pt]
  \hline\\[-10pt]
  OA-b4-uv9  & $10^4$ & $\pit$ & 180 & 1.0 & \no & \yes& \yes& $10^{-9}$ \\[-2pt]
  OA-b4      & $10^4$ & $\pit$ & 180 & 1.0 & \no & \yes& \yes& $10^{-8}$ \\[-2pt]
  OA-b4-uv7  & $10^4$ & $\pit$ & 180 & 1.0 & \no & \yes& \yes& $10^{-7}$ \\[+2pt]
  OA-b5      & $10^5$ & $\pit$ & 180 & 1.0 & \no & \yes& \yes& $10^{-8}$ \\[-2pt]
  OA-b6      & $10^6$ & $\pit$ & 180 & 1.0 & \no & \yes& \yes& $10^{-8}$ \\[-2pt]
  OA-b8      & $10^8$ & $\pit$ & 180 & 1.0 & \no & \yes& \yes& $10^{-8}$ \\[+2pt]
  OA-b4-hr   & $10^4$ & $\pit$ & 360 & 1.0 & \no & \yes& \yes& $10^{-8}$ \\[-2pt]
  OA-b4-fd   & $10^4$ & $\pi$  & 360 & 1.0 & \no & \yes& \yes& $10^{-8}$ \\[+2pt]
  OA-b4-lop  & $10^4$ & $\pit$ & 180 & 0.1 & \no & \yes& \yes& $10^{-8}$ \\[-2pt]
  OA-b4-ohm  & $10^4$ & $\pit$ & 180 & 1.0 & \yes& \yes& \yes& $10^{-8}$ \\[-2pt]
  OA-b4-noir & $10^4$ & $\pit$ & 180 & 1.0 & \no & \no & \no & ---       \\[+2pt]
  \hline
\end{tabular}
  \end{center}
  \footnotesize All simulations are labeled according to the strength of the net-vertical magnetic field, expressed in terms of the midplane value $\bp(t_0)$, prefixed with the letter `b'. The nominal value $\bp=10^4$ corresponds to $B_z\simeq 100\mG$ at $1\au$. Opacity, $\kR$, is in $\cm^2\g^{-1}$. \yes/\no~= enabled/disabled. Accretion rates, $\dot{M}$ (in $\Msun\yr^{-1}$) refer to the stellar FUV flux used as input for the thermochemistry.

\end{table}

We conducted a set of axisymmetric simulations, with key model parameters listed in \Table{models}. We begin by discussing the fiducial scenario OA-b4, which is largely identical to the one used in our previous study, but notably has a ten times higher net-vertical magnetic flux (i.e., in terms of the plasma parameter), corresponding to $B_z \simeq 100\mG$ at a radius of $\Rc=1\au$, which is arguably more consistent with estimates from the Semarkona meteorite \citep{2014Sci...346.1089F} than those previously used.
Moreover, we now assume a flared disk with $\bar{q}=-3/4$ by default, whereas the fiducial disk of \gt had $\bar{q}=-1$.
For the ionization model, we again assume a dust-to-gas mass ratio of $10^{-3}$, that is, a moderate depletion (via growth into larger grains, which are less efficient in absorbing free charges) compared to the generic value for interstellar grains. In addition, we reduce the X-ray flux to its nominal value (as opposed to enhanced by a factor of five as in \gt). Note that, following \citet{2017ApJ...845...75B}, we now also introduce a geometric correction in $\Sigma_{\rm X,abs}$ when attenuating the direct X-ray flux (see \Sec{XR} for details).

\subsection{The fiducial scenario} \label{sec:fiducial} 

\begin{figure}
  \center\includegraphics[width=\columnwidth]{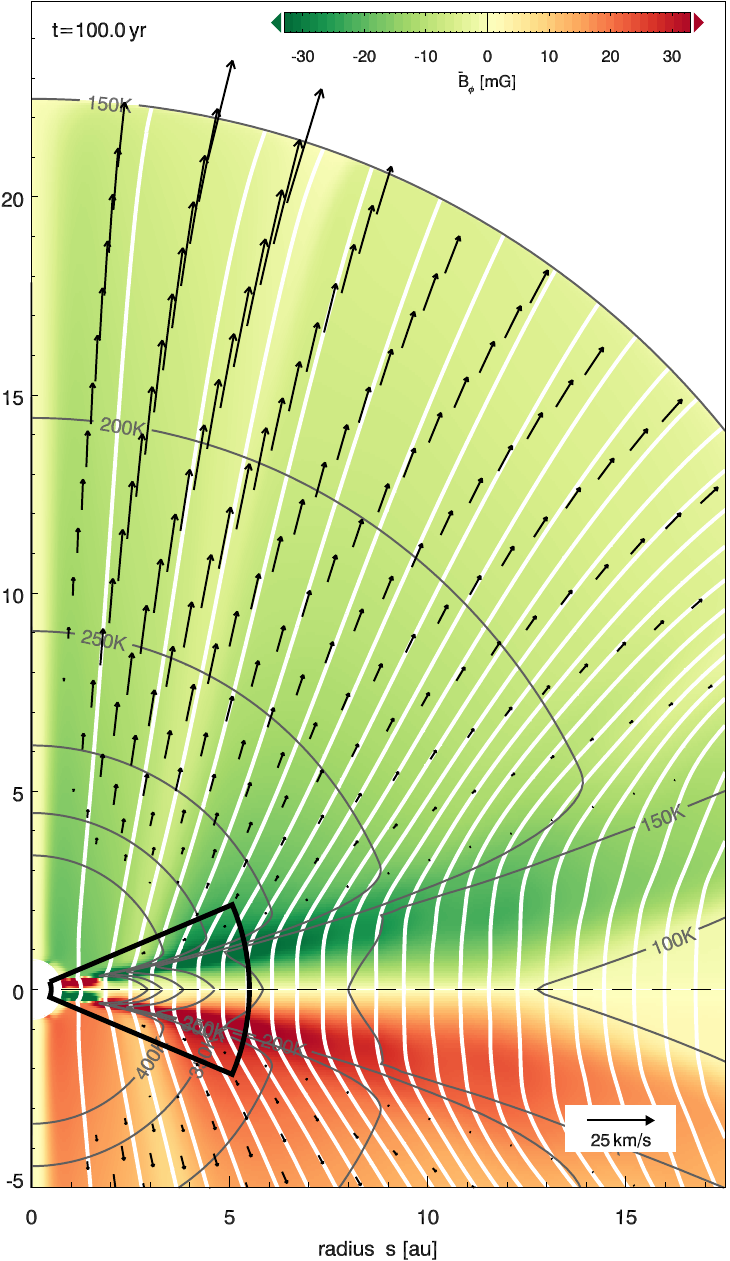}
  \caption{Poloidal structure for model OA-b4, including projected field lines (white) velocity vectors (black), and isocontours of the radiation temperature (gray). The azimuthal magnetic field (color) has been restricted to values $|B_\phi| \le 33\mG$ for clarity; peak values are $\sim 500\mG$. The region with $z\le 0$ has been mirrored at $z=0$, and we suppress ``back flow'' (i.e., regions with $v_r<0$ near the axis). The thick black outline shows the extent of our previous simulations in \gt, illustrating the changed domain size.}
  \label{fig:OA-b4-vis2D}
\end{figure}

We begin by showing the overall field topology, which is plotted in \Fig{OA-b4-vis2D} in terms of projected field lines (white), along with the poloidal velocity field (black arrows) at time $t=100\yr$. As discussed in detail in \gt, our simulations quickly adjust to a quasi-stationary state, and only a small level of residual time evolution is seen, mainly pertaining to the readjustment of the horizontal field in the presence of current sheets. \Figure{OA-b4-vis2D} moreover shows isocontours of the radiation temperature (gray lines), which largely follows the gas temperature, with the exception of the irradiated disk surface at $z\simeq 4\,H$, expressed in terms of the gas scale height.\footnote{Note that, when quoting pressure scale heights, we always refer to the value, $H_0$, of the isothermal-on-cylinders initial model.}
For clarity, we have suppressed ``back flow'' (with $v_r<0$), that is, regions where material is in free-fall towards the central star. The issue is related to the exclusion of the inner sphere (with $r\le r_{\rm in}=0.75\au$) from the simulation domain. As a consequence, we lack the inner part of the disk where the field-lines would be anchored that are supposed to accelerate the material near the axis to form a protostellar jet.
Also note that most of our simulations only evolve the hemisphere with $z>0$ (see the third column in \Tab{models}), and that we simply restore the missing part by means of the assumed symmetry in order to produce more intuitive figures.

\begin{figure*}
  \center\includegraphics[width=0.8\textwidth]{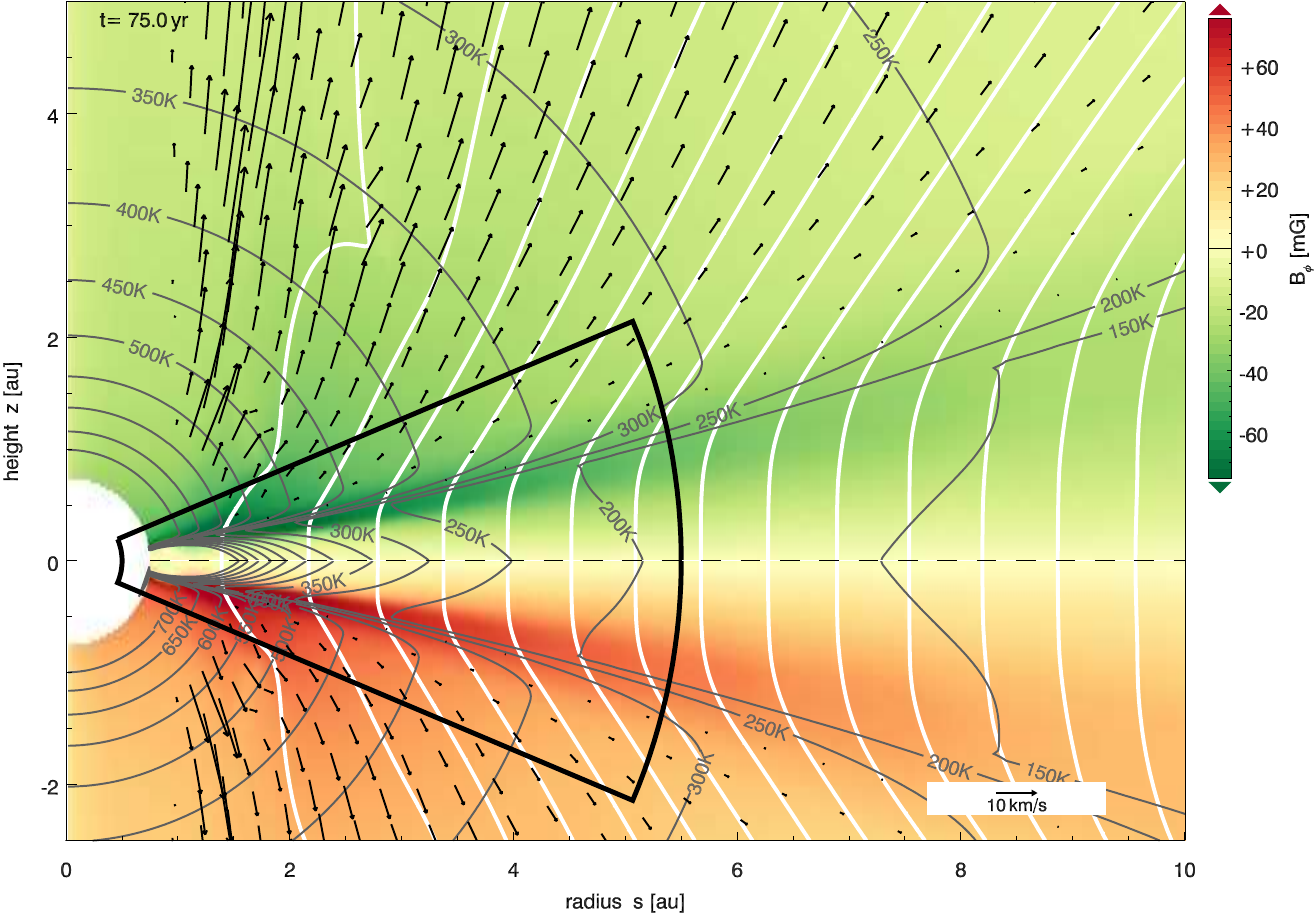}
  \caption{Same as \Fig{OA-b4-vis2D}, but showing a close-up of model OA-b4-hr with twice the linear grid resolution. Note the different scale of the color bar; compared to model OA-b4, peak values are now only $\sim 100\mG$.}\smallskip
  \label{fig:OA-b4-hr-vis2D}
\end{figure*}

The extended scope of our new simulations becomes apparent when looking at the small segment highlighted by a black outline in \Fig{OA-b4-vis2D}, which corresponds to the simulation domain (with $r\in[0.5,5.5]$, and covering $\pm8\,H$ in $\theta$) adopted previously.
A noticeable difference to the similar plot shown in the lower panel of figure~5 in \gt is the reversed radial sequence of the field belts.
For our new model OA-b4 (and, in addition, for model OA-b5, which more closely resembles \gt), we observe belts of positive (negative) $B_\phi$ for $\Rc\simlt 2\au$, which connect to the wind with negative (positive) azimuthal field via a current sheet in the upper (lower) half of the disk.
In contrast, the field belts in the \gt model had the \emph{same} polarity as the wind inside of $\Rc\simeq1.5\au$, and displayed current sheets \emph{outside} of that radius.
This illustrates that the formation of field belts (described in detail in sect.~3.5 of \gt) and associated current sheets is likely sensitive to the initial conditions, and potentially to the location of the inner domain boundary, and/or numerical resolution.

Note, moreover, that the sharp field reversal at $z=0$ for $\Rc\simlt 2\au$ in model OA-b4 is a consequence of the imposed mirror symmetry about the midplane, as the low Ohmic conductivity in this region is unlikely to support a current sheet at this location.
As can be seen in \Fig{OA-b4-hr-vis2D}, this structure is absent in the high-resolution model OA-b4-hr, which features twice the grid resolution of our fiducial model OA-b4, but is otherwise identical.


\begin{table*}\begin{center}
\caption{Summary of simulation results.
  \label{tab:results}}
\begin{tabular}{p{2.25cm}cccccccc}\hline\hline\\[-10pt]
  & $\zb$
  & $\Rc_{\rm A}$
  & $\Mx$
  & $c_B$
  & $\dot{M}_{\rm wind}$
  & $\dot{M}_{\rm accr}$
  \\[2pt]
  & $(\,H_0\,)$
  & $(\,\Rc_{\rm b}\,)$
  & $(\,10^{\,\mi4}\,p_0\,)$
  & $(\,10^{\,\mi2}\au/\!\yr\,)$
  & $(\,10^{\,\mi7}\!\!\Msun/\!\yr\,)$
  & $(\,10^{\,\mi7}\!\!\Msun/\!\yr\,)$
  \\[2pt]\hline\\[-6pt]
  %
  %
  OA-b4       & $6.49\pm 0.26$ & $1.56\pm 0.13$ & $2.81\pm 0.15$ & $0.35\pm 0.03$
              & $0.31\pm 0.01$ & $1.13\pm 0.06$ \\[-2pt]
  OA-b5       & $6.46\pm 0.09$ & $1.10\pm 0.28$ & $0.59\pm 0.04$ & $0.10\pm 0.01$
              & $0.12\pm 0.02$ & $1.09\pm 0.07$ \\[-2pt]
  OA-b6       & $6.72\pm 1.80$ & $0.92\pm 0.01$ & $0.00\pm 0.00$ & $0.16\pm 0.01$
              & $0.11\pm 0.05$ & $1.05\pm 0.05$ \\[+4pt]
  OA-b4-uv9   & $6.46\pm 0.23$ & $1.62\pm 0.05$ & $2.45\pm 0.06$ & $0.37\pm 0.03$
              & $0.25\pm 0.01$ & $1.13\pm 0.06$ \\[-2pt]
  OA-b4-uv8$^{\,\star}$
              & $6.49\pm 0.26$ & $1.56\pm 0.13$ & $2.81\pm 0.15$ & $0.35\pm 0.03$
              & $0.31\pm 0.01$ & $1.13\pm 0.06$ \\[-2pt]
  OA-b4-uv7   & $6.48\pm 0.28$ & $1.50\pm 0.07$ & $3.16\pm 0.22$ & $0.34\pm 0.05$
              & $0.39\pm 0.01$ & $1.14\pm 0.07$ \\[+4pt]
  \multirow{2}{*}{OA-b4-fd$^\dagger$}
              & $8.91\pm 1.15$ & $0.94\pm 0.01$ & $0.41\pm 0.24$ &
                \multirow{2}{*}{$0.83\pm 0.24$}
              & $0.14\pm 0.01$ & \multirow{2}{*}{$0.76\pm 0.00$} \\[-2pt]
              & $8.26\pm 0.71$ & $1.69\pm 0.00$ & $2.83\pm 0.05$ &
              & $0.17\pm 0.01$ & \\[+4pt]
  OA-b4-hr    & $6.42\pm 0.14$ & $1.66\pm 0.01$ & $2.66\pm 0.04$ & $0.66\pm 0.08$
              & $0.33\pm 0.01$ & $0.16\pm 0.00$ \\[-2pt]
  OA-b4-lop   & $6.52\pm 0.01$ & $1.67\pm 0.00$ & $2.26\pm 0.01$ & $0.57\pm 0.13$
              & $0.31\pm 0.04$ & $1.03\pm 0.00$ \\[-2pt]
  OA-b4-ohm   & $6.49\pm 0.26$ & $1.57\pm 0.13$ & $2.82\pm 0.17$ & $0.46\pm 0.15$
              & $0.31\pm 0.01$ & $1.13\pm 0.06$ \\[-2pt]
  OA-b4-noir  & $6.49\pm 0.44$ & $1.44\pm 0.05$ & $3.98\pm 0.39$ & $0.49\pm 0.22$
              & $0.59\pm 0.02$ & $1.12\pm 0.06$ \\[+4pt]
  \hline
\end{tabular}
\\[+4pt]
$^\star$) identical to model OA-b4,~
$^\dagger$) top/bottom hemisphere listed individually
\end{center}
  \parbox[t]{2\columnwidth}{\footnotesize Wind base, $\zb$, Alfv{\'e}n radius, $\Rc_{\rm A}$, Maxwell stress at the wind base, $\Mx$, and field-line migration speed, $c_{\rm B}$, are measured for a field line starting at $\Rc_0=5\au$. Mean values and deviations are computed using approximately ten snapshots between $t=75$--$100\yr$, except for $c_B$, which is an average value between $t=17.5$--$117.5\yr$.}
  \medskip
\end{table*}


\subsection{A preliminary parameter survey} \label{sec:param} 

The complexity of the model naturally bears with it a variety of input parameters that will in one way or another affect the precise outcome of the simulation.
Here we restrict ourselves to a coarse first survey of the most fundamental effects, namely the amount of net magnetic flux, and the level of incident FUV radiation.

\subsubsection{Measurement protocol} 

To facilitate a comparison between the various models, we have compiled a summary of simulation results in \Table{results}, where we list time-averaged properties.
As in \gt, the location of the wind base, $\zb$, is derived from the criterion $v_\phi(\zb) \ge \vK$ \citep{1993ApJ...410..218W}, which is approximately independent of radius when expressed in local scale heights, $H_0$.
The Alfv{\'e}n radius, $\Rc_{\rm A}$, where the poloidal component of the flow exceeds the Alfv{\'e}n speed, is obtained by tracing a field line starting at $z_0 = 0$ and a fiducial value of $s_0 = 5\au$; it is quoted with respect to the location of the wind base, $\Rc_{\rm b}$, which is motivated by the observation that, due to the poor coupling below the wind base, $\zb$, the magnetic lever arm with respect to $\Rc_0$ is not meaningful (also see the discussion in \bai).
The vertical-azimuthal component of the Maxwell stress at the location of the wind base, $\Mx$, quantifies the amount of angular momentum extracted by the wind, and is normalized by the thermal pressure, $p_0$, at the foot point of the field line.
We quantify the evolution of the vertical flux by means of a field-line migration speed, $c_B$, which is obtained at $z=0.1\au$ (see \Fig{OA-b4-flux-transport}) for a fiducial radius of $\Rc=5\au$. The mass loss rate, $\dot{M}_{\rm wind}$ is defined as the surface integral
\beq
  \dot{M}_{\rm wind} \equiv 2\pi
    \int_{r_1}^{r_2} \rho\, v_\theta\,r\sin\theta\,\dd r
    \;\Big|_{\theta_1}^{\theta_2}\,,
\eeq
with $r_1\eq2.5\au$, $r_2\eq22.5\au$, and evaluated at the wind base, i.e., with $\theta_1$ and $\theta_2$ corresponding to $\pm \zb$. Conversely, the mass accretion rate is defined as
\beq
  \dot{M}_{\rm accr} \equiv 2\pi
  \int_{\theta_1}^{\theta_2} \rho\, v_r\,r^2\sin\theta\,\dd\theta
  \;\Big|_{r_1}\,,
\eeq
applying an average within the interval $r\in [1.5,2.5] \au$ to obtain an estimate for its variance.

It is important to note that the unambiguous determination of the wind-driven mass accretion rate is somewhat hindered by the presence of poloidal circulation, which we attribute to the rudimentary development of the vertical shear instability \citep*[VSI,][]{2013MNRAS.435.2610N}.
While this hydrodynamic instability had been intentionally suppressed in \gt by prescribing a slow-acting cooling prescription, the characteristics of the RMHD simulations put us into the parameter regime \citep[see][]{2015ApJ...811...17L} where VSI can, in principle, operate.
As demonstrated by \citet{2018MNRAS.474.3110L}, the VSI can easily be stabilized by magnetic tension forces.
These are, however, subject to the development of small-scale fluctuations in the field, which are naturally precluded in regions where significant AD and/or Ohmic dissipation is present.
As a further note of caution, we stress that the grid resolution of the current radially global RMHD models is not adequate to follow the evolution of the VSI in detail \citep[but see][for high-resolution radially restricted non-ideal MHD simulations]{2019arXiv191202941C}.

\subsubsection{Effect of varying the field strength} \label{sec:nvf} 

Despite our ignorance about the actual magnitude of magnetic fields in the T~Tauri phase of protoplanetary disks, it can be argued that there exists a reasonably narrow corridor of realistic values \citep[also see the parametric study of][]{2019arXiv191104510R}.
We have run three models, OA-b4, OA-b5, and OA-b6 with initial midplane plasma parameters of $\bp\eq10^{4}$, $10^{5}$, and $10^{6}$, respectively.
These models share a similar location, $\zb\simeq 6.5\,H$, for the wind base, and have similar accretion rates of $\dot{M}_{\rm accr}\simeq 10^{-7}\Mdot$, which evidently only very weakly depend on the field strength.
Moreover, they produce outflow rates ranging from $\dot{M}_{\rm wind}\simeq 0.1 - 0.3\ee{-7}\Mdot$.
Compared with its isothermal counterpart OA-b5-flr ---see table~2 in \gt--- the corresponding radiative model, OA-b5, has a four times higher mass loss rate.
This appears to indicate that, for TACOs, the magnetic field strength plays a limited role in determining the mass loading of the wind \citep[also see fig.~12 and discussion in][]{2016ApJ...818..152B}.

This is not to say that the magnetocentrifugal mechanism does not scale with the magnetic field strength --- clearly, both the lever arm
$\lambda_{\rm A}\equiv \Rc_{\rm A}^2/s_{\rm b}^2$, and the wind stress, $\Mx$, depend on the amount of net-vertical flux. Rather, this points to the importance of radiation and thermodynamics in determining the wind properties (cf.~the isothermal assumption).
Moreover, the outward flux transport (measured via $c_B$) appears to increase for stronger fields, which qualitatively agrees with \citet{2013ApJ...769...76B}.

We note in passing that, for the purpose of defining a baseline model for the synthetic observations presented in \Section{synthetic}, we have performed a model OA-b8 with essentially negligible vertical flux. This model only develops a weak photoevaporative outflow, which is why we refrain from further discussing it in this section and refer the reader to the broad parameter survey presented in \citet{2019arXiv191104510R} instead.

\subsubsection{Dependence on the incident UV flux} \label{sec:uv} 

To assess the importance of the thermochemical heating in the wind launching process, we have performed a series of models with varying FUV flux.
These are derived from model OA-b4 (which has an accretion rate of $10^{-8}\Mdot$) and carry suffixes -uv9, -uv8, and -uv7, respectively, indicating mass accretion rates of $\dot{M}_{\rm FUV}= 10^{-9}$, $10^{-8}$, and $10^{-7}\Mdot$ that are converted into FUV luminosities according to
\beq
  L_{\rm FUV} = 10^{-2}\; \frac{M_\star}{M_\odot}\; \frac{R_\odot}{R_\star}
                \;\frac{\dot{M}_{\rm FUV}}{10^{-8}\Mdot}\;L_\odot\,;
\eeq
see also eqn.~(6) in \citet{2009ApJ...705.1237G}. Resultant changes in the thermochemical heating rate manifest most prominently in $\dot{M}_{\rm wind}$, which increases by 25\% when increasing the FUV luminosity by an order of magnitude.
In contrast, we do not find the mass accretion rate, $\dot{M}_{\rm accr}$, to depend on the amount of thermochemical heating at all.
Moreover, the flux migration speed, magnetic lever arm, and the wind stress only seem to weakly depend on the amount of thermochemical heating, while the location of the wind base remains unaffected.

\subsection{Kinematics of the emerging wind solution} 

\begin{figure}
  \center\includegraphics[width=\columnwidth]{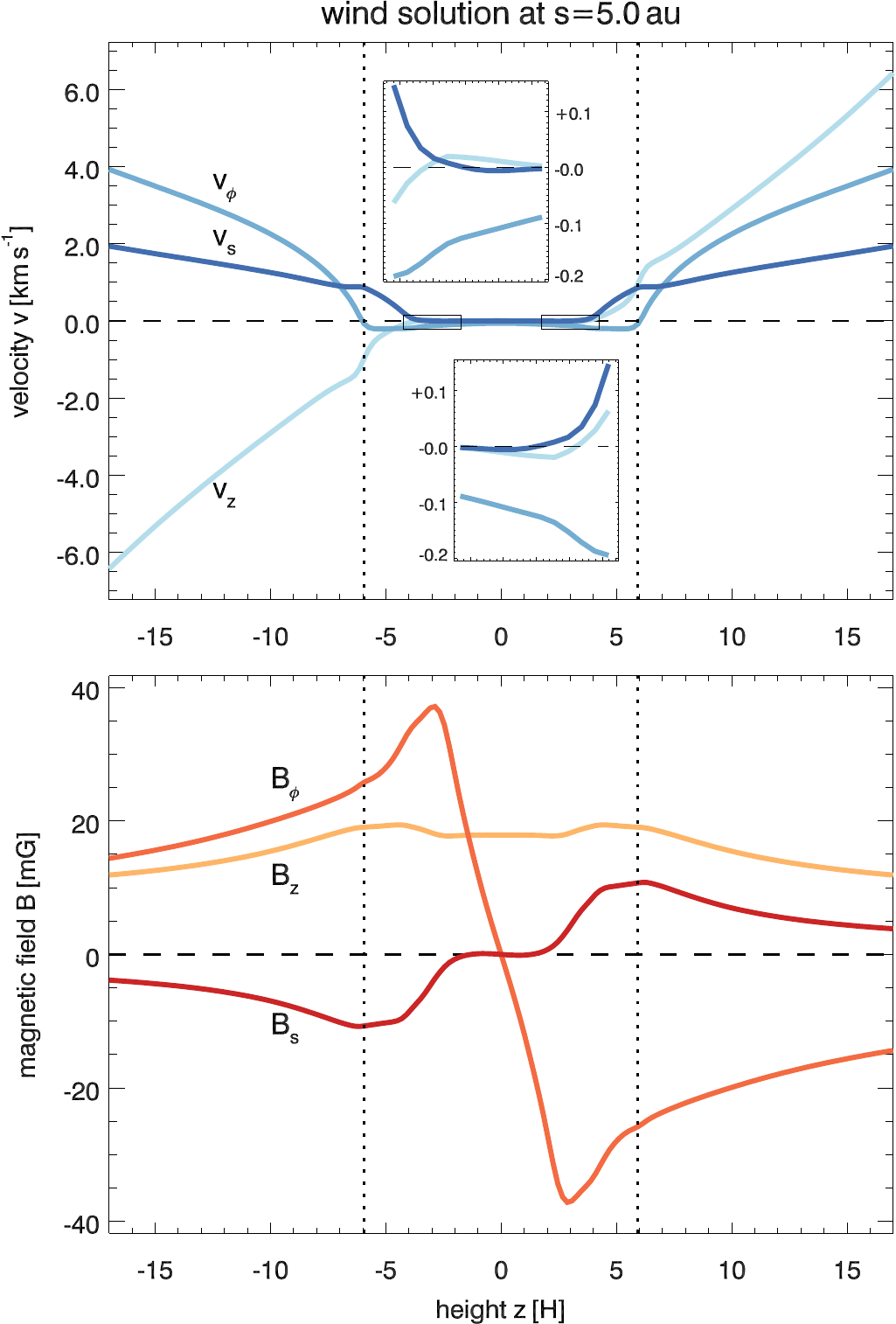}
  \caption{Vertical profiles of the velocity (upper panel) and magnetic field components (lower panel) at a constant cylindrical radius, $\Rc=5\au$, for model OA-b4. Note that we are plotting the cylindrical components of the vectors (e.g., $v_\Rc$, $v_\phi$, and $v_z$). Dotted lines indicate the wind base, i.e., where $v_\phi$ becomes super-Keplerian. Inset panels show zoom-ins of the regions indicated by the black boxes.}
  \label{fig:OA-b4-wind}
\end{figure}

In \Figure{OA-b4-wind}, we plot the emerging wind solution at a constant cylindrical radius $\Rc=5\au$, which compares well to the similar figure~6 in \gt, that is, in the part of the disk \emph{without} a strong current sheet.
Despite the stronger net-vertical field, the wind base is marginally higher up in the disk, at $\zb\simeq 6.5\,H$ (also see \Tab{results}) compared to $\zb\simeq 5.25\,H$ as found previously.
Albeit barely discernible in the figure, the radial velocity, $v_\Rc$, dips below zero just below the launching point (see insets in \Fig{OA-b4-wind}) owing to the torque exerted by the wind.
At the same time, the rebound of the wind acceleration is seen in the vertical velocity, $v_z$, that changes its sign in this region.
In contrast to \gt, the azimuthal velocity $v_\phi$ describing the deviation from the Keplerian rotation profile now supersedes the radial velocity $v_\Rc$.
This may indicate that the collimation of the outflow in the previous set of simulations was hindered by the limited vertical domain size.
Note that in model OA-b4, the wind robustly reaches super-Keplerian rotation, which is in contrast to the similar model `B40' (see right panel of their fig.~18) of \citet{2017ApJ...845...75B}, who do \emph{not} find an unambiguous centrifugal outflow.

\begin{figure*}
  \center
  \includegraphics[width=0.468\textwidth]{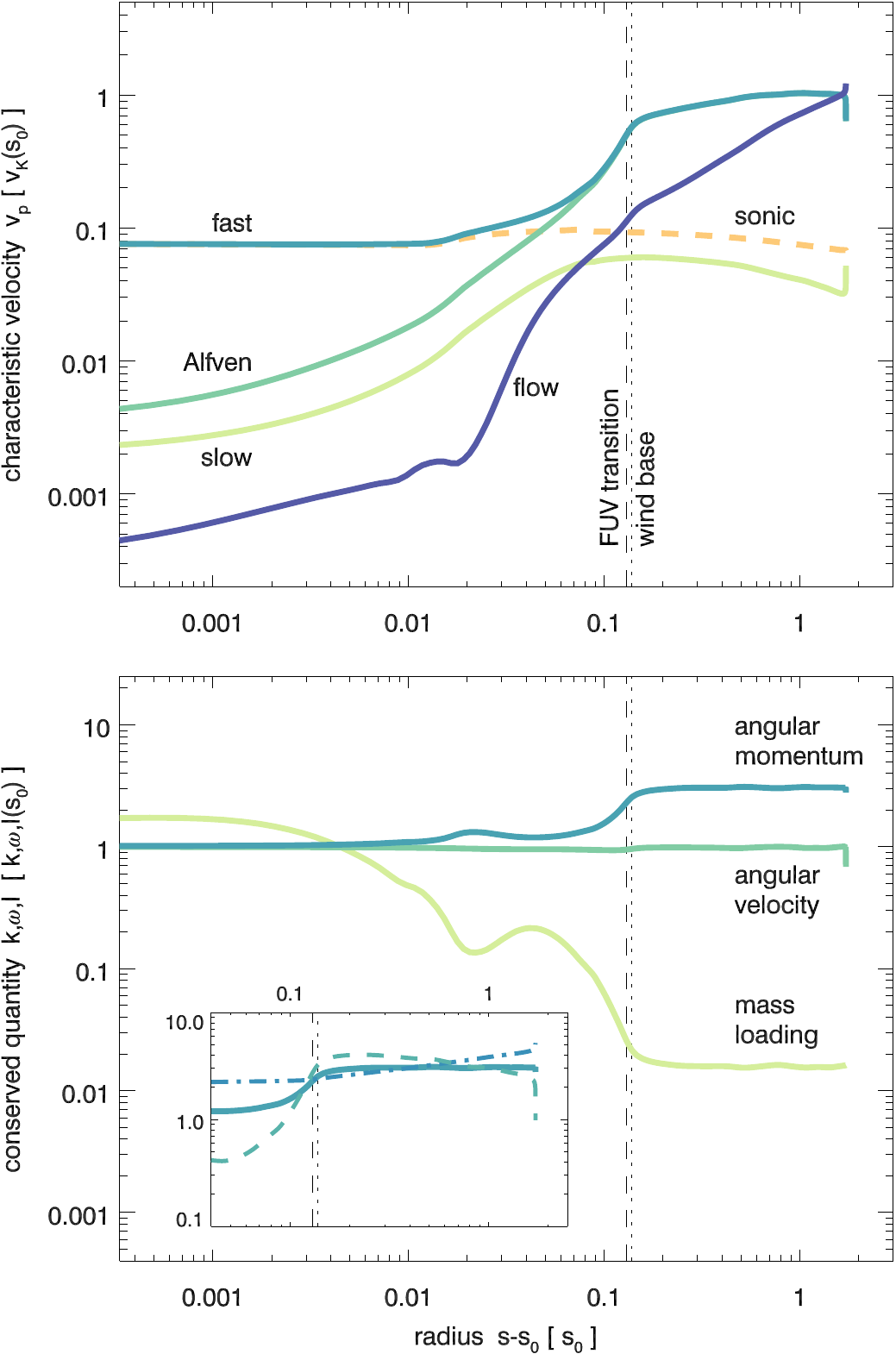}\qquad
  \includegraphics[width=0.468\textwidth]{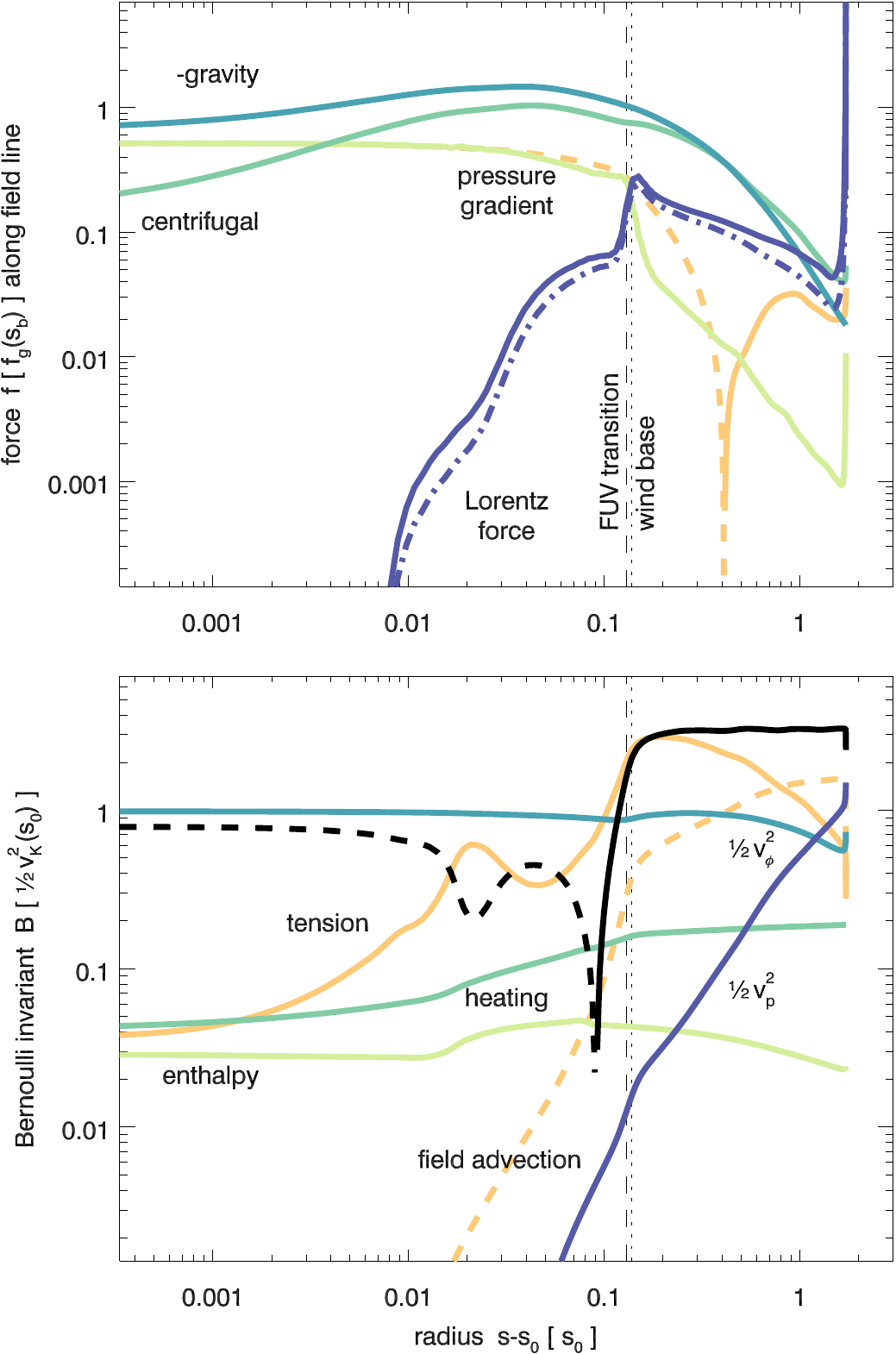}
  \caption{\emph{Top left:} Characteristics (solid lines) and sound speed (dashed) along a field line originating at $(\Rc_0,\,z_0)=(5,0)$. \emph{Top right:} The various forces are projected onto the field line and normalized by the gravitational force at the wind base. The orange line shows the effective centrifugal force (i.e., excess over central gravity), where the dashed part corresponds to negative values. \emph{Bottom left:} conserved quantities ---see \Eqn{invar}--- $k/k_0$ (mass loading), $\omega/\omega_0$ (angular velocity), and $l/l_0$ (specific ang.\ mom.) along the same field line; for the latter, the inset shows kinetic (dot--dash) and magnetic (dashed) contributions. \emph{Bottom right:} The Bernoulli invariant is shown in black and changes sign near the slow transition. The different components of the Bernoulli invariant are also shown. Vertical lines in all panels indicate the transition into the FUV ionization layer (dashed), and the wind base (dotted), beyond which the flow invariants are preserved.}
  \label{fig:OA-b4-speeds-forces}
\end{figure*}

Previous simulations in the local framework \citep{2013A&A...552A..71F}, but notably also the global models of \blf and \gt, have found that the point where the outflow passes through the fast magnetosonic point tends to lie uncomfortably close to the edge of the computational domain.
As can be seen in the top left panel of \Fig{OA-b4-speeds-forces}, where we plot the poloidal flow velocity along a field line starting at $\Rc_0=5\au$ near the midplane, this is also still the case for run OA-b4.\footnote{Note that the sudden dip in the fast speed at the domain boundary is likely a result of setting the azimuthal magnetic field to zero there (even though we plot \emph{poloidal} quantities, here).}

As previously found, the FUV transition (dashed line in \Fig{OA-b4-speeds-forces}) closely coincides with the wind base (dotted line), defined as the point where $v_\phi\ge \vK$.
This illustrates that determining the FUV ionization layer from first principles is key in obtaining a realistic wind structure.
Notably, the flow crosses the slow magnetosonic surface below the FUV transition layer, and it is here that the mass flux of the wind is determined, i.e.,
\beq
  \frac{1}{2\pi} \frac{\dd \dot{M}_{\rm wind}}{\dd\log \Rc} \approx \left. 2\rho\, \cS \frac{B_z}{B_\Rc}\right|_{\rm sms}\,,
\eeq
depending on the local field-line inclination angle, sound speed, and mass density \citep{1995MNRAS.275..244L,2011ppcd.book..283K}.
This transition occurs in a region where the field line's inclination is still affected significantly by ambipolar diffusion, hinting at a potentially sensitive dependence of the mass loading on the microphysics.

We now briefly discuss the classic invariants (along a field line) in the theory of wind kinematics \citep[see, e.g., section~2 in][]{2007prpl.conf..277P,1996ASIC..477..249S}.
Importantly, the invariants have been derived in the limit of steady-state \emph{ideal} MHD, where the magnetic flux is ``frozen'' into the fluid.
Combining this concept with mass conservation along the field line, one obtains the so-called mass loading, $k$.
Similarly, employing the induction equation, one can find the angular velocity of magnetic flux surfaces, $\omega$.
The specific angular momentum, $l$, is derived by relating the inertia term with the Lorentz force in the azimuthal component of the momentum equation.
In summary, we have
\beq
  k \equiv \frac{\rho\Vp}{\Bp}\,,\quad
  \omega \equiv \frac{v_\phi}{\Rc} - \frac{k B_\phi}{\rho\Rc}\,,\quad
  l \equiv \Rc v_\phi - \frac{\Rc B_\phi}{\mu_0\,k}\,,\quad  \label{eq:invar}
\eeq
which we plot in \Fig{OA-b4-speeds-forces} (lower left panel).
From \Eqn{invar}, one can see that the magnetic contributions to the invariants depend on the mass loading, $k$.

The invariants are normalized with their respective values at the foot point of the field line and qualitatively agree reasonably well with figure~5 in \bai.
The angular velocity of the flux surfaces appears to be conserved very well, even in the regions of the flow where Ohmic resistivity and ambipolar diffusion dominate.
Conversely, both angular momentum and mass are \emph{not} conserved below the wind base, which is hardly surprising, given that flux freezing should be a poor approximation in this region.
In the FUV-ionized upper layers, all invariants remain reasonably constant.
As pointed out by \blf --- see their figure~10 --- by means of the magnetic torque, the magnetocentrifugal acceleration mechanism leaves its signature in the kinetic (i.e., $\Rc v_\phi$) and magnetic (i.e., $\Rc B_\phi/k$) contributions to the total specific angular momentum.
We plot both terms separately in the inset of \Fig{OA-b4-speeds-forces}, where the kinetic part (dot-dashed line) is clearly boosted in the FUV layer, at the expense of the magnetic part (dashed line).

\subsubsection{Poloidal force balance} 

To better understand the wind launching mechanism, it is instructive to look at the various forces and, in particular, their tangential components along a projected field line.
Force balance is then expressed as
\beq
  \frac{\dd \Vp}{\dd t} =
    - \frac{1}{\rho} \frac{\dd p}{\dd\xi}
    + \frac{v_\phi^2}{\Rc} \frac{\dd\Rc}{\dd\xi}
    - \frac{\dd \Phi}{\dd \xi}
    - \frac{B_\phi}{\rho\Rc\,\mu_0} \frac{\dd (\Rc B_\phi)}{\dd\xi}\,,
    \label{eq:wind_forces}
\eeq
where the terms on the RHS are (i) the pressure gradient, (ii) centrifugal force, (iii) gravity, and (iv) tangential Lorentz force,\footnote{Note that the Lorentz force, $(\nabla\tms\B)\tms\B/\mu_0$, is of course perpendicular to the field. That is to say that the term ``tangential'' refers to the projection of the field line onto the poloidal plane.} and where $\dd/\dd\xi$ denotes derivatives with respect to the spatial coordinate $\xi$, along the projected field line.

We plot the respective terms in the upper right panel of \Fig{OA-b4-speeds-forces}, along a field line starting from $\Rc_0\eq5\au$, where we normalize all terms with the strength of the gravitational term at the location of the wind base.
Below this line, the effective centrifugal force --- i.e., the sum of terms (ii) and (iii) --- is balanced by the pressure gradient term, illustrating the thermal character of the wind.
Unlike \bai, who find that the effective centrifugal force is always negative, we observe that once the magnetocentrifugal mechanism kicks in (i.e., at $\Rc-\Rc_0 \simeq 0.4\,\Rc_0$), there is an excess over the central potential (see the dashed and solid orange lines in the upper right panel of \Fig{OA-b4-speeds-forces}).
Above the wind base, the pressure gradient force is superseded by the Lorentz force, which nicely illustrates the dual character of TACOs. Complementary to the last term in \Eqn{wind_forces}, one can define the azimuthal component,
\beq
F_\phi \equiv \frac{\Bp}{\rho\Rc\,\mu_0} \frac{\dd (\Rc B_\phi)}{\dd\xi}
= -\frac{\Bp}{B_\phi}F_\parallel\,,  \label{eq:f_phi}
\eeq
of the Lorentz force \citep[see eqn.~(29) in][]{2007A&A...469..811Z}, that is responsible for the magnetocentrifugal acceleration adding to the excess of angular momentum. We superimpose \Eqn{f_phi} with a dash-dot line in the upper right panel of \Fig{OA-b4-speeds-forces}, where we see that it is comparable in amplitude with the tangential part of the Lorentz force related to pressure gradients in the azimuthal field component -- highlighting the role of the magnetocentrifugal effect in TACOs.

\subsubsection{Energy budget along field line} 

Prompted by similar analyses in \citet{2009ApJ...691L..49S} and \blf, we now take a look at the energetics of the outflow in terms of the Bernoulli invariant
\beq
  \mathcal{B} \equiv \frac{v_\phi^2}{2} + \frac{\Vp^2}{2} + \Phi
    + \frac{B_\phi^2}{\mu_0\,\rho}
    - \frac{v_\phi B_\phi\Bp}{\mu_0\,\rho\Vp} + \mathcal{H} - \mathcal{Q}\,,
    \label{eq:bernoulli}
\eeq
with
\beq
  \mathcal{H} \equiv \frac{\gamma}{\gamma-1}\,\frac{p}{\rho}\,,\text{ and}\quad
  \mathcal{Q} \equiv \mathcal{H} \,-\! \int\frac{\nabla p}{\rho}\,\dd\xi\,,
\eeq
the contributions from the flow enthalpy and heating effects, respectively.
The fourth and fifth term on the RHS of \Eqn{bernoulli} are the advective transport of magnetic energy, and energy transport via field-line tension, respectively (note the minus sign in the latter).
As in \blf, we normalize $\mathcal{B}$ by subtracting the value of $\mathcal{Q}$ at the outer domain boundary, even though this makes little difference in our case.

We plot the Bernoulli invariant (along with its constituents) in the bottom right panel of \Fig{OA-b4-speeds-forces}, where one can see that, after changing its sign near the slow magnetosonic transition, it remains nearly constant above the wind base.
In contrast to \blf, the heating term, $\mathcal{Q}$, remains rather sub-dominant compared to the dominant field-line tension component.
This moderate influence of thermal effects in our model may partly be due to our assumption of a two-temperature system (i.e., gas + radiation), meaning that the dust- and gas-temperatures are taken to be identical. This may, however, be a poor approximation high up in the disk \citep[see][]{2018ApJ...857...57N}.
We plan to avoid this restriction in favor of a three-temperature description (with distinct gas-, dust- and radiation-temperatures) in the future.

\subsubsection{Heating mechanisms} 

The acceleration of material by a combined magnetocentrifugal and thermal wind naturally begs the question of how the vertical temperature structure of the disk compares to the location of the FUV ionization transition and the base of the wind.

\begin{figure}
  \center\includegraphics[width=\columnwidth]{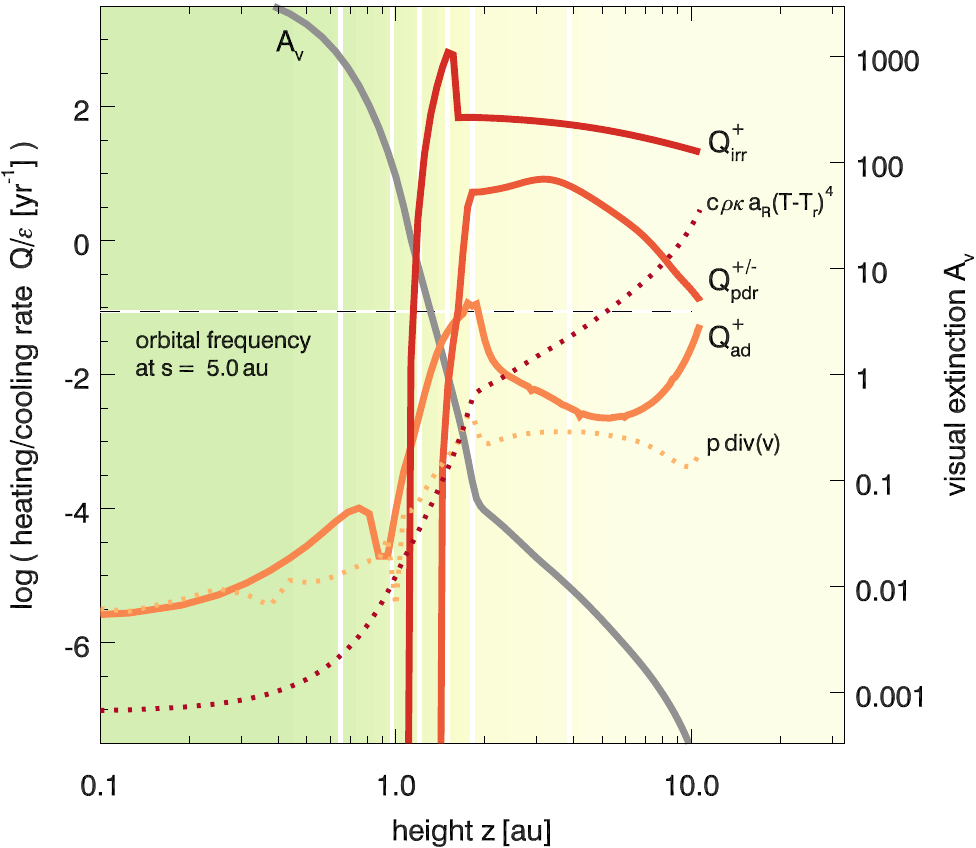}
  \caption{Heating rates at $\Rc\!=\!5\au$ for the three dominant heating sources, $\Qirr$, $\Qpdr$, and $\Qad$ for model OA-b4-uv7, with an incident FUV flux equivalent of $\dot{M}=10^{-7}\Mdot$. Vertical lines indicate decades in the visual extinction coefficient, $A_{\rm v}$. Dotted lines show the magnitude of the remaining two source terms in the internal energy equation.}
  \label{fig:OA-b4-uv7-heating}
\end{figure}

In \Figure{OA-b4-uv7-heating}, we plot heating rates, $\mathcal{Q}/\epsilon$, for the three dominant contributors in our model, that is:
\begin{enumerate}\itemsep0pt
 \item stellar irradiation heating, $\Qirr$,
 \item thermochemical heating/cooling, $\Qpdr$, and
 \item ambipolar dissipation heating, $\Qad$.
\end{enumerate}
Ohmic dissipation heating is comparatively unimportant, with peak values $\mathcal{Q}/\epsilon\simeq 10^{-6}$ near the midplane, which is likely related to the very moderate curvature of the field near $z\eq0$, where it is difficult for the kinematics of the outflow to bend the field lines (see \Fig{OA-b4-hr-vis2D}).

The dark red dotted line in \Fig{OA-b4-uv7-heating} shows the radiation matter coupling\footnote{Note that the plotted expression does not, in fact, contain the contribution from $\Qirr$ included during the implicit update.}, which acts as a sink term here, and which generally dominates over the adiabatic expansion of the outflow (orange dotted).
Whether a source term will significantly contribute to the heating of the disk is partly determined by optical depth effects.
Our current treatment is rather limited in this regard, since we do not distinguish separate opacities for the stellar irradiation and the thermal re-emission from the dust.

Nevertheless, as can be seen in \Fig{OA-b4-uv7-heating}, $\Qirr$ naturally peaks at $A_{\rm v}\simeq 1$, and we will use this as a first rough reference marker.
At the radius of $\Rc=5\au$, the thermochemical heating only contributes significantly at very low column densities/visual extinction.
Ambipolar diffusion heating, $\Qad$, has a similar domain as irradiation heating, but at a rate that is approximately ten times slower than the orbital timescale, at $\Rc=5\au$.
Closer towards the star, where the magnetic energy density gradually increases in absolute terms, $\Qad$ can become comparable to $\Qpdr$.
Simulations with more realistic opacities may be able to determine whether MHD dissipation heating can substantially contribute to the thermal structure of the disk on secular timescales.
In practical terms, the model OA-b4-ohm with MHD dissipation terms enabled produced nearly indistinguishable results compared with the fiducial model OA-b4.

\subsection{Evolution of the magnetic flux} \label{sec:flux} 

The net-vertical flux plays a central role in determining the wind stress and lever arm. Thus, we now turn to studying the evolution of the flux itself.
We remark that including the Hall effect (albeit with a simplified ionization prescription) has been found to have a significant effect on the radial transport of the vertical magnetic field \citep{2017ApJ...836...46B,2019MNRAS.tmp.1545L}.
This redistribution of flux was attributed to a global manifestation of the HSI, and a similar result has been found in Hall-MHD simulations of cloud collapse by \citet{2017PASJ...69...95T}.
At least in the aligned-field case, however, the HSI transport saturates to the same effective migration timescale as in the Hall-free case \citep{2017ApJ...836...46B}, which warrants revisiting the problem with our more detailed ionization model.

\begin{figure}
  \center\includegraphics[width=0.95\columnwidth]{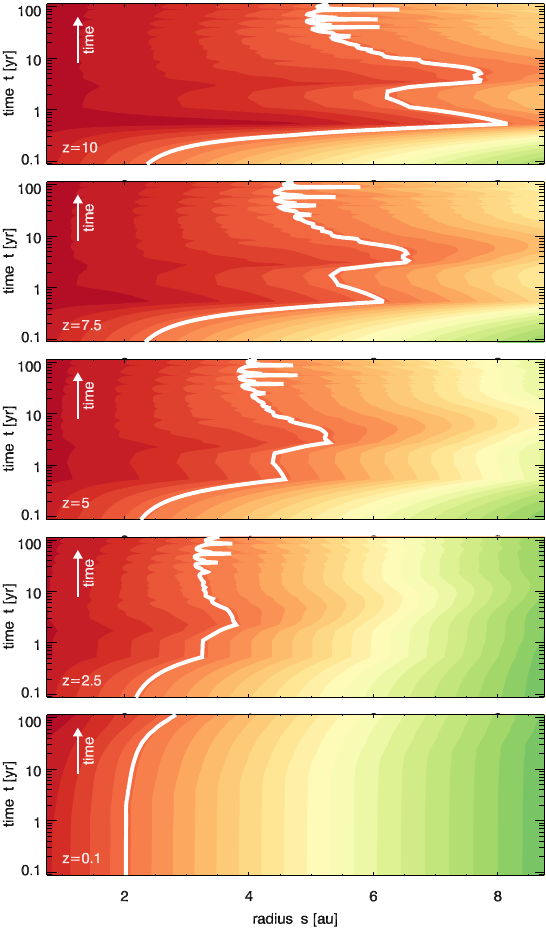}
  \caption{Space-time plots of horizontal slices of $\Psi\equiv \Rc\,A_\phi$ at $z\eq0.1$, 2.5, 5, 7.5 and $10\au$ (bottom to top panels), illustrating the radial wandering of field lines at various heights. White lines exemplify a flux surface initially at $s=2\au$.}
  \label{fig:OA-b4-wandering}
\end{figure}

We begin by describing the qualitative evolution of the magnetic flux.
The initial configuration consists of a vertical magnetic field with constant plasma-$\beta$ parameter ($\bp=10^4$ for model OA-b4) in the disk midplane, that is, concentric cylindrical flux surfaces.
As discussed in detail in sect.~3.4.1 of \gt, the field quickly re-adjusts because of radial magnetic pressure gradients, leading to the typical outward-bent field configuration required for the magnetocentrifugal mechanism to operate.
This adjustment phase is nicely captured in \Fig{OA-b4-wandering}, where we present space-time plots of magnetic flux surfaces at various horizontal slices through the disk.\footnote{That is, isocontours of $\Psi\equiv \Rc\,A_\phi$, where $A_\phi$ is the toroidal component of the vector potential -- see \Sec{vp} for details.}
The timescale, $\tau\simeq 0.5\yr$ for this initial adjustment is roughly independent of radius (abscissa) and height (top to bottom panels).
The exemplary flux surface initially at $\Rc=2\au$ (white line) becomes displaced as the disk evolves by $\simeq 1.5$, 2.5, 4.0, $6.0\au$ at $z=2.5$, 5, 7.5, and $10\au$, respectively,
illustrating the increasing imbalance of the vertically constant $\nicefrac{1}{2}\,B_z^2/\mu_0$ with respect to the decreasing gas pressure, $p(z)$.

The initial relaxation is followed by a phase of collimation, that is, inward migration of flux surfaces high up in the disk atmosphere.
This process is modulated by a second adjustment phase at $t=2-3\yr$, which appears to propagate upward and outward through the disk.
At late times ($t\simgt 20\yr$), the general trend towards collimation stalls at all heights and radii in the disk.
This phase is accompanied by bursts of outward displacement, reflecting time-dependent behavior of the outflow at small radii, where material that falls back onto the disk surface competes with the wind, leading to time-variability and distortion of field lines as disturbances travel along the outflow.
The associated distortion of the flow is also clearly seen in the snapshot from model OA-b4-hr displayed in \Fig{OA-b4-hr-vis2D}, where the innermost field line shows a pronounced kink.
In view of recent observations of a clumpy outflow in EX Lup \citep{2018ApJ...859..111H}, a closer study of this time variability (which moreover modulates the mass loading of the wind) and its precise origin certainly appears warranted.

\begin{figure}
  \center\includegraphics[width=\columnwidth]{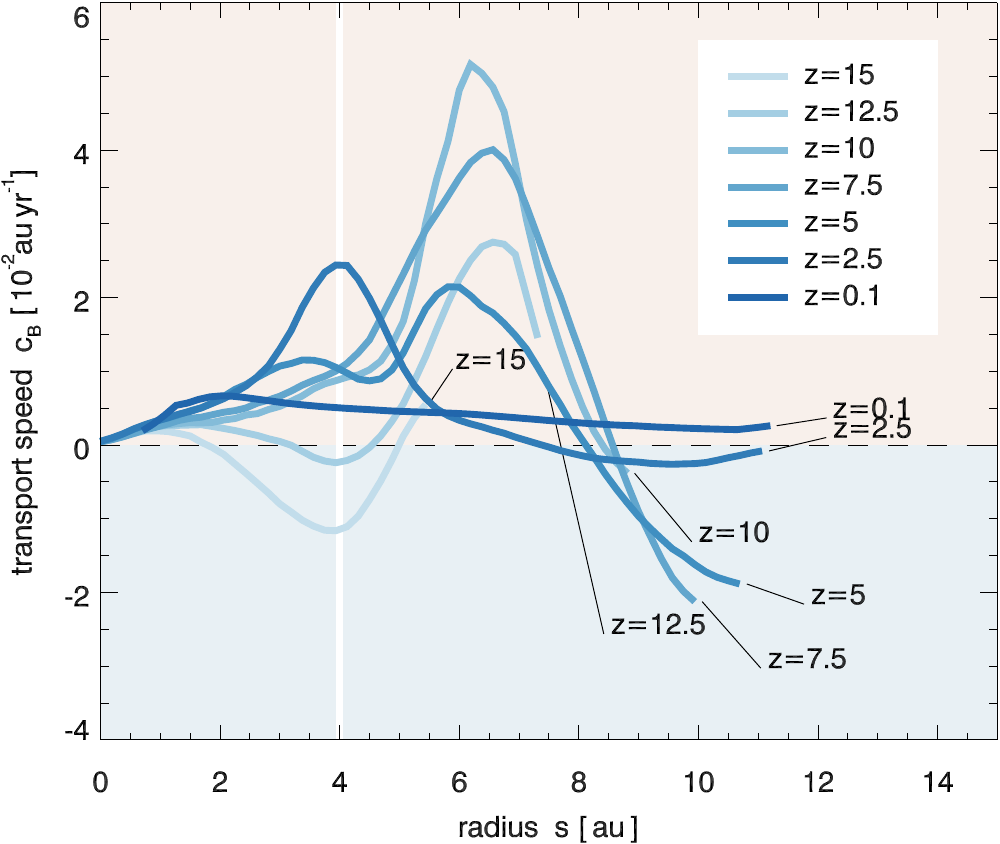}
  \caption{Average radial field-line migration speed (in units of $1\au$ per $100\yr$) shown at various heights, $z$, in the disk. The white line marks the location of the slice shown in \Fig{OA-b4-flux-transport-terms}.}
  \label{fig:OA-b4-flux-transport}
\end{figure}

Based on the space-time plot shown in \Fig{OA-b4-wandering}, and starting from $t\eq17.5\yr$, i.e., after the initial readjustment of the field, we attempt to infer an average radial transport velocity, $c_B$, which we plot in \Fig{OA-b4-flux-transport} as a function of radius, for various heights in the disk.
With the exception of $z\geq 12.5$, where collimation dominates, we generally find an effective \emph{outward} transport of magnetic flux with a magnitude of a few au per century.
For the midplane, we find $c_B\simeq 0.35\ee{-2}\au\yr^{-1}$ (see also \Tab{results}), which corresponds to one percent of the Keplerian velocity, and which is slightly larger than the value of $4\ee{-3}\,\vK$ quoted by \citet{2017ApJ...836...46B} for their Fid0 run with AD only.

\begin{figure}
  \center\includegraphics[width=\columnwidth]{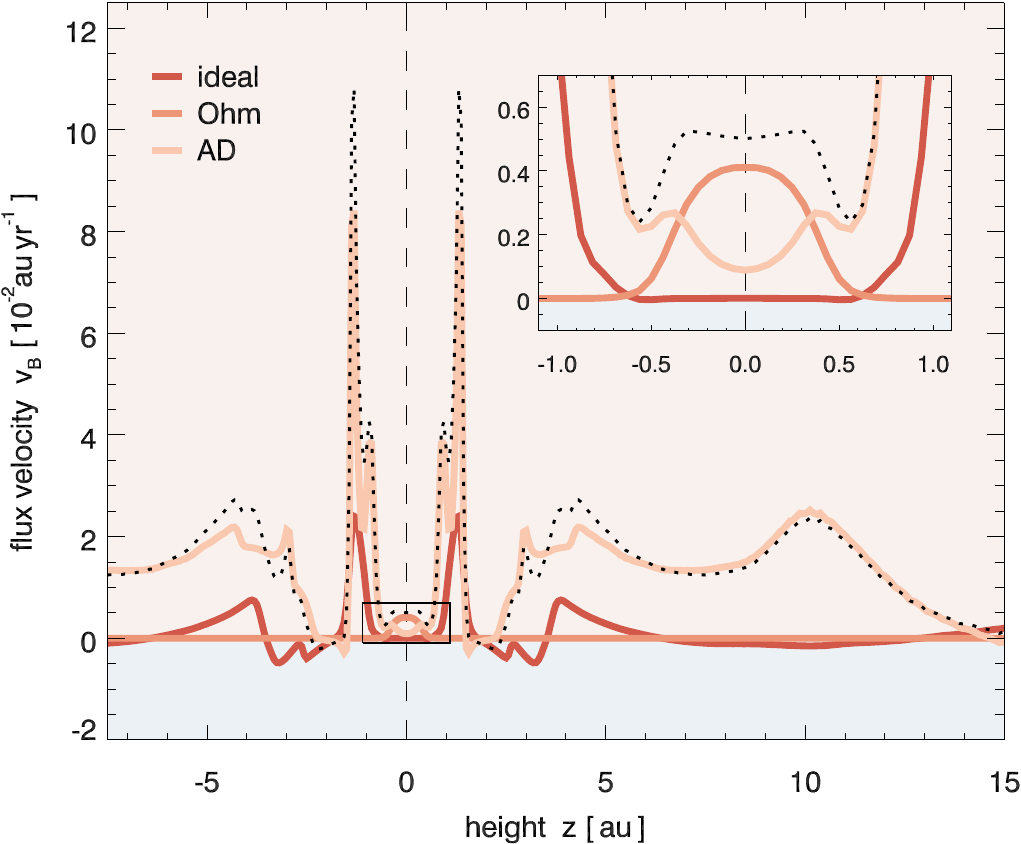}
  \caption{Flux transport terms at $\Rc\eq4\au$. The resulting effective transport velocity, $v_B$, is indicated by a dotted line. The inset shows a magnification for $z\in[-1,1]$.}
  \label{fig:OA-b4-flux-transport-terms}
\end{figure}

The mechanism leading to the radially outward transport can be identified by means of an effective flux velocity, $v_{\Rc,B}\equiv \partial_t A_\phi/B_z$, which can be decomposed as
\beq
  v_{\Rc,B} = \frac{E_\phi}{B_z}
  = \frac{v_\Rc B_z - v_z B_\Rc}{B_z}
  + \Big(\etao + \etad \frac{B_z^2}{B^2}\Big)\frac{J_\phi}{B_z}\,,
  \label{eq:effective_v}
\eeq
where the first term is due to (ideal MHD) field-line advection, and the second describes the slippage of field lines caused by the non-ideal MHD terms.

We plot the ideal-MHD transport term along with the Ohmic and AD contributions in \Fig{OA-b4-flux-transport-terms} for a vertical slice at $\Rc=4\au$, corresponding to the location of the white vertical line in \Fig{OA-b4-flux-transport}.
The resulting flux transport velocity, $v_B$, is shown by a dotted line, and is in decent agreement with the transport speed, $c_B$ from the previous figure.
Advection of field is greatest in a narrow layer around $|z|\simeq 1.5\au$, but is still much weaker than AD. In general, field transport is caused mainly by AD, with the exception of near $z\simeq0$, where Ohmic dissipation takes over.

\subsection{Vertical symmetry and azimuthal flux} 

While we have up to now focused on models with an enforced symmetry with respect to the disk midplane, we now turn our attention to the case where both hemispheres are included in the simulation.
Removing the imposed symmetry allows for a net-azimuthal field to build up in the midplane region of the disk.
As reported in \bai and \blf, such a toroidal disk field is typically found in simulations including Hall-MHD, primarily as a result of the HSI.
In the absence of the HSI, there is no dedicated instability mechanism that can produce a strong radial/toroidal net flux. By-chance fluctuations can, however, create a slight asymmetry that, due to the effect of differential rotation, can become more pronounced when allowed to diffuse into the disk midplane \citep*[see, e.g.,][]{2007ApJ...659..729T}.

\begin{figure}
  \center\includegraphics[width=0.9\columnwidth]{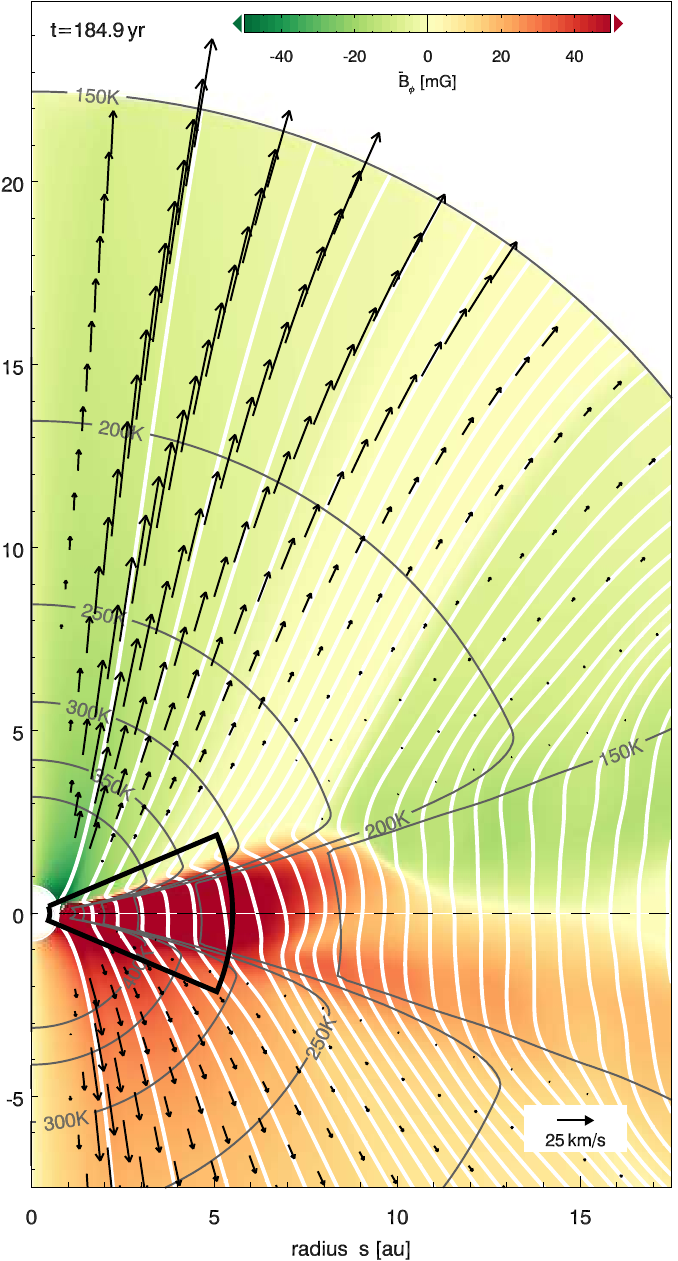}
  \caption{Same as \Fig{OA-b4-vis2D}, but now for model OA-b4-fd, that is, without reflection symmetry about the midplane. Peak field values in this simulation are $B_\phi\simeq 650\mG$.}
  \label{fig:OA-b4-fd-vis2D}
\end{figure}

In \Figure{OA-b4-fd-vis2D}, we show a snapshot of model OA-b4-fd (for ``full domain'') after $t=185\yr$, that indeed shows a pronounced up-down asymmetry.
The amount of azimuthal field is considerably larger than for similar models in \gt, where no symmetry was implied in any of the models.
The precise reason for the more severe symmetry breaking (which is not observed in the similar run `B40' of \bai) is unknown and prompts further investigation with a particular focus on the role played by the inner radial boundary.

\begin{figure}
  \center\includegraphics[width=\columnwidth]{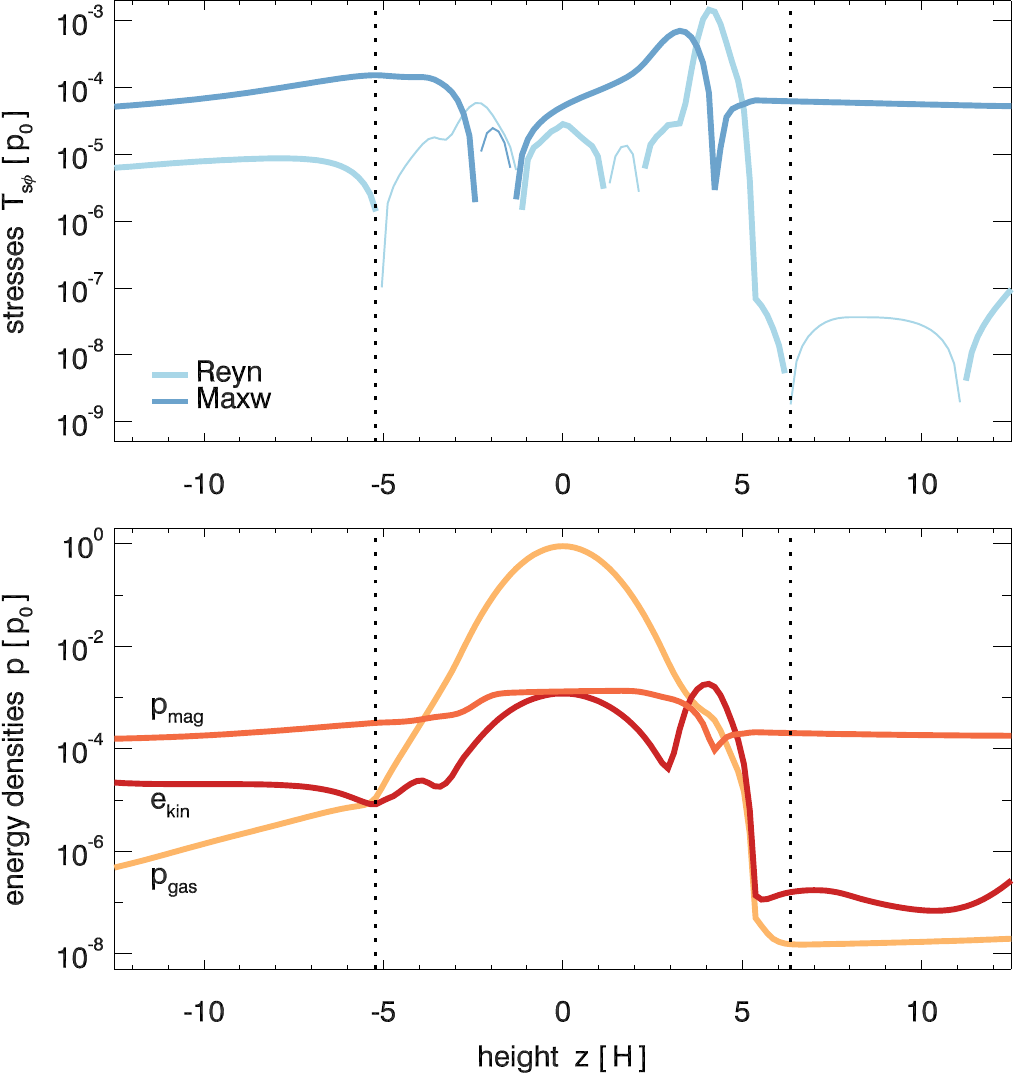}
  \caption{Profiles of the radial-azimuthal components of the Maxwell- and Reynolds stresses (upper panel) and energy densities (lower panel) at $\Rc=2\au$ for run OA-b4-fd, averaged over $t=125$--$175\yr$. In the upper panel, thick (thin) lines indicate positive (negative) values.}
  \label{fig:OA-b4-fd-stresses}
\end{figure}

The positive polarity of the azimuthal field in the midplane implies a current sheet above the disk, where the disk field connects to the wind with $B_\phi<0$.
As a consequence, the wind base is pushed upward in the disk compared to the lower half, even though the inferred mass loss rates are comparable on both sides (see \Tab{results}, where values are listed for the upper/lower disk half separately).
While the Alfv{\'e}n radius in the lower hemisphere is comparable to model OA-b4 (with an assumed symmetry about the midplane), the additional inward bending at the current sheet reduces the lever arm to a value comparable to the magnetically weaker OA-b6 model (see \Tab{results}) in the upper hemisphere.

Large-scale coherent planar fields may drive accretion in PPDs via the radial-azimuthal component of their associated Maxwell stress, and might have important implications for planet migration theory \citep{2017MNRAS.472.1565M,2018MNRAS.477.4596M}. Thus, we plot vertical profiles of the Reynolds and Maxwell stresses in \Figure{OA-b4-fd-stresses}, where substantial values are obtained near $z=+4\au$, i.e., at the location of the current sheet.
Integrating the stresses vertically inside $|z|\le \zb$ (as indicated by dotted lines in \Fig{OA-b4-fd-stresses}), we find an average Reynolds stress of $\mathcal{R}_{\Rc\phi}\eq(0.87\pm 0.02)\ee{-4}\,p_0$ as well as a Maxwell stress of $\mathcal{M}_{\Rc\phi}\eq(1.34\pm  0.02)\ee{-4}\,p_0$, translating into an estimated accretion rate of $\dot{M}_{\rm d}=0.74\ee{-8}\Mdot$ at $2\au$, which amounts to ten percent of the measured $\dot{M}_{\rm accr}=0.76\ee{-7}\Mdot$.
As compared to the other simulations (see \Tab{results}), this highlights the importance of not only the vertical field, but also the laminar in-plane field in setting the accretion rate of the system.

\subsection{Sensitivity to thermodynamic aspects} \label{sec:other} 

Given the importance of the disk thermal structure on the mass loading of the TACO, we briefly discuss three simulation sets where we control different aspects of the radiative physics entering the equation.

Model OA-b4-lop (standing for ``low opacity'') has $\kR\eq 0.1\opa$, that is, an opacity coefficient reduced by a factor of ten compared to the fiducial setup.
As can be seen from \Tab{results}, the key diagnostics of the wind are very similar to the standard case, indicating that the outcome does not appear to be critically sensitive to the assumed constant opacity coefficient.

As discussed earlier, in model OA-b4-ohm we explicitly enable the Ohmic and ambipolar dissipational heating terms.
This is in contrast to the standard model sets, where the non-ideal MHD heating terms were deliberately disabled.
Comparing the corresponding results (shown in \Tab{results}) with our fiducial run demonstrates that ambipolar dissipation heating is irrelevant for setting the parameters of the outflow, e.g., via affecting pressure gradient forces or the ionization level.
The inclusion of the non-ideal heating terms does, however, appear to modestly affect the timescale of the flux evolution, even though it is not clear by which means this is achieved.

\begin{figure*}
  \center\includegraphics[width=0.9\textwidth]{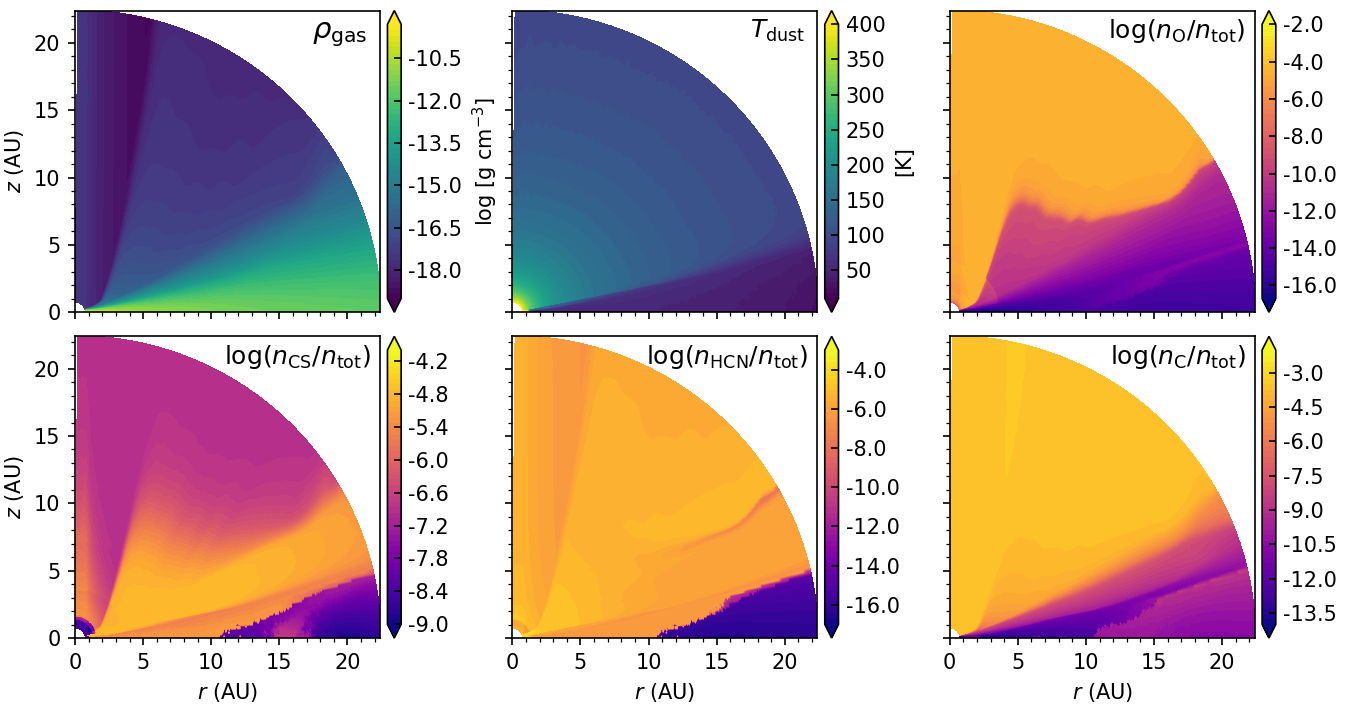}
  \caption{Gas density from model OA-b4, plus the dust temperatures and chemical abundances of select species relative to $n_\mathrm{tot} \equiv n_\HH + 2\,n_{\HH_2}$ obtained by post-processing the model with \rad and \krome, respectively.}
  \label{fig:OA-b4-abundances}
\end{figure*}

\begin{figure*}
  \center\includegraphics[width=0.9\textwidth]{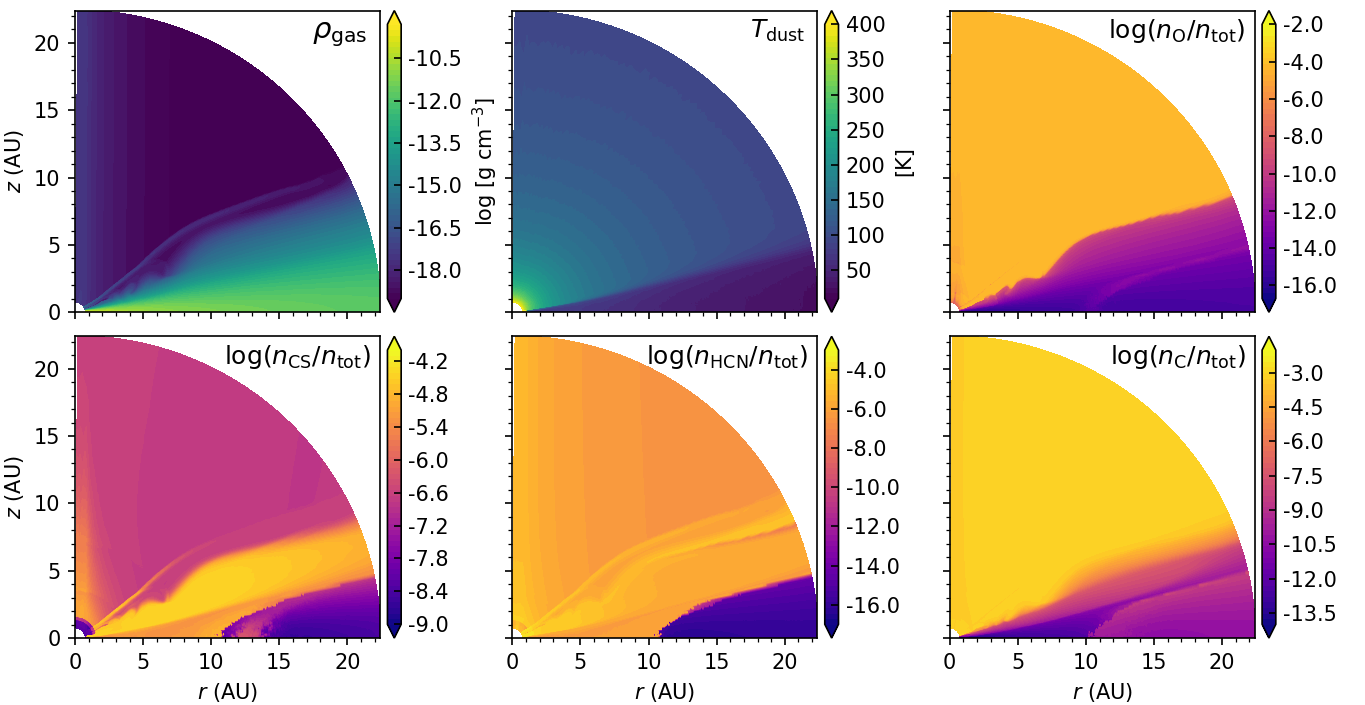}
  \caption{Same as \Figure{OA-b4-abundances}, but for model OA-b8.}
  \label{fig:OA-b8-abundances}
\end{figure*}

Finally, model OA-b4-noir (for ``no irradiation'') is a model that neither has stellar irradiation heating to the surface nor thermochemical heating driven by the FUV flux from the star. It does, however, still include the contribution of FUV and X-rays going into the ionization chemistry.
As such, the model serves to expose the (potentially spurious) effect of initial and boundary conditions (in particular, for the diffuse radiation field) on the emerging thermal structure of our disk model.
Notably, this model creates a noticeably (by a factor of about two) larger mass loss rate into the wind.
Otherwise, the model is again fairly comparable to the default parameter set, OA-b4, which may indicate that the temperature structure is at least in part determined by the inner radial boundary condition of the radiation energy density rather than the irradiation heating term.


\section{Synthetic observations} \label{sec:synthetic}

\begin{figure*}
  \center\includegraphics[width=0.90\textwidth]{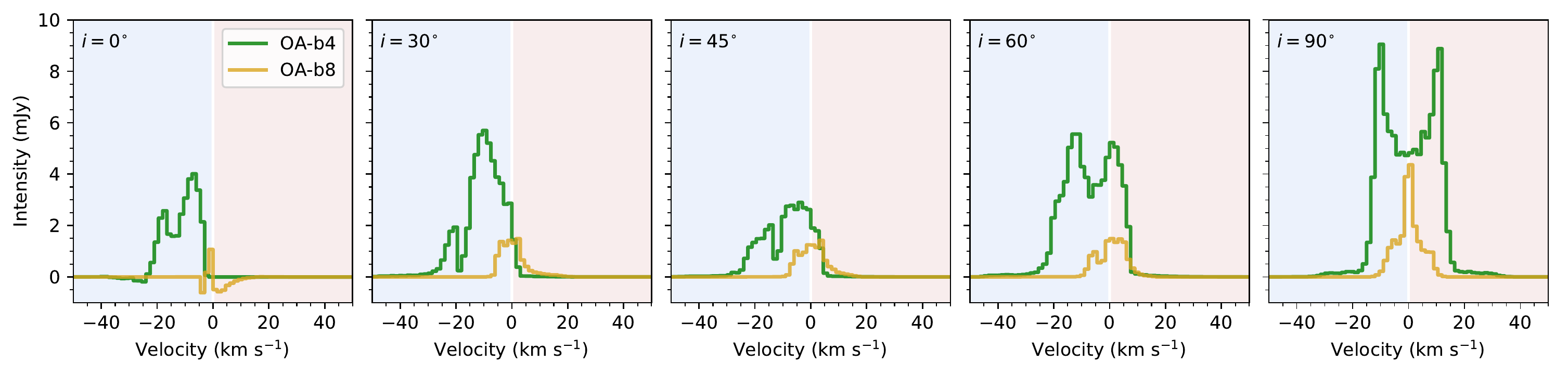}
  \caption{Synthetic line profiles of the atomic oxygen transition at $63.2\um$ (${}^3$P$_1$--${}^3$P$_2$) and at inclination angles of $0\degr$, $30\degr$, $45\degr$, $60\degr$, and $90\degr$. Green (yellow) lines denote model OA-b4 (OA-b8).}
  \label{fig:O-syn-spectra}
\end{figure*}

\begin{figure*}
  \center\includegraphics[width=0.95\textwidth]{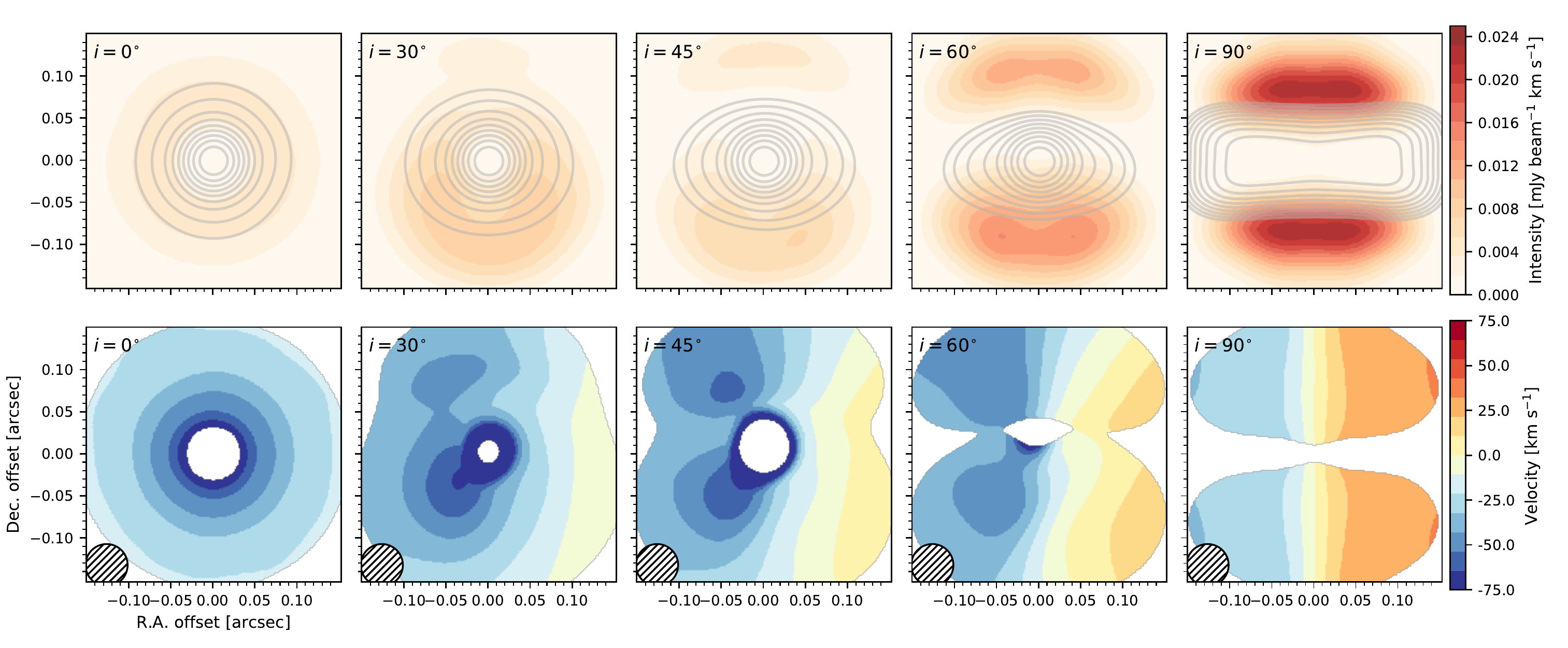}
  \includegraphics[width=0.95\textwidth]{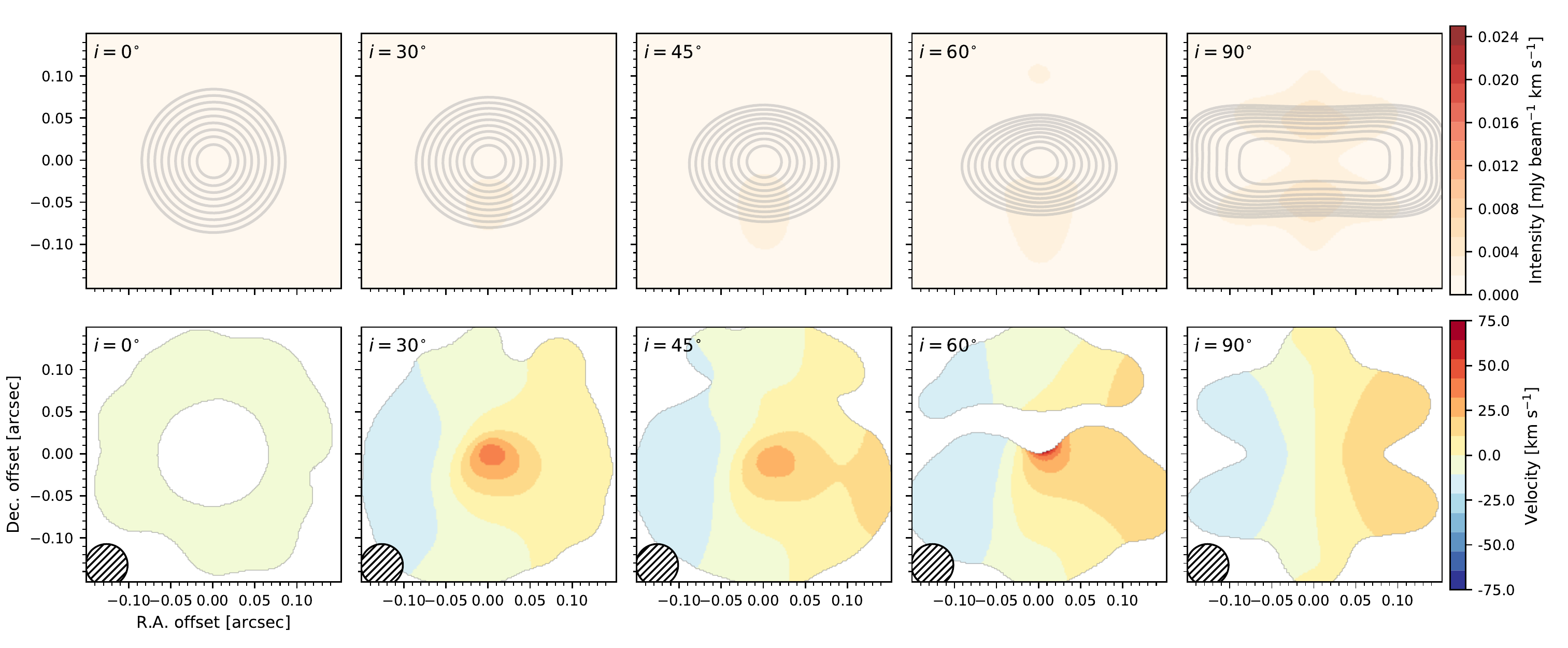}
  \caption{Synthetic moment maps of the atomic oxygen transition at $63.2\um$ (${}^3$P$_1$--${}^3$P$_2$). The top two (bottom two) rows show the moment 0/1 maps for model OA-b4 (OA-b8) at inclination angles of $0\degr$, $30\degr$, $45\degr$, $60\degr$, and $90\degr$. Note the clear, blue-shifted asymmetry in the moment 1 maps of the model with an outflow (OA-b4). This feature persists at all inclinations notwithstanding $90\degr$ (edge-on). Note also that all moment 1 maps are clipped where the intensity three times the root-mean-square (RMS) value.}
  \label{fig:O-syn-momentmaps}
\end{figure*}

Pursuing synthetic observations of line emission naturally starts by determining suitable chemical tracers.
Although the simplified thermochemical prescription that we evolve along with the dynamical simulations already provides us with a first guess at a set of relevant chemical species (see \Sec{pdr} and Appendix~\ref{app:simple_pdr}), we turn instead to post-processing the simulations with \krome and a much larger chemical network (see \Sec{lime}) in order to search for distinct signatures of outflow from other chemical species.

In \Figure{OA-b4-abundances}, we show the relative abundances of select chemical species as calculated by \krome for model OA-b4, along with the dust temperature as calculated by \rad and the gas density from \nir.
These species were selected based on their abundances, the presence of emission lines accessible by ALMA or SOFIA (the Stratospheric Observatory For Infrared Astronomy), and whether or not \lime predicts significant emission in those lines. Furthermore, we prefer atoms/molecules and lines for which there is a substantial difference between models OA-b4 and OA-b8 (i.e.\ that distinguishes between the two models).
In \Figure{OA-b8-abundances}, we show the abundances of the same species for model OA-b8, which has a $100\times$ weaker initial magnetic field than model OA-b4.
As can be seen by comparing Figs.\ \ref{fig:OA-b4-abundances} and \ref{fig:OA-b8-abundances}, while model OA-b4 has a clear and significant collimated outflow, model OA-b8 has only a weak photoevaporative wind.
In the following, we will exploit this difference to look for signatures between a disk with a collimated magnetic wind versus a weak radial photoevaporative wind.

\begin{figure*}
  \center\includegraphics[width=0.90\textwidth]{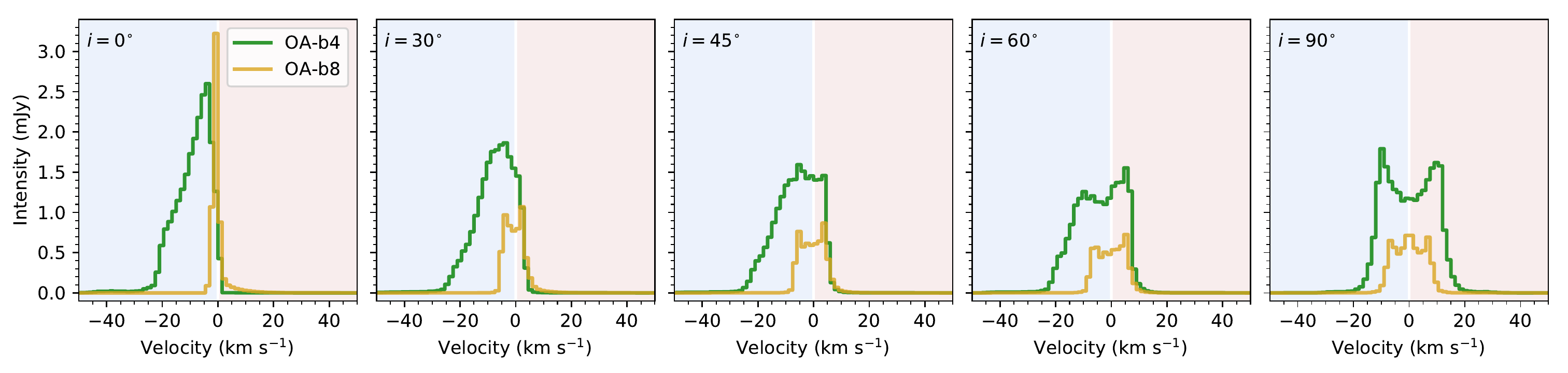}
  \caption{Synthetic line profiles of the atomic carbon transition at $492.16\GHz$ (${}^3$P$_1$--${}^3$P$_0$) and at inclination angles of $0\degr$, $30\degr$, $45\degr$, $60\degr$, and $90\degr$. Green (yellow) lines denote model OA-b4 (OA-b8).}
  \label{fig:C-syn-spectra}
\end{figure*}

\begin{figure*}
  \center\includegraphics[width=0.90\textwidth]{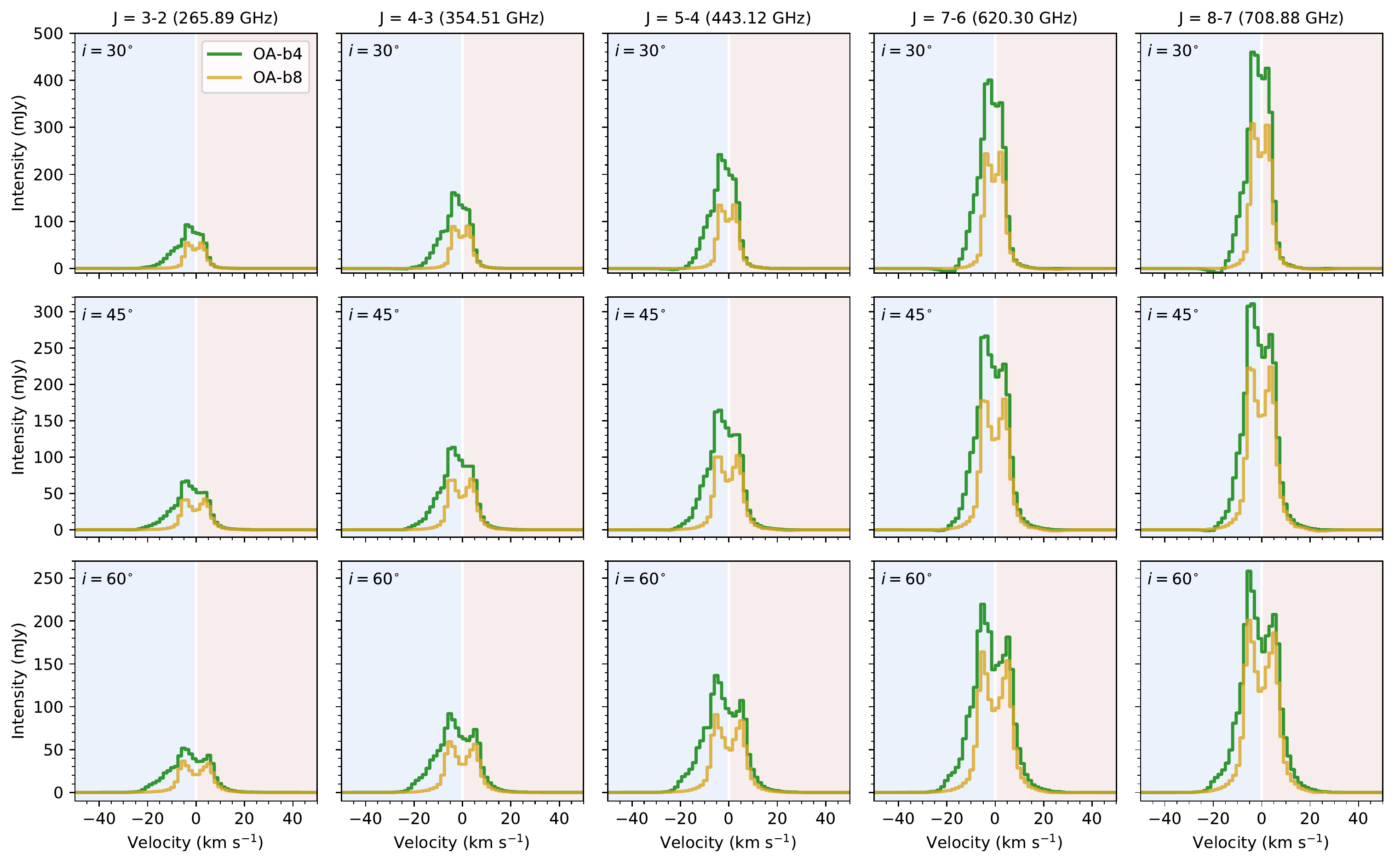}
  \caption{Synthetic line profiles of HCN transitions J=3--2 ($265.89\GHz$), J=4--3 ($354.51\GHz$), J=5--4 ($443.12\GHz$), J=7--6 ($620.30\GHz$), and J=8--7 ($708.88\GHz$) at inclinations of $30\degr$, $45\degr$, and $60\degr$ (top to bottom rows). Green (yellow) lines denote model OA-b4 (OA-b8). Note the varying $y$-axis limits between rows (i.e., inclinations).}
  \label{fig:HCN-syn-spectra}
\end{figure*}

\Figure{O-syn-spectra} shows the integrated spectral profiles of the atomic oxygen emission line at $63.2\um$ in models OA-b4 and OA-b8 as a function of velocity, integrated over the disk, for different inclinations.
The profiles were produced using \lime, assuming a distance from the observer of 140 pc, after subtracting the continuum in the image plane using CASA \citep{2007ASPC..376..127M}. Note that here, as well as in the similar Figs.~\ref{fig:C-syn-spectra} and \ref{fig:HCN-syn-spectra}, intensities are predicted assuming \emph{no} extinction along the line-of-sight between the observer and the source.

The model with a collimated outflow (OA-b4) shows a characteristic blue-shift in the line profile relative to the model with a weak radial wind (OA-b8) that is nearly independent of inclination angle.
The [\ion{O}{1}] line traces the outflowing warm gas, and so it gives the emission line a characteristic blue shift which is visible at all angles except $90\degr$ (edge-on).
Despite the moderate intensity of the line ($\sim 2-9\mJy$), the differences between outflow and no-outflow models are very clear and this line could be used to distinguish between the two scenarios.

The $63.2\um$ [\ion{O}{1}] line is indeed known to be an important coolant in PDRs and circumstellar disks \citep[e.g.][]{2013PASP..125..477D}, and its brightness is expected to correlate with FUV and X-ray luminosity in photoevaporative winds \citep{2008ApJ...683..287G}.
In our case, however, it is instead the presence of warm oxygen in the outflow that results in a brighter line luminosity at $63.2\um$ (relative to OA-b8), and allows us to clearly distinguish between outflow and no-outflow models.

These effects are also visible in the moment 0 (integrated intensity) and moment 1 (intensity-weighted velocity) maps shown in \Fig{O-syn-momentmaps}. The substantial outflow velocity in model OA-b4 results in a clear characteristic asymmetry in the moment 1 maps, providing an additional observational diagnostic for the presence of a collimated outflow.

\begin{figure*}
  \center\includegraphics[width=0.95\textwidth]{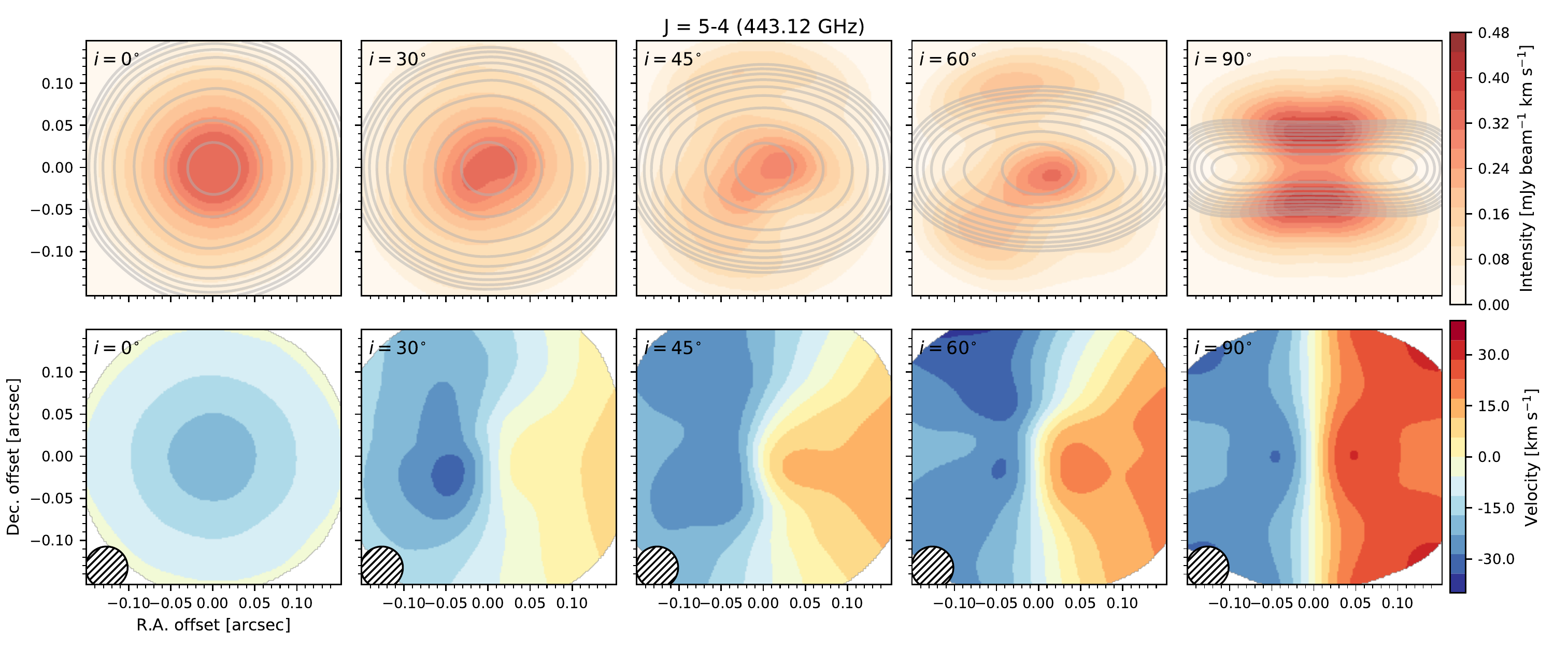}
  \includegraphics[width=0.95\textwidth]{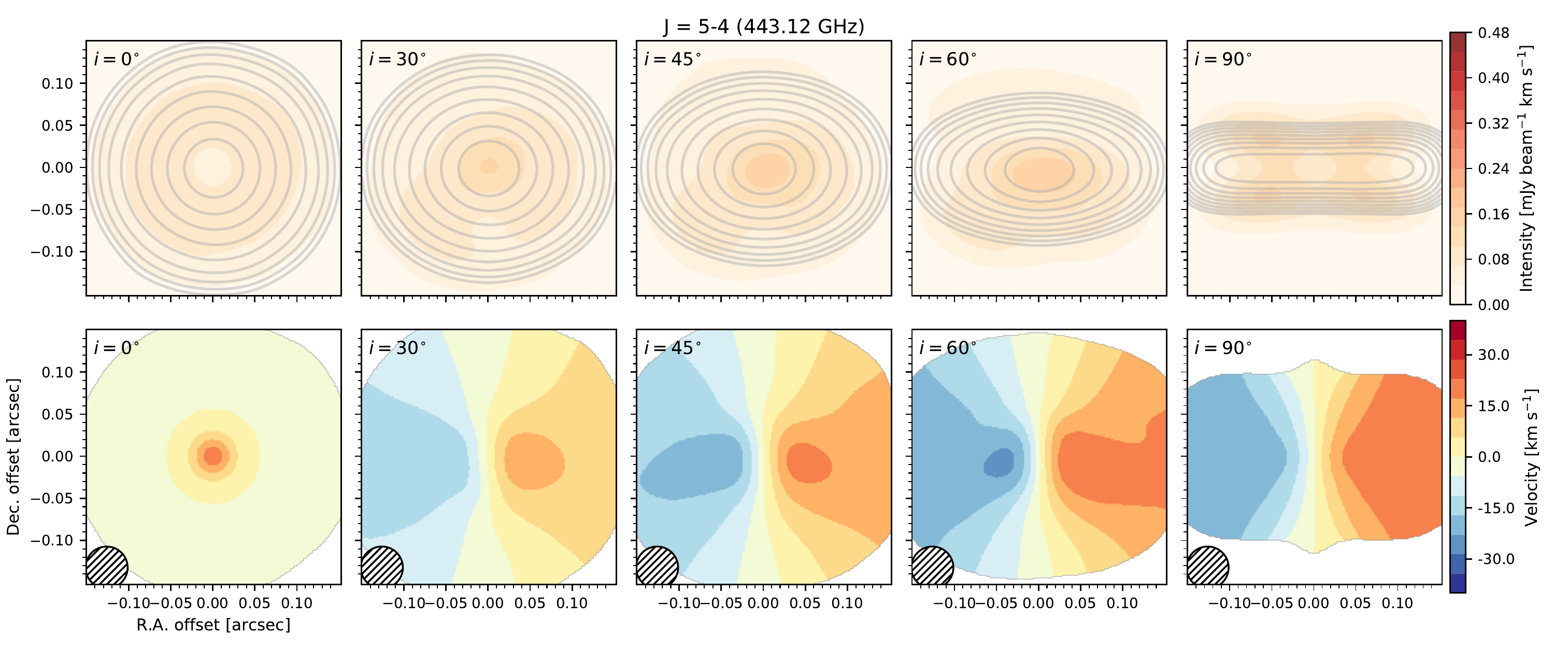}
  \caption{Synthetic moment maps of the J=5--4 ($443.12\GHz$) transition of HCN. The top two (bottom) rows show the moment 0 and moment 1 maps for model OA-b4 (OA-b8) at inclinations of $0\degr$, $30\degr$, $45\degr$, $60\degr$, and $90\degr$.}
  \label{fig:HCN-syn-momentmaps}
\end{figure*}

Although the $63.2\um$ [\ion{O}{1}] moment maps predict a clear difference between the model with a magnetocentrifugal outflow (OA-b4) and that with a weaker, photoevaporative outflow (OA-b8), the line luminosity is too low to be detectable with current facilities (e.g., SOFIA), even under these ideal conditions. Thus, we turn to exploring other, brighter emission lines that are detectable, in particular, with ALMA.

We considered atoms and molecules with emission lines typically associated with PDRs or outflows and that are predicted to be relatively abundant in our post-processed models. These include C, C$^+$, CO, as well as CN, CS, HCN, and SiO. However, here we choose to focus on not only the brightest lines, but also those that show a clear difference between outflow and no-outflow models. For example, we find that atomic ions (like C$^+$) are not abundant enough to produce significant emission in our models. Furthermore, although we predict there is appreciable flux in the sub-mm lines of CN and CS, i.e., $\cal{O}$($10\mJy$) and $\cal{O}$($10-100\mJy$), respectively, the line formation region is similar in both models, and thus the resulting spectra and moment maps do not clearly distinguish between outflow and no-outflow models.

Instead, in \Fig{C-syn-spectra}, we show the integrated spectral profiles for the atomic carbon line at $492.16\GHz$ as a function of inclination. Like the $63.2\um$ [\ion{O}{1}] line, the [\ion{C}{1}] $492\GHz$ emission also shows a characteristic blue-shift in the line profile at most inclinations. Like [\ion{O}{1}] at $63.2\um$, the [\ion{C}{1}] line at $492\GHz$ also traces the warm outflowing gas. What differs from [\ion{O}{1}], however, is that the double-peak profile is all but washed out in the outflow model. Furthermore, although the brightness of the $492\GHz$ line is only moderate, the differences between the outflow and no-outflow models are clear, and the sensitivity of ALMA is good enough that these differences should be detectable with a reasonable amount of array time (see also \citealt{2020MNRAS.492.5030H}). The moment 0 and 1 maps for [\ion{C}{1}] $492\GHz$ are, meanwhile, qualitatively similar to the [\ion{O}{1}] moment maps, and so, for brevity, are not included here.

\Figure{HCN-syn-spectra} shows integrated spectral profiles for different transitions of HCN, a molecule that is commonly detected by ALMA in and around protoplanetary disks.
As can be seen, both outflow and no-outflow models predict significant line intensities, in particular at higher energy transitions.
Although the outflow model (OA-b4) has a slightly greater absolute intensity than the no-outflow model (OA-b8), given the difficulty in comparing absolute fluxes between different objects, however, we instead look for differences in the spectra as a function of velocity in order to distinguish between the character of the models.
As with atomic carbon and oxygen, the outflow model shows a significant, asymmetric, blue-shifted feature in the HCN lines at all inclinations. In contrast, the profile remains symmetric across the systemic velocity in the no-outflow model.
We find that, in the no-outflow model, the HCN emission is produced predominantly at the disk surface, while, in the outflow model, significant emission is produced in the gas that has been launched into the outflow.

\Figure{HCN-syn-momentmaps} shows the moment maps of the J=5-4 transition of HCN as a function of inclination; the moment maps for different transitions are qualitatively similar and so we omit them here.
Although the moment 0 maps for models OA-b4 and OA-b8 have different absolute intensities, the morphologies are similar to one another, and shows that the HCN emission is concentrated in the inner disk.
The moment 1 maps, however, clearly show a blue-shifted velocity asymmetry in OA-b4 relative to OA-b8 (notwithstanding $i = 90\degr$), corresponding to the asymmetry seen in the spectra.

The spectra and moment maps for models OA-b4 and OA-b8 for CO and SiO lines within the ALMA bands are qualitatively quite similar to those for HCN, and therefore we choose not to show those results here.
Indeed, we purposefully choose HCN over CO because, even though it is less abundant, this is made up for with much larger Einstein A coefficients at comparable line frequencies.

The synthetic observations presented here demonstrate that collimated outflows from disks can produce clear, characteristic features in the intensity and velocity structure of some atomic and molecular lines -- for instance, HCN, but neither CN, nor CS.
The outflows and disk surface in our models are warm, that is, $\mathcal{O}(200\K)$, which then correlates with higher energy transitions being generally brighter.
However, many observational outflow studies with ALMA focus on lower energy transitions and/or on CO and its isotopologues \citep[e.g.,][]{2018A&A...620L...1G,2019ApJ...883....1Z}, whereas our results suggest to instead look at higher frequency transitions in other species (e.g., C; \citealt{2015ApJ...802L...7T,2016A&A...588A.108K}, or HCN).


\section{Summary and conclusions} \label{sec:summary}

In summary, we have presented a set of extended simulations, improving our model over \gt both in terms of the computational domain, which now covers the radial range of the PPD relevant for comparison with observations, as well as the thermodynamics, including a simple treatment of thermo-chemical heating and cooling effects, as well as redistribution of energy by radiative transfer.

We have studied how the relative mass budgets, radially through accretion, and vertically through mass loss in the wind, depend on the assumed energy input into the system in the FUV band.
We find that while the wind mass loss indeed scales with the thermochemical energy input, the overall mass accretion rate appears to be largely unaffected --- motivating future investigation of the baseline mass accretion rate set by disk interior processes such as VSI.

Compared to \gt, we have analyzed the emerging wind kinematics in more detail with respect to characteristic speeds, flow invariants, as well as the detailed forces acting along field lines.
We have robustly identified the magnetocentrifugal mechanism as the driver of the (super-Keplerian) outflow.
This is in contrast to \bai and \citet{2019ApJ...874...90W}, who promote wind launching primarily by vertical pressure gradients (both magnetic and thermal) with minimal magnetic lever arms, and do not find super-Keplerian rotation in the outflow.
The origin of this discrepancy is at present unclear; in view of the immense complexity of the current simulations, a detailed comparison of assumptions and specific treatments is warranted.

Despite the comparatively small effect of the improved thermodynamics in our current model, we have looked into the role played by the various heating mechanisms and have found that Ohmic and AD heating remain sub-dominant compared to stellar irradiation and thermochemical heating.

An important question arising in the formation and long-term evolution of PPDs is that of the evolution of the entrained vertical magnetic flux.
Employing a vector-potential formulation, we have shown that the magnetic flux in our models escapes the disk radially on secular timescales, which we attribute to the combined effect of Ohmic (near the midplane) and ambipolar (away from the midplane) diffusion of the field lines.

Most of the models presented here employ an enforced mirror symmetry with respect to the midplane, precluding the build-up of a net-azimuthal field.
In contrast to this, our model that evolved both hemispheres of the disk did develop a pronounced asymmetry between the upper and lower disk halves as well as a significant azimuthal disk field -- something that is typically a consequence of including the Hall effect, and was not observed in \gt. A possible explanation may be the corrugation of the midplane by the VSI \citep{2019arXiv191202941C}.

Going beyond a direct diagnostic approach, the inclusion of chemical abundances into the simulation -- in the form of a minimal ``black-box'' PDR library advanced along with the RMHD solver, and also via more detailed post-processing -- has enabled us to derive synthetic maps for different atomic and molecular emission lines.
The synthetic spectral profiles and moment maps for several HCN lines, the $63.2\um$ [\ion{O}{1}] line, and the $492\GHz$ [\ion{C}{1}] line show a characteristic blue-shifted velocity asymmetry in the model with a collimated outflow versus the model without (Figs.\ \ref{fig:O-syn-momentmaps} and \ref{fig:HCN-syn-momentmaps}).
The spectral profiles also demonstrate that outflowing material can produce a P~Cygni-like net absorption ([\ion{O}{1}]; see \Fig{O-syn-spectra}), or that the forward velocity of the outflowing gas can mask the signature of gas rotation in the outflow (see \Fig{C-syn-spectra}).
Together, the spectral line intensities and moment 1 maps provide a means to explore outflows at their base, right above the disk surface, and potentially allow for observationally distinguishing between disks with and without collimated outflows.
While the presented results merely represent a demonstration, refined outputs of this type will provide useful priors in the endeavor to robustly observe collimated, centrifugally-driven disk outflows, and to delineate them from their purely photoevaporative counterparts.


\acknowledgments

The authors thank Ewine van~Dishoeck for helping to initiate this research and for useful discussions,
and the anonymous referee for providing a timely and concise report.
OG acknowledges helpful discussions with Christian Fendt, Ruben Krasnopolsky, Xuening Bai, Jeremy Goodman, Wilhelm Kley, Mart{\'i}n Pessah, and Tobias Heinemann.
JPR acknowledges helpful discussions with Tommaso Grassi, Ilse Cleeves, and Nick Ballering.
The research leading to these results has received funding from the European Research Council under the European Union's Horizon 2020 research and innovation programme (grant agreement No 638596).
This research was supported in part by the National Science Foundation under Grant No.\ NSF PHY17-48958.
NJT's work was carried out at the Jet Propulsion Laboratory, California Institute of Technology, under a contract with NASA and with the support of Exoplanets Research Program grant 17-XRP17\_2-0081.
This research was supported by the Munich Institute for Astro- and Particle Physics (MIAPP) of the DFG cluster of excellence ``Origin and Structure of the Universe''.
OG and NJT thank the CAS Research Group ``The Ionisation Structure of Planet Forming Discs and their Atmospheres'' of PI Barbara Ercolano for its hospitality.
The Centre for Star and Planet Formation is funded by the Danish National Research Foundation (DNRF97).
JPR was supported, in part, by the Virginia Initiative on Cosmic Origins and, in part, by the NSF under grant nos.~AST-1910106 and AST-1910675.
This work used a modified version of the \nir code based on v3.5 developed by Udo Ziegler at the Leibniz Institute for Astrophysics Potsdam (AIP).
We acknowledge that the results of this research have been achieved using the PRACE Research Infrastructure resource MareNostrum-4 based in Spain at the Barcelona Supercomputing Center (BSC).
Numerical computations were moreover performed on the \texttt{astro} partition of the Steno cluster at the University of Copenhagen HPC center, funded in part by Villum Fonden (VKR023406).


\software{\nir \citep{2004JCoPh.196..393Z,2011JCoPh.230.1035Z,2016A&A...586A..82Z}, \rad \citep{2012ascl.soft02015D}, \krome \citep{2014MNRAS.439.2386G}, \lime \citep{2010A&A...523A..25B}, CASA \citep{2007ASPC..376..127M}, Matplotlib \citep{2007CSE.....9...90H}}


\appendix

\section{Relaxation test for the discrete ordinates scheme} \label{sec:relax}

\begin{figure}[hb]
  \center\includegraphics[width=0.5\columnwidth]{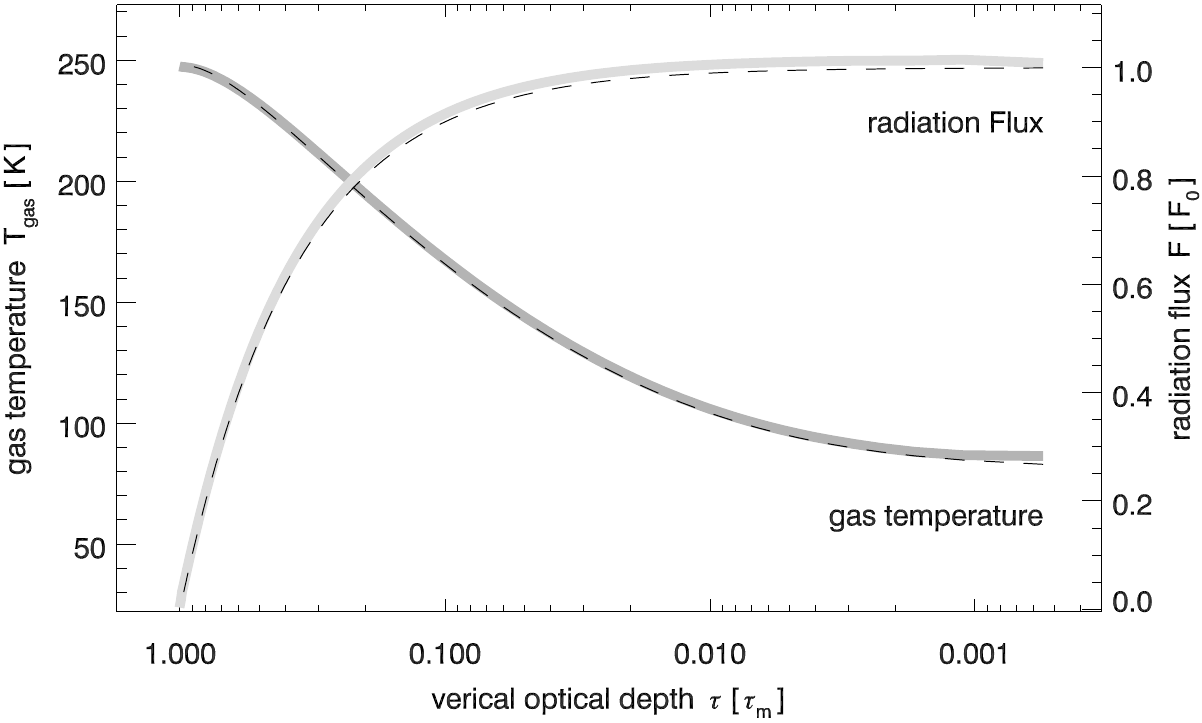}
  \caption{Results from the relaxation test from section 4.1 of \citet{2007ApJ...665.1254B}.}
  \label{fig:BDNL-atmosphere}
\end{figure}

We here briefly describe the benchmark solution of a plane-parallel radiative atmosphere that we used to check our implementation of the discrete ordinates scheme \citep[see][and references therein]{2007ApJ...665.1254B}.
Consider a one-dimensional atmosphere with constant acceleration, $\mathbf{g}\equiv -g\,\zz$, and a prescribed heating rate
\beq
  Q^+(z) \equiv \pi F_0 \frac{\rho\,\kappa}{\tau_{\rm m}}\,,
\eeq
where $F_0\equiv \frac{c}{4\pi}\,\aR\,T_{\rm eff}^4$ is the flux at $\tau=0$, with $T_{\rm eff}=100\K$ the effective temperature of the atmosphere. It can be shown that, once hydrostatic and radiative equilibrium are reached, the steady-state vertical radiative flux is given by
\beq
  F_z(\tau) = F_0 \left( 1 - \frac{\tau}{\tau_{\rm m}}\right)\,,
  \label{eq:Fz}
\eeq
with $\tau(z) \equiv \int_{\infty}^{z}\rho(z')\,\kappa\;\dd z'$, and where the optical depth to the midplane, $\tau_{\rm m}\equiv\tau(0)$ is treated as a free parameter. Using the Eddington approximation to relate the flux to the mean intensity, and assuming $\tau_{\rm m}\simgt 10$, one can derive an analytic temperature profile of the form
\beq
  T^4(\tau) = \frac{3}{4}\,T_{\rm eff}^4
  \left( 1 + \frac{\sqrt{3}}{3\tau} - \frac{\tau}{2\tau_{\rm m}} \right)\,\tau\,.
  \label{eq:T4}
\eeq
Assuming a constant opacity $\kappa=\kappa_0$, we initialize our simulation with an isothermal atmosphere of (constant) temperature $T=T_{\rm m}\equiv T(\tau_m)$, as given by \Eqn{T4}, that is
\beq
  \rho(z) = \rho_0\,\exp{(-z\,g/\cS^2)}\,,\qquad
  \epsilon(z) = \epsilon_0\,\exp{(-z\,g/\cS^2)}\,,\qquad
  \eR(z) = \aR\,T_m^4\,,
\eeq
where, $\epsilon_{\rm m}=p_{\rm m}/(\gamma-1)$ with $\gamma=5/3$, and $\cS^2=p_{\rm m}/\rho_{\rm m}=T_m\,k_{\rm B}(\bar{\mu}\,m_{\rm H})^{-1}$, with $\bar{\mu}=1$.

Initial conditions for $\rho_0$, $g_0$, and $\kappa_0$ are chosen such that the resulting atmosphere covers approximately three decades in optical depth over the adopted vertical extent of the simulation domain.
Boundary conditions are mirror symmetric at the disk midplane, i.e., we only model one half of the entire disk.
At the surface, boundary conditions are of the standard ``outflow'' type, where we moreover extrapolate
\beq
  \rho(z+\Delta z) \,=\, \rho(z)\,\exp{(-\Delta z g/\cS^2)}\,,
  \qquad\text{and}\qquad
  \epsilon(z+\Delta z) \,=\, \epsilon(z)\,\frac{\rho(z+\Delta z)}{\rho(z)}
\eeq
according to the requirement of hydrostatic equilibrium, while maintaining a zero gradient in the temperature. We keep $\eR(z_1)$, at the upper domain boundary, $z_1$, at near-zero temperature to force the FLD flux used at low optical depth into the free-streaming limit.

To help the model settle into a hydrostatic balance, we apply a constant kinematic viscosity of substantial amplitude. To avoid viscous heating having an effect on the final solution, we however suppress the dissipational heating (in the very same way as we describe for the MHD heating terms in \Sec{thermo}).
As the density profile adjusts, we moreover automatically tune the opacity coefficient, $\kappa_0$, via a damped feedback loop, thus relaxing the integrated optical depth to the target value of $\tau_{\rm m}=100$. We do the same for the gravitational acceleration parameter, $g$, which we tune according to the requirement $p_m=g\,\Sigma_{\rm m}$, where $\Sigma_{\rm m}$ is the column density at the midplane (i.e., half the total surface density). We show the relaxed solution in \Fig{BDNL-atmosphere}, which is in good agreement with the analytic profiles from \Eqn{Fz} and \Eqn{T4}, plotted as dashed lines. The small discrepancy at low optical depth may be related to our choice of  boundary condition for the thermal energy density, which was derived assuming a vanishing gradient in the gas temperature. This is in contrast to the analytic temperature profile, which only asymptotically reaches $\partial_z T=0$.


\section{Calculation of gas heating/cooling rates}
\label{app:simple_pdr}

\begin{figure}
 \center\includegraphics[width=0.55\columnwidth]{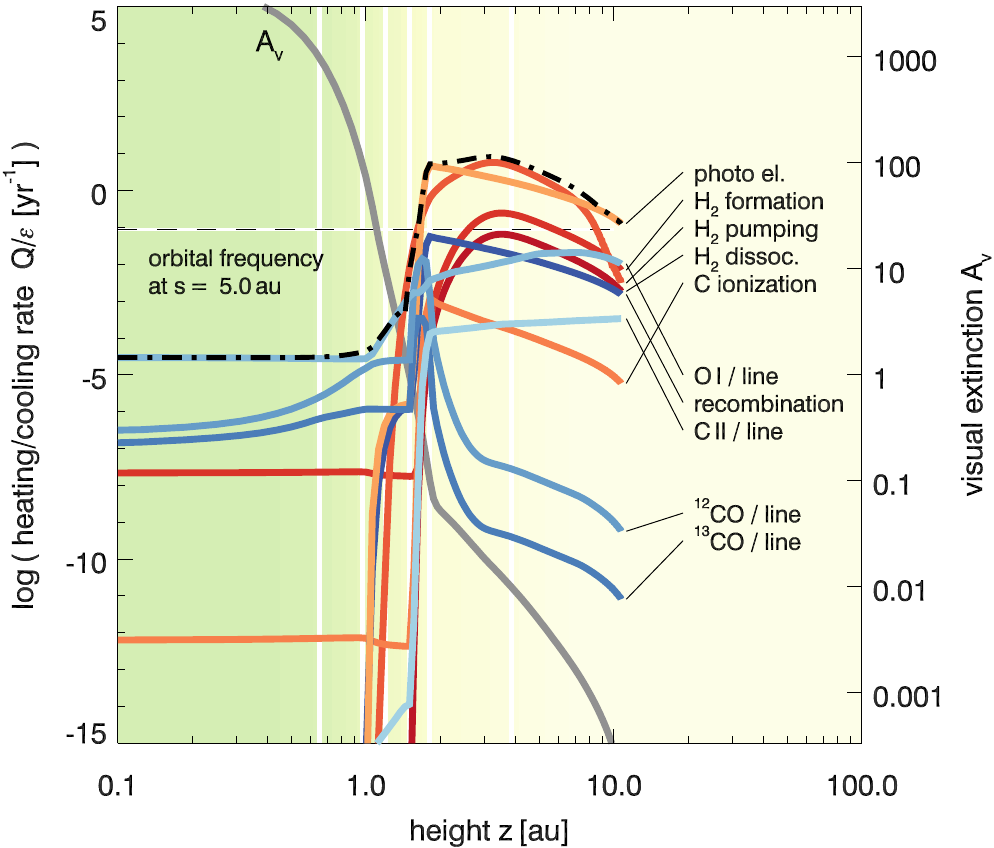}
  \caption{PDR heating (red tones) and cooling rates (blue tones) at $\Rc\!=\!5\au$ for model OA-b4-uv7, with an incident FUV flux equivalent of $\dot{M}=10^{-7}\Mdot$. The dot-dashed line shows the resulting effective heating/cooling rate, $\Qpdr$. Vertical lines indicate decades in the visual extinction coefficient, $A_{\rm v}$.}
  \label{fig:OA-b4-uv7-pdr-effects}
\end{figure}

The net heating/cooling rate in FUV irradiated gas is set by the sum of many different processes (see \Fig{OA-b4-uv7-pdr-effects} for an illustration of the most dominant ones in the case of model OA-b4-uv7). If the gas heating rates exceed the cooling rates, then the gas temperature, $T$, will exceed the dust temperature, $T_{\rm dust}$. For example, line cooling by atomic fine structure lines (e.g., [\ion{O}{1}], [\ion{C}{2}]) is important at the surface of the irradiated gas, while cooling by molecular lines (e.g., CO) is important deeper in the disk. The chemical composition of the gas will thus affect the heating and cooling rates \citep[e.g.,][]{1985ApJ...291..722T,1997ARA&A..35..179H,1999RvMP...71..173H}. Herein, we describe our implementation of a simplified thermochemistry module that includes the most important heating and cooling rates in a photon-dominated region (PDR) setting. We moreover demonstrate that our module is in good agreement with the full PDR codes benchmarked in \cite{2007A&A...467..187R}, while being significantly less computationally expensive.

\subsection{Chemistry} 
\label{app:chemistry}

The hydrogen-chemistry in a PDR is dominated by the photo-dissociation of H$_2$ gas plus the formation of H$_2$ on grain surfaces. Since temperatures in dense and strongly irradiated gas can reach several thousand degrees, collisional dissociation of H$_2$ must also be accounted for. The hydrogen chemical network employed here thus consists of the following reactions:
\begin{eqnarray*}
& {\rm H} + {\rm H} \rightarrow {\rm H}_2 & \quad(\text{formation}) \\
& {\rm H}_2 + \gamma_{\rm FUV} \rightarrow {\rm H} + {\rm H} & \quad(\text{photo-dissociation}) \\
& {\rm H}_2 + {\rm H} \rightarrow {\rm H} + {\rm H} + {\rm H} & \quad(\text{collisional dissociation})
\end{eqnarray*}

Carbon is assumed to be in the form of either C$^+$, C, or CO, following the simple network suggested by \cite{2008ApJ...683..238K}. This network includes the photodissociation of CO, the photoionization of C and the formation of CO initiated by the reaction of C$^+$ with H$_2$. We supplement the network with the recombination of C$^+$ to form C. The carbon chemical network is thus:
\begin{eqnarray*}
& {\rm C}^+ + {\rm H}_2 \rightarrow {\rm CH}_2^+ \rightarrow  {\rm CH}_x + {\rm O} \rightarrow {\rm CO} & \quad(\text{formation}) \\
& {\rm CO} + \gamma_{\rm FUV} \rightarrow {\rm C} + {\rm O} & \quad(\text{photo-dissociation}) \\
& {\rm C} + \gamma_{\rm FUV} \rightarrow {\rm C}^+ & \quad(\text{photo-ionization}) \\
& {\rm C}^+ + e^- \rightarrow {\rm C} & \quad(\text{recombination})
\end{eqnarray*}

The rate coefficients for the aforementioned reactions are taken according to \cite{2007A&A...467..187R}:
\begin{eqnarray}
k_{{\rm H}_2, \text{form}} &~=~& 3.0 \times 10^{-18} \sqrt{T} \quad {\rm cm}^3 {\rm s}^{-1}; \\
k_{{\rm H}_2, \text{disso}} &~=~& 2.6 \times 10^{-11} \chi e^{-3.0 {\rm A}_{\rm V}} \beta_{{\rm H}_2} \quad {\rm s}^{-1}; \\
k_{{\rm H}_2, \text{coll}} &~=~&  4.7 \times 10^{-7} \left(\frac{T}{300\, {\rm K}} \right)^{-1} e^{-5.5 \times 10^4\, {\rm K} / T} \quad {\rm cm}^3 {\rm s}^{-1}; \\
k_{{\rm CO}, \text{form}} &~=~& 5.0 \times 10^{-16} \xi + 10^{-17} n_{\rm H} / n_{{\rm H}_2} \quad {\rm cm}^{3} {\rm s}^{-1};  \\
\xi &~=~& \frac{ 5 \times 10^{-10} n_{\rm O}}{5 \times 10^{-10} n_{\rm O} + 2.5 \times 10^{-10} \chi e^{-2.0{\rm A}_{\rm V}}}; \\
k_{{\rm CO}, \text{disso}} &~=~& 1.0 \times 10^{-10} \chi e^{-2.5 {\rm A}_{\rm V}} \beta_{\rm CO} + 2.73 \times 10^{-15} \left(\frac{T}{300 {\rm K}}\right)^{1.17} \quad {\rm s}^{-1}; \\
k_{{\rm C}, \text{ion}} &~=~& 1.5 \times 10^{-10} \chi e^{-3.0 {\rm A}_{\rm V}} + 10^{-17} \quad {\rm s}^{-1}; \\
k_{{\rm C}^+, \text{recomb}} &~=~& 4.7 \times 10^{-12} \left(\frac{T_{\rm gas}}{300\, {\rm K}} \right)^{-0.6} \quad {\rm cm}^3 {\rm s}^{-1}.
\end{eqnarray}

The strength of the FUV field, $\chi$, irradiated from one side ($2\pi$ steradians), is given in units of the ``Draine-field'' (\citealt{1978ApJS...36..595D}; $2.7 \tms 10^{-3}$ erg s$^{-1}$\,cm$^{-2}$). Other parameters are the abundance of each species, $n_{\rm X}$ (cm$^{-3}$), the visual extinction A$_{\rm V}$, the H$_2$ and CO self-shielding factors $\beta_{{\rm H}_2}$ and $\beta_{{\rm CO}}$, and the formation efficiency of CO, $\xi$, for the reaction of C$^+ +\,$H$_2$. The rates are based primarily on the UMIST RATE99 database \citep{2000A&AS..146..157L} but did not change strongly in updated versions \citep{2007A&A...466.1197W,2013A&A...550A..36M}. The CO formation efficiency is derived following \cite{2005A&A...438..923N}, assuming that not all reactions of C$^+$ with H$_2$ lead to CO because intermediate molecules (e.g., CH$_x$) may be photodissociated. Meanwhile, the H$_2$ self-shielding factor from \cite{1996ApJ...468..269D} and the CO self/mutual-shielding factor derived by \cite{2012A&A...538A...2P} from \cite{1996A&A...311..690L} are used.

In order to approximately reproduce the low level of CO in regions with a low H/H$_2$ ratio, a small contribution independent of $\xi$ was empirically added to $k_{{\rm CO}, \text{form}}$. The parameter $k_{{\rm C},\text{ion}}$ is also modified to account for cosmic-ray ionization and charge exchange with other species. Photo-dissociation of CO due to cosmic rays is also included in $k_{{\rm CO}, \text{disso}}$ following the UMIST database \citep{2013A&A...550A..36M}. These modifications yield more realistic abundances in regions where CO or C$^+$ are depleted by several orders of magnitude compared to the carbon gas-phase abundance. Since these species are not important cooling agents in these regions, the modifications do not significantly affect the gas temperature or net heating/cooling rate.

To solve for the abundance of the species considered here (H, H$_2$, C$^+$, C, CO, O and e$^-$), we assume that the chemistry is in steady-state. This is a reasonable assumption, given that chemical time-scales in the upper layers of PPDs, where the gas and dust temperature decouple, are short \citep[e.g.,][]{2009A&A...501..383W} and certainly always shorter than the dynamical timescales near the disk midplane. It is assumed that the electron abundance is equal to the abundance of C$^+$ (i.e., $n_{e^-} \eq n_{{\rm C}^+}$) and that oxygen is either in the form of CO or atomic oxygen, i.e., $n_{\rm CO} + n_{\rm O} = n\, x_{\rm O}$, where $x_{\rm O}$ is the gas-phase fractional abundance of oxygen and $n$ is the total number density of the gas.

\subsection{Heating and cooling rates} 
\label{app:heatcool}
We explicitly include heating and cooling rates for the most important processes in the current context. These include line cooling by atomic fine structure lines ([\ion{C}{2}], [\ion{O}{1}], [\ion{C}{1}]), molecular lines (CO and $^{13}$CO), gas-grain thermal accommodation, Ly$\alpha$ cooling, cooling by the meta-stable line of \ion{O}{1} at 6300 \AA, photoelectric heating on small grains or polycyclic aromatic hydrocarbons (PAHs), recombination cooling of electrons with charged grains and PAHs, cosmic-ray heating, heating by C ionization and several heating/cooling processes involving H$_2$.

\subsubsection{Line cooling}

The line cooling rates for [\ion{C}{2}], [\ion{O}{1}], [\ion{C}{1}], CO and $^{13}$CO, are calculated using an escape probability approach \citep[e.g.,][]{2007A&A...468..627V,2009ApJ...700..872B}. The rate equations  are solved for the normalized population, $x_i$, of level $i$ ($\sum_i x_i =1$):
\beq
\sum_{i \neq j} x_j P_{ij} - x_i \sum_{i \neq j} P_{ij} = 0\,,
\eeq
using the rate coefficients
\beq
P_{ij} = \left\{
  \begin{array}{ll}
  A_{ij}\beta_{ij}+B_{ij} \beta_{ij} \langle J'_{ij} \rangle + C_{ij}, & E_i > E_j\,; \\
  B_{ij}\beta_{ij} \langle J'_{ij} \rangle + C_{ij}, & E_i < E_j\,.
  \end{array}
\right.
\eeq
The rate coefficient of a transition from level $i$ with energy $E_i$ to a level $j$ with energy $E_j$ is given by the Einstein coefficients, $A_{ij}$, $B_{ij}$, the cosmic microwave background radiation field, $\langle J'_{ij} \rangle$, collisional rates, $C_{ij}$, and the probability for a photon to escape, $\beta_{ij}$. We implement the escape probability function for a plane-parallel slab as derived by \cite{1980A&A....91...68D}:
\beq
\beta_{ij}(\tau_{ij}) = \begin{cases}
               \dfrac{1 - e^{-2.34\tau_{ij}}}{4.68\tau_{ij}}, & \tau_{ij} < 7\,; \\[2.0ex]
               \dfrac{1}{4\tau_{ij} \sqrt{\log{\tau_{ij} / \sqrt{\pi}}}}, & \tau_{ij} \geq 7\,,
              \end{cases}
\eeq
where the line center opacity is given by:
\beq
\tau_{ij} = \frac{ A_{ij} c^3 N_{\rm mol} }{ 8 \pi \nu_{ij}^3 \Delta {\rm v} } \frac{2 \sqrt{ \log{2}}}{\sqrt{\pi}} \left( x_j \frac{g_i}{g_j} - x_i \right)\,,
\eeq
and $N_{\rm mol}$ is the atomic/molecular column density, $\nu_{ij}$ is the frequency of the transition, $\Delta {\rm v}$ is the FWHM line width, $c$ is the speed of light and $g_i$, $g_j$ are the statistical weights.

The above system of equations is solved iteratively until convergence and, once the level population has been found, the cooling rate is calculated from
\beq
\Lambda_{\rm line}= n_{\rm X} \sum_{i > j}  h \nu_{ij}  \beta_{ij}  \left[ x_i A_{ij} + (x_i B_{ij} -x_j B_{ji}) \langle J'_{ij} \rangle \right]\,,
\eeq
where $h$ is Planck's constant.

The calculation of the cooling rates is generally too computationally intense to couple with a hydrodynamical simulation in real time, and so we have pre-calculated a grid of cooling rates for each species ($\Lambda_{\rm line}/n_{\rm X}$, erg s$^{-1}$) as a function of the collision partner density ($n_{\rm H}, n_{{\rm H}_2}, n_{e^-}$), the ratio $N_{\rm mol}/\Delta {\rm v}$, and the gas temperature. The cooling rate for a set of given conditions is then interpolated from this grid. Atomic and molecular data from the LAMDA database \citep{2005A&A...432..369S} and a CO/$^{13}$CO ratio of 70 \citep{1994ARA&A..32..191W} are used for the calculations.

\subsubsection{Gas-grain thermal accommodation}

The energy exchange between gas and grains is implemented following \cite{1989ApJ...342..306H}:
\beq
\Lambda_\mathrm{g-g} = 1.2 \times 10^{-31} \, n^2 \, \sqrt{T / 1000\, {\rm K}} \, \left(1-0.8 e^{-75\, {\rm K} / T}\right) \left(T - T_{\rm dust}\right) \quad {\rm erg}\,{\rm s}^{-1} {\rm cm}^{-3}\,.
\eeq
Note, however, that while this term is used in the PDR benchmarks presented below, the simulations in the main body of the paper assume $T = T_{\rm dust}$ and so the thermal accommodation term is always zero.

\subsubsection{Ly$\alpha$ line cooling}

Cooling through Ly$\alpha$ emission is accounted for using the rate from \citet{1989ApJ...338..197S}:
\beq
\Lambda_{{\rm Ly}\alpha} = 7.3 \times 10^{-19} \, n_{\rm H} \, n_{\rm e^-} \, e^{-118,400\, {\rm K}/T} \quad {\rm erg}\, {\rm s}^{-1} {\rm cm}^{-3}\,.
\eeq

\subsubsection{\ion{O}{1} 6300 \AA\ line cooling}

Cooling via this neutral oxygen meta-stable line is included using the rate from \citet{1989ApJ...338..197S}:
\beq
\Lambda_{\rm OI,6300} = 1.8 \times 10^{-24} \, n_{\rm O} \, n \, e^{-22,800\, {\rm K}/T} \quad {\rm erg}\, {\rm s}^{-1} {\rm cm}^{-3}\,.
\eeq

\subsubsection{Photoelectric heating}

Photoelectric heating by small grains and PAHs is one of the main heating mechanisms in dense PDRs and the surface layers of PPDs. We implement the rate following \cite{1994ApJ...427..822B}. As discussed in \cite{2006A&A...451..917R}, the rate assumes impinging FUV radiation over $4\pi$ steradians and intensity equal to the so-called ``Habing'' field (\citealt{1968BAN....19..421H}; $1\,G_0 = 1.6 \times 10^{-3}$ erg\, s$^{-1}$ cm$^{-2}$). We thus account for the conversion from Draine to Habing units, $G_0 = 1.71 \times 0.5 \times \chi$, and use the radiation field attenuated by the dust ($G_0^{\rm att}$). The heating rate is thus:
\beq
\Gamma_{\rm PE} = 10^{-24}\, \epsilon \, n \, G_0^{\rm att} \quad {\rm erg}\, {\rm s}^{-1} {\rm cm}^{-3}\,,
\eeq
where the efficiency is given by
\beq
\epsilon = \frac{4.87 \times 10^{-2}}{1 + 4 \times 10^{-3} \, \gamma^{0.74}} + \frac{3.65 \times 10^{-2} \left(T / 10^4\,{\rm K}\right)^{0.7}}{1 + 2 \times 10^{-4} \, \gamma}\,,
\eeq
and $\gamma=G_0^{\rm att} \sqrt{T}/ n_{\rm e^-}$ is the grain charge parameter.

\subsubsection{Recombination cooling}

Cooling by recombination of electrons with dust grains is implemented following \cite{1994ApJ...427..822B}:
\beq
\Lambda_{\rm recomb} = 3.49 \times 10^{-30} \, T^{0.944} \, \gamma^{0.735/T^{0.068}} \, n \, n_{\rm e^-} \quad {\rm erg}\, {\rm s}^{-1} {\rm cm}^{-3}\,.
\eeq

\subsubsection{Cosmic-ray heating}

The heating rate due to cosmic-ray ionization of H and H$_2$ is taken from \cite{2004A&A...428..511J}:
\beq
\Gamma_{\rm CR} = \zeta_{\rm CR} \left(5.5 \times 10^{-12} \, n_{{\rm H}_2}+ 2.5 \times 10^{-11} \, n_{\rm H}\right) \quad {\rm erg}\, {\rm s}^{-1} {\rm cm}^{-3}\,,
\eeq
with a cosmic-ray ionization rate of $\zeta_{\rm CR} = 5 \times 10^{-17}$ s$^{-1}$.

\subsubsection{Heating via the ionization of carbon}

The ionization of neutral carbon heats the gas at a rate of \citep{2004A&A...428..511J}:
\beq
\Gamma_{\rm C,ion} = 1.6 \times 10^{-12} \, k_{\rm C, ion} \, n_{\rm C}  \quad {\rm erg}\, {\rm s}^{-1} {\rm cm}^{-3}\,.
\eeq

\subsubsection{H$_2$ heating and cooling}

Molecular hydrogen H$_2$ can heat or cool the gas in different ways. For example, H$_2$ can (i) heat the gas through FUV pumping followed by collisional de-excitation, (ii) cool through line emission, (iii) heat chemically during formation or (iv) heat by photodissociation.

Processes (i) and (ii) are line processes, which we implement following \citet{2006A&A...451..917R}:
\beq
\Gamma_{{\rm H}_2{\rm -line}} = n_{{\rm H}_2}\,\frac{9.4\times 10^{-22}\,\chi^{{\rm att,H}_2}}{1+\left(\frac{1.9\times 10^{-6}+
 4.7\times 10^{-10}\,\chi^{{\rm att,H}_2}}{n \gamma_{\rm coll}}\right)}  \quad {\rm erg}\, {\rm s}^{-1} {\rm cm}^{-3}
\eeq
and
\beq
\Lambda_{{\rm H}_2{\rm -line}} = n\,n_{{\rm H}_2}\,9.1\times 10^{-13}\,\gamma_{\rm coll}\,e^{-6592 \rm{K}/T} \frac{8.6\times 10^{-7}+2.6\times 10^{-11}\,\chi^{{\rm att,H}_2}}{\gamma_{\rm coll}\,n+8.6\times 10^{-7}+2.6\times 10^{-11}\,\chi^{{\rm att,H}_2}} \quad {\rm erg}\ {\rm s}^{-1} {\rm cm}^{-3}\,.
\eeq
The strength of the FUV field, corrected for dust absorption and H$_2$ self-shielding, is $\chi^{{\rm att,H}_2}$. The effective collision rate is $\gamma_{\rm coll}=5.4 \times 10^{-13} \sqrt{T}$ s$^{-1}$ cm$^{-3}$.

Heating from formation and photodissociation is implemented using the rates provided by \cite{2004A&A...428..511J}:
\begin{align}
\Gamma_{{\rm H}_2{\rm -form}} &= 2.4 \times 10^{-12} \, k_{{\rm H}_2, {\rm form}}\, n \, n_{\rm H}  \quad {\rm erg}\, {\rm s}^{-1} {\rm cm}^{-3}; \\
\Gamma_{{\rm H}_2{\rm -disso}} &= 6.4 \times 10^{-13} \, k_{{\rm H}_2, {\rm disso}} \, n_{{\rm H}_2}  \quad {\rm erg}\, {\rm s}^{-1} {\rm cm}^{-3}.
\end{align}

\subsection{Validation} 
\label{app:pdr_benchmark}
The simplified thermochemistry module was tested against the benchmark problems of \citet{2007A&A...467..187R}. In that study, different PDR codes were tasked with calculating the gas temperature in a plane-parallel slab of gas with constant density and irradiated by a FUV source from one side. The density values were chosen to be $n=10^3$ cm$^{-3}$ and $n=10^{5.5}$ cm$^{-3}$, while the FUV field strength was $\chi=10$ or $\chi=10^5$ (ISRF).

For this benchmark, the dust temperature is obtained using the prescription in \cite{1991ApJ...377..192H}.
To calculate the gas temperature, we solve for equilibrium in the heating/cooling rates in each cell, that is, $\sum_i \Gamma_i = \sum_i \Lambda_i$, using a simple bisection algorithm. The calculation is run until convergence between heating and cooling rates down to an absolute tolerance of $10^{-10}$.

The gas temperature calculated by the thermochemical module compares well with the results of more complete PDR codes \citep{2007A&A...467..187R}, despite the simplified chemistry used here. Our calculation is several orders of magnitude faster than a complete PDR code, which is a necessary precondition to couple it with a hydrodynamical simulation. On a standard laptop (e.g., Intel Core i5 $1.6\GHz$), the calculation of a slab with 500 points in ${\rm A}_{\rm V}$ takes less than 1/100th of a second, in contrast to a more complete PDR code that could take between a few minutes and an hour.

\begin{figure*}[htb]
  \includegraphics[width=1.0\textwidth]{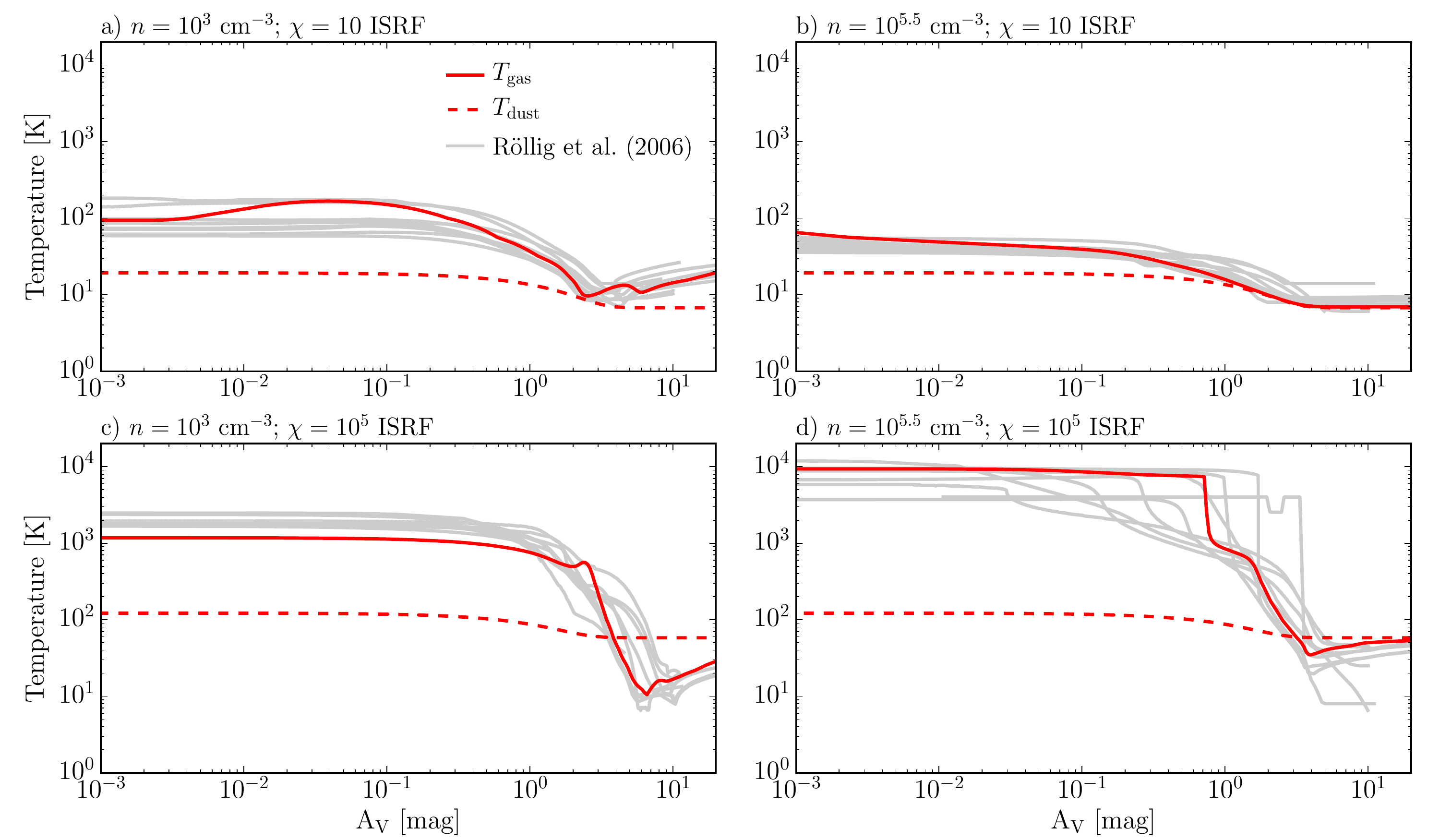}
  \caption{The gas (solid) and dust (dashed) temperatures as a function of visual extinction (${\rm A}_{\rm V}$) as calculated by the simplified thermochemistry module for the PDR benchmarks of \cite{2007A&A...467..187R}. The gray lines are results from the PDR codes featured in the aforementioned study (the data is publicly available from \texttt{http://zeus.ph1.uni-koeln.de/site/pdr-comparison/}). Note the scatter between different PDR codes is considerable.\bigskip}
  \label{fig:pdr_benchmark}
\end{figure*}



\begin{thebibliography}{}
\expandafter\ifx\csname natexlab\endcsname\relax\def\natexlab#1{#1}\fi
\providecommand{\url}[1]{\href{#1}{#1}}

\bibitem[{{Alexander} {et~al.}(2014){Alexander}, {Pascucci}, {Andrews},
  {Armitage}, \& {Cieza}}]{2014prpl.conf..475A}
{Alexander}, R., {Pascucci}, I., {Andrews}, S., {Armitage}, P., \& {Cieza}, L.
  2014, in Protostars and Planets VI, ed. Henrik Beuther et al., (Tucson, AZ:
  Univ. of Arizona Press), 475

\bibitem[{{Asplund} {et~al.}(2009){Asplund}, {Grevesse}, {Sauval}, \&
  {Scott}}]{2009ARA&A..47..481A}
{Asplund}, M., {Grevesse}, N., {Sauval}, A.~J., \& {Scott}, P. 2009,
  \href{http://dx.doi.org/10.1146/annurev.astro.46.060407.145222}{\color{magenta}\araa},
  \href{https://ui.adsabs.harvard.edu/abs/2009ARA&A..47..481A}{\color{cyan}47},
  481

\bibitem[{{Bai}(2013)}]{2013ApJ...772...96B}
{Bai}, X.-N. 2013,
  \href{http://dx.doi.org/10.1088/0004-637X/772/2/96}{\color{magenta}\apj},
  \href{http://adsabs.harvard.edu/abs/2013ApJ...772...96B}{\color{cyan}772}, 96

\bibitem[{{Bai}(2016)}]{2016ApJ...821...80B}
---. 2016,
  \href{http://dx.doi.org/10.3847/0004-637X/821/2/80}{\color{magenta}\apj},
  \href{http://adsabs.harvard.edu/abs/2016ApJ...821...80B}{\color{cyan}821}, 80

\bibitem[{{Bai}(2017)}]{2017ApJ...845...75B}
---. 2017,
  \href{http://dx.doi.org/10.3847/1538-4357/aa7dda}{\color{magenta}\apj},
  \href{http://adsabs.harvard.edu/abs/2017ApJ...845...75B}{\color{cyan}845}, 75

\bibitem[{{Bai} \& {Goodman}(2009)}]{2009ApJ...701..737B}
{Bai}, X.-N., \& {Goodman}, J. 2009,
  \href{http://dx.doi.org/10.1088/0004-637X/701/1/737}{\color{magenta}\apj},
  \href{http://adsabs.harvard.edu/abs/2009ApJ...701..737B}{\color{cyan}701},
  737

\bibitem[{{Bai} \& {Stone}(2013)}]{2013ApJ...769...76B}
{Bai}, X.-N., \& {Stone}, J.~M. 2013,
  \href{http://dx.doi.org/10.1088/0004-637X/769/1/76}{\color{magenta}\apj},
  \href{http://adsabs.harvard.edu/abs/2013ApJ...769...76B}{\color{cyan}769}, 76

\bibitem[{{Bai} \& {Stone}(2017)}]{2017ApJ...836...46B}
---. 2017,
  \href{http://dx.doi.org/10.3847/1538-4357/836/1/46}{\color{magenta}\apj},
  \href{http://adsabs.harvard.edu/abs/2017ApJ...836...46B}{\color{cyan}836}, 46

\bibitem[{{Bai} {et~al.}(2016){Bai}, {Ye}, {Goodman}, \&
  {Yuan}}]{2016ApJ...818..152B}
{Bai}, X.-N., {Ye}, J., {Goodman}, J., \& {Yuan}, F. 2016,
  \href{http://dx.doi.org/10.3847/0004-637X/818/2/152}{\color{magenta}\apj},
  \href{http://adsabs.harvard.edu/abs/2016ApJ...818..152B}{\color{cyan}818},
  152

\bibitem[{{Bakes} \& {Tielens}(1994)}]{1994ApJ...427..822B}
{Bakes}, E.~L.~O., \& {Tielens}, A.~G.~G.~M. 1994,
  \href{http://dx.doi.org/10.1086/174188}{\color{magenta}\apj},
  \href{http://adsabs.harvard.edu/abs/1994ApJ...427..822B}{\color{cyan}427},
  822

\bibitem[{{Balbus} \& {Hawley}(1991)}]{1991ApJ...376..214B}
{Balbus}, S.~A., \& {Hawley}, J.~F. 1991,
  \href{http://dx.doi.org/10.1086/170270}{\color{magenta}\apj},
  \href{http://adsabs.harvard.edu/abs/1991ApJ...376..214B}{\color{cyan}376},
  214

\bibitem[{{Banzatti} {et~al.}(2019){Banzatti}, {Pascucci}, {Edwards}, {Fang},
  {Gorti}, \& {Flock}}]{2019ApJ...870...76B}
{Banzatti}, A., {Pascucci}, I., {Edwards}, S., {et~al.} 2019,
  \href{http://dx.doi.org/10.3847/1538-4357/aaf1aa}{\color{magenta}\apj},
  \href{https://ui.adsabs.harvard.edu/abs/2019ApJ...870...76B}{\color{cyan}870},
  76

\bibitem[{{Banzatti} {et~al.}(2015){Banzatti}, {Pinilla}, {Ricci},
  {Pontoppidan}, {Birnstiel}, \& {Ciesla}}]{2015ApJ...815L..15B}
{Banzatti}, A., {Pinilla}, P., {Ricci}, L., {et~al.} 2015,
  \href{http://dx.doi.org/10.1088/2041-8205/815/1/L15}{\color{magenta}\apjl},
  \href{http://adsabs.harvard.edu/abs/2015ApJ...815L..15B}{\color{cyan}815},
  L15

\bibitem[{{Bethell} \& {Bergin}(2011)}]{2011ApJ...739...78B}
{Bethell}, T.~J., \& {Bergin}, E.~A. 2011,
  \href{http://dx.doi.org/10.1088/0004-637X/739/2/78}{\color{magenta}\apj},
  \href{http://adsabs.harvard.edu/abs/2011ApJ...739...78B}{\color{cyan}739}, 78

\bibitem[{{B{\'e}thune} {et~al.}(2016){B{\'e}thune}, {Lesur}, \&
  {Ferreira}}]{2016A&A...589A..87B}
{B{\'e}thune}, W., {Lesur}, G., \& {Ferreira}, J. 2016,
  \href{http://dx.doi.org/10.1051/0004-6361/201527874}{\color{magenta}\aap},
  \href{http://adsabs.harvard.edu/abs/2016A%26A...589A..87B}{\color{cyan}589},
  A87

\bibitem[{{B{\'e}thune} {et~al.}(2017){B{\'e}thune}, {Lesur}, \&
  {Ferreira}}]{2017A&A...600A..75B}
---. 2017,
  \href{http://dx.doi.org/10.1051/0004-6361/201630056}{\color{magenta}\aap},
  \href{http://adsabs.harvard.edu/abs/2017A%26A...600A..75B}{\color{cyan}600},
  A75

\bibitem[{{Bitsch} {et~al.}(2013){Bitsch}, {Crida}, {Morbidelli}, {Kley}, \&
  {Dobbs-Dixon}}]{2013A&A...549A.124B}
{Bitsch}, B., {Crida}, A., {Morbidelli}, A., {Kley}, W., \& {Dobbs-Dixon}, I.
  2013,
  \href{http://dx.doi.org/10.1051/0004-6361/201220159}{\color{magenta}\aap},
  \href{http://adsabs.harvard.edu/abs/2013A%26A...549A.124B}{\color{cyan}549},
  A124

\bibitem[{{Bohren} \& {Huffman}(1983)}]{1983asls.book.....B}
{Bohren}, C.~F., \& {Huffman}, D.~R. 1983, {Absorption and scattering of light
  by small particles} (New York: Wiley)

\bibitem[{{Boley} {et~al.}(2007){Boley}, {Durisen}, {Nordlund}, \&
  {Lord}}]{2007ApJ...665.1254B}
{Boley}, A.~C., {Durisen}, R.~H., {Nordlund}, {\r{A}}., \& {Lord}, J. 2007,
  \href{http://dx.doi.org/10.1086/519767}{\color{magenta}\apj},
  \href{https://ui.adsabs.harvard.edu/abs/2007ApJ...665.1254B}{\color{cyan}665},
  1254

\bibitem[{{Brauer} {et~al.}(2017){Brauer}, {Wolf}, \&
  {Flock}}]{2017A&A...607A.104B}
{Brauer}, R., {Wolf}, S., \& {Flock}, M. 2017,
  \href{http://dx.doi.org/10.1051/0004-6361/201731140}{\color{magenta}\aap},
  \href{https://ui.adsabs.harvard.edu/abs/2017A&A...607A.104B}{\color{cyan}607},
  A104

\bibitem[{{Brinch} \& {Hogerheijde}(2010)}]{2010A&A...523A..25B}
{Brinch}, C., \& {Hogerheijde}, M.~R. 2010,
  \href{http://dx.doi.org/10.1051/0004-6361/201015333}{\color{magenta}\aap},
  \href{http://adsabs.harvard.edu/abs/2010A%26A...523A..25B}{\color{cyan}523},
  A25

\bibitem[{{Bruderer} {et~al.}(2009{\natexlab{a}}){Bruderer}, {Benz}, {Doty},
  {van Dishoeck}, \& {Bourke}}]{2009ApJ...700..872B}
{Bruderer}, S., {Benz}, A.~O., {Doty}, S.~D., {van Dishoeck}, E.~F., \&
  {Bourke}, T.~L. 2009{\natexlab{a}},
  \href{http://dx.doi.org/10.1088/0004-637X/700/1/872}{\color{magenta}\apj},
  \href{https://ui.adsabs.harvard.edu/abs/2009ApJ...700..872B}{\color{cyan}700},
  872

\bibitem[{{Bruderer} {et~al.}(2009{\natexlab{b}}){Bruderer}, {Doty}, \&
  {Benz}}]{2009ApJS..183..179B}
{Bruderer}, S., {Doty}, S.~D., \& {Benz}, A.~O. 2009{\natexlab{b}},
  \href{http://dx.doi.org/10.1088/0067-0049/183/2/179}{\color{magenta}\apjs},
  \href{http://adsabs.harvard.edu/abs/2009ApJS..183..179B}{\color{cyan}183},
  179

\bibitem[{{Bruls} {et~al.}(1999){Bruls}, {Vollm{\"o}ller}, \&
  {Sch{\"u}ssler}}]{1999A&A...348..233B}
{Bruls}, J.~H.~M.~J., {Vollm{\"o}ller}, P., \& {Sch{\"u}ssler}, M. 1999, \aap,
  \href{http://adsabs.harvard.edu/abs/1999A%26A...348..233B}{\color{cyan}348},
  233

\bibitem[{{Carr} {et~al.}(2018){Carr}, {Najita}, \&
  {Salyk}}]{2018RNAAS...2c.169C}
{Carr}, J.~S., {Najita}, J.~R., \& {Salyk}, C. 2018,
  \href{http://dx.doi.org/10.3847/2515-5172/aadfe7}{\color{magenta}\rm RNAAS},
  \href{https://ui.adsabs.harvard.edu/abs/2018RNAAS...2c.169C}{\color{cyan}2},
  169

\bibitem[{{Chiang} \& {Goldreich}(1997)}]{1997ApJ...490..368C}
{Chiang}, E.~I., \& {Goldreich}, P. 1997, \apj,
  \href{http://adsabs.harvard.edu/abs/1997ApJ...490..368C}{\color{cyan}490},
  368

\bibitem[{{Cui} \& {Bai}(2019)}]{2019arXiv191202941C}
{Cui}, C., \& {Bai}, X.-N. 2019, arXiv e-prints, arXiv:1912.02941

\bibitem[{{de Jong} {et~al.}(1980){de Jong}, {Boland}, \&
  {Dalgarno}}]{1980A&A....91...68D}
{de Jong}, T., {Boland}, W., \& {Dalgarno}, A. 1980, \aap,
  \href{http://adsabs.harvard.edu/abs/1980A%26A....91...68D}{\color{cyan}91},
  68

\bibitem[{{Del Zanna} {et~al.}(2001){Del Zanna}, {Velli}, \&
  {Londrillo}}]{2001A&A...367..705D}
{Del Zanna}, L., {Velli}, M., \& {Londrillo}, P. 2001,
  \href{http://dx.doi.org/10.1051/0004-6361:20000455}{\color{magenta}\aap},
  \href{https://ui.adsabs.harvard.edu/abs/2001A%26A...367..705D}{\color{cyan}367},
  705

\bibitem[{{Dent} {et~al.}(2013){Dent}, {Thi}, {Kamp}, {Williams}, {Menard},
  {Andrews}, {Ardila}, {Aresu}, {Augereau}, {Barrado y Navascues}, {Brittain},
  {Carmona}, {Ciardi}, {Danchi}, {Donaldson}, {Duchene}, {Eiroa}, {Fedele},
  {Grady}, {de Gregorio-Molsalvo}, {Howard}, {Hu{\'e}lamo}, {Krivov},
  {Lebreton}, {Liseau}, {Martin-Zaidi}, {Mathews}, {Meeus}, {Mendigut{\'\i}a},
  {Montesinos}, {Morales-Calderon}, {Mora}, {Nomura}, {Pantin}, {Pascucci},
  {Phillips}, {Pinte}, {Podio}, {Ramsay}, {Riaz}, {Riviere-Marichalar},
  {Roberge}, {Sand ell}, {Solano}, {Tilling}, {Torrelles}, {Vandenbusche},
  {Vicente}, {White}, \& {Woitke}}]{2013PASP..125..477D}
{Dent}, W.~R.~F., {Thi}, W.~F., {Kamp}, I., {et~al.} 2013,
  \href{http://dx.doi.org/10.1086/670826}{\color{magenta}\pasp},
  \href{https://ui.adsabs.harvard.edu/abs/2013PASP..125..477D}{\color{cyan}125},
  477

\bibitem[{{Desch} \& {Turner}(2015)}]{2015ApJ...811..156D}
{Desch}, S.~J., \& {Turner}, N.~J. 2015,
  \href{http://dx.doi.org/10.1088/0004-637X/811/2/156}{\color{magenta}\apj},
  \href{http://adsabs.harvard.edu/abs/2015ApJ...811..156D}{\color{cyan}811},
  156

\bibitem[{{Dorschner} {et~al.}(1995){Dorschner}, {Begemann}, {Henning},
  {Jaeger}, \& {Mutschke}}]{1995A&A...300..503D}
{Dorschner}, J., {Begemann}, B., {Henning}, T., {Jaeger}, C., \& {Mutschke}, H.
  1995, \aap,
  \href{https://ui.adsabs.harvard.edu/abs/1995A&A...300..503D}{\color{cyan}300},
  503

\bibitem[{{Draine}(1978)}]{1978ApJS...36..595D}
{Draine}, B.~T. 1978,
  \href{http://dx.doi.org/10.1086/190513}{\color{magenta}\apjs},
  \href{http://adsabs.harvard.edu/abs/1978ApJS...36..595D}{\color{cyan}36}, 595

\bibitem[{{Draine} \& {Bertoldi}(1996)}]{1996ApJ...468..269D}
{Draine}, B.~T., \& {Bertoldi}, F. 1996,
  \href{http://dx.doi.org/10.1086/177689}{\color{magenta}\apj},
  \href{http://adsabs.harvard.edu/abs/1996ApJ...468..269D}{\color{cyan}468},
  269

\bibitem[{{Dra{\.z}kowska} \& {Alibert}(2017)}]{2017A&A...608A..92D}
{Dra{\.z}kowska}, J., \& {Alibert}, Y. 2017,
  \href{http://dx.doi.org/10.1051/0004-6361/201731491}{\color{magenta}\aap},
  \href{https://ui.adsabs.harvard.edu/abs/2017A&A...608A..92D}{\color{cyan}608},
  A92

\bibitem[{{Dullemond} {et~al.}(2012){Dullemond}, {Juhasz}, {Pohl}, {Sereshti},
  {Shetty}, {Peters}, {Commercon}, \& {Flock}}]{2012ascl.soft02015D}
{Dullemond}, C.~P., {Juhasz}, A., {Pohl}, A., {et~al.} 2012, {RADMC-3D: A
  multi-purpose radiative transfer tool},  Astrophysics Source Code Library

\bibitem[{{Evans} \& {Hawley}(1988)}]{1988ApJ...332..659E}
{Evans}, C.~R., \& {Hawley}, J.~F. 1988,
  \href{http://dx.doi.org/10.1086/166684}{\color{magenta}\apj},
  \href{http://adsabs.harvard.edu/abs/1988ApJ...332..659E}{\color{cyan}332},
  659

\bibitem[{{Flaherty} {et~al.}(2018){Flaherty}, {Hughes}, {Teague}, {Simon},
  {Andrews}, \& {Wilner}}]{2018ApJ...856..117F}
{Flaherty}, K.~M., {Hughes}, A.~M., {Teague}, R., {et~al.} 2018,
  \href{http://dx.doi.org/10.3847/1538-4357/aab615}{\color{magenta}\apj},
  \href{http://adsabs.harvard.edu/abs/2018ApJ...856..117F}{\color{cyan}856},
  117

\bibitem[{{Frank} {et~al.}(2014){Frank}, {Ray}, {Cabrit}, {Hartigan}, {Arce},
  {Bacciotti}, {Bally}, {Benisty}, {Eisl{\"o}ffel}, {G{\"u}del}, {Lebedev},
  {Nisini}, \& {Raga}}]{2014prpl.conf..451F}
{Frank}, A., {Ray}, T.~P., {Cabrit}, S., {et~al.} 2014, in Protostars and
  Planets VI, ed. Henrik Beuther et al., (Tucson, AZ: Univ. of Arizona Press),
  451

\bibitem[{{Fromang} {et~al.}(2013){Fromang}, {Latter}, {Lesur}, \&
  {Ogilvie}}]{2013A&A...552A..71F}
{Fromang}, S., {Latter}, H., {Lesur}, G., \& {Ogilvie}, G.~I. 2013,
  \href{http://dx.doi.org/10.1051/0004-6361/201220016}{\color{magenta}\aap},
  \href{http://adsabs.harvard.edu/abs/2013A%26A...552A..71F}{\color{cyan}552},
  A71

\bibitem[{{Fromang} \& {Nelson}(2006)}]{2006A&A...457..343F}
{Fromang}, S., \& {Nelson}, R.~P. 2006,
  \href{http://dx.doi.org/10.1051/0004-6361:20065643}{\color{magenta}\aap},
  \href{https://ui.adsabs.harvard.edu/abs/2006A&A...457..343F}{\color{cyan}457},
  343

\bibitem[{{Fu} {et~al.}(2014){Fu}, {Weiss}, {Lima}, {Harrison}, {Bai}, {Desch},
  {Ebel}, {Suavet}, {Wang}, {Glenn}, {Le Sage}, {Kasama}, {Walsworth}, \&
  {Kuan}}]{2014Sci...346.1089F}
{Fu}, R.~R., {Weiss}, B.~P., {Lima}, E.~A., {et~al.} 2014,
  \href{http://dx.doi.org/10.1126/science.1258022}{\color{magenta}Science},
  \href{http://adsabs.harvard.edu/abs/2014Sci...346.1089F}{\color{cyan}346},
  1089

\bibitem[{{Gardiner} \& {Stone}(2008)}]{2008JCoPh.227.4123G}
{Gardiner}, T.~A., \& {Stone}, J.~M. 2008,
  \href{http://dx.doi.org/10.1016/j.jcp.2007.12.017}{\color{magenta}\rm JCoPh},
  \href{http://adsabs.harvard.edu/abs/2008JCoPh.227.4123G}{\color{cyan}227},
  4123

\bibitem[{{Gnedin} \& {Abel}(2001)}]{2001NewA....6..437G}
{Gnedin}, N.~Y., \& {Abel}, T. 2001,
  \href{http://dx.doi.org/10.1016/S1384-1076(01)00068-9}{\color{magenta}NewA},
  \href{http://adsabs.harvard.edu/abs/2001NewA....6..437G}{\color{cyan}6}, 437

\bibitem[{{Gorti} {et~al.}(2009){Gorti}, {Dullemond}, \&
  {Hollenbach}}]{2009ApJ...705.1237G}
{Gorti}, U., {Dullemond}, C.~P., \& {Hollenbach}, D. 2009,
  \href{http://dx.doi.org/10.1088/0004-637X/705/2/1237}{\color{magenta}\apj},
  \href{https://ui.adsabs.harvard.edu/abs/2009ApJ...705.1237G}{\color{cyan}705},
  1237

\bibitem[{{Gorti} \& {Hollenbach}(2008)}]{2008ApJ...683..287G}
{Gorti}, U., \& {Hollenbach}, D. 2008,
  \href{http://dx.doi.org/10.1086/589616}{\color{magenta}\apj},
  \href{https://ui.adsabs.harvard.edu/abs/2008ApJ...683..287G}{\color{cyan}683},
  287

\bibitem[{{Grassi} {et~al.}(2017){Grassi}, {Bovino}, {Haugb{\o}lle}, \&
  {Schleicher}}]{2017MNRAS.466.1259G}
{Grassi}, T., {Bovino}, S., {Haugb{\o}lle}, T., \& {Schleicher}, D.~R.~G. 2017,
  \href{http://dx.doi.org/10.1093/mnras/stw2871}{\color{magenta}\mnras},
  \href{http://adsabs.harvard.edu/abs/2017MNRAS.466.1259G}{\color{cyan}466},
  1259

\bibitem[{{Grassi} {et~al.}(2014){Grassi}, {Bovino}, {Schleicher}, {Prieto},
  {Seifried}, {Simoncini}, \& {Gianturco}}]{2014MNRAS.439.2386G}
{Grassi}, T., {Bovino}, S., {Schleicher}, D.~R.~G., {et~al.} 2014,
  \href{http://dx.doi.org/10.1093/mnras/stu114}{\color{magenta}\mnras},
  \href{http://adsabs.harvard.edu/abs/2014MNRAS.439.2386G}{\color{cyan}439},
  2386

\bibitem[{{Gressel}(2017)}]{2017JPhCS.837a2008G}
{Gressel}, O. 2017, in Journal of Physics Conference Series, Vol. 837, 012008

\bibitem[{{Gressel} {et~al.}(2012){Gressel}, {Nelson}, \&
  {Turner}}]{2012MNRAS.422.1140G}
{Gressel}, O., {Nelson}, R.~P., \& {Turner}, N.~J. 2012,
  \href{http://dx.doi.org/10.1111/j.1365-2966.2012.20701.x}{\color{magenta}\mnras},
  \href{http://adsabs.harvard.edu/abs/2012MNRAS.422.1140G}{\color{cyan}422},
  1140

\bibitem[{{Gressel} {et~al.}(2013){Gressel}, {Nelson}, {Turner}, \&
  {Ziegler}}]{2013ApJ...779...59G}
{Gressel}, O., {Nelson}, R.~P., {Turner}, N.~J., \& {Ziegler}, U. 2013,
  \href{http://dx.doi.org/10.1088/0004-637X/779/1/59}{\color{magenta}\apj},
  \href{http://adsabs.harvard.edu/abs/2013ApJ...779...59G}{\color{cyan}779}, 59

\bibitem[{{Gressel} {et~al.}(2015){Gressel}, {Turner}, {Nelson}, \&
  {McNally}}]{2015ApJ...801...84G}
{Gressel}, O., {Turner}, N.~J., {Nelson}, R.~P., \& {McNally}, C.~P. 2015,
  \href{http://dx.doi.org/10.1088/0004-637X/801/2/84}{\color{magenta}\apj},
  \href{http://adsabs.harvard.edu/abs/2015ApJ...801...84G}{\color{cyan}801}, 84

\bibitem[{{G{\"u}del} {et~al.}(2018){G{\"u}del}, {Eibensteiner}, {Dionatos},
  {Audard}, {Forbrich}, {Kraus}, {Rab}, {Schneider}, {Skinner}, \&
  {Vorobyov}}]{2018A&A...620L...1G}
{G{\"u}del}, M., {Eibensteiner}, C., {Dionatos}, O., {et~al.} 2018,
  \href{http://dx.doi.org/10.1051/0004-6361/201834271}{\color{magenta}\aap},
  \href{https://ui.adsabs.harvard.edu/abs/2018A&A...620L...1G}{\color{cyan}620},
  L1

\bibitem[{{Gullbring} {et~al.}(1998){Gullbring}, {Hartmann}, {Briceno}, \&
  {Calvet}}]{1998ApJ...492..323G}
{Gullbring}, E., {Hartmann}, L., {Briceno}, C., \& {Calvet}, N. 1998,
  \href{http://dx.doi.org/10.1086/305032}{\color{magenta}\apj},
  \href{http://adsabs.harvard.edu/abs/1998ApJ...492..323G}{\color{cyan}492},
  323

\bibitem[{{Habing}(1968)}]{1968BAN....19..421H}
{Habing}, H.~J. 1968, \bain,
  \href{http://adsabs.harvard.edu/abs/1968BAN....19..421H}{\color{cyan}19}, 421

\bibitem[{{Haisch} {et~al.}(2001){Haisch}, {Lada}, \&
  {Lada}}]{2001ApJ...553L.153H}
{Haisch}, Jr., K.~E., {Lada}, E.~A., \& {Lada}, C.~J. 2001,
  \href{http://dx.doi.org/10.1086/320685}{\color{magenta}\apjl},
  \href{http://adsabs.harvard.edu/abs/2001ApJ...553L.153H}{\color{cyan}553},
  L153

\bibitem[{{Hales} {et~al.}(2018){Hales}, {P{\'e}rez}, {Saito}, {Pinte}, {Knee},
  {de Gregorio-Monsalvo}, {Dent}, {L{\'o}pez}, {Plunkett}, {Cort{\'e}s},
  {Corder}, \& {Cieza}}]{2018ApJ...859..111H}
{Hales}, A.~S., {P{\'e}rez}, S., {Saito}, M., {et~al.} 2018,
  \href{http://dx.doi.org/10.3847/1538-4357/aac018}{\color{magenta}\apj},
  \href{https://ui.adsabs.harvard.edu/abs/2018ApJ...859..111H}{\color{cyan}859},
  111

\bibitem[{{Hartmann}(1998)}]{1998apsf.book.....H}
{Hartmann}, L. 1998, {Accretion Processes in Star Formation} (Cambridge, New
  York: Cambridge Univ. Press)

\bibitem[{{Hartmann} {et~al.}(1998){Hartmann}, {Calvet}, {Gullbring}, \&
  {D'Alessio}}]{1998ApJ...495..385H}
{Hartmann}, L., {Calvet}, N., {Gullbring}, E., \& {D'Alessio}, P. 1998,
  \href{http://dx.doi.org/10.1086/305277}{\color{magenta}\apj},
  \href{http://adsabs.harvard.edu/abs/1998ApJ...495..385H}{\color{cyan}495},
  385

\bibitem[{{Hartmann} {et~al.}(2017){Hartmann}, {Ciesla}, {Gressel}, \&
  {Alexander}}]{2017SSRv..212..813H}
{Hartmann}, L., {Ciesla}, F., {Gressel}, O., \& {Alexander}, R. 2017,
  \href{http://dx.doi.org/10.1007/s11214-017-0406-0}{\color{magenta}\ssr},
  \href{http://adsabs.harvard.edu/abs/2017SSRv..212..813H}{\color{cyan}212},
  813

\bibitem[{{Hartmann} {et~al.}(2016){Hartmann}, {Herczeg}, \&
  {Calvet}}]{2016ARA&A..54..135H}
{Hartmann}, L., {Herczeg}, G., \& {Calvet}, N. 2016,
  \href{http://dx.doi.org/10.1146/annurev-astro-081915-023347}{\color{magenta}\araa},
  \href{https://ui.adsabs.harvard.edu/abs/2016ARA&A..54..135H}{\color{cyan}54},
  135

\bibitem[{{Hawley} \& {Balbus}(1991)}]{1991ApJ...376..223H}
{Hawley}, J.~F., \& {Balbus}, S.~A. 1991,
  \href{http://dx.doi.org/10.1086/170271}{\color{magenta}\apj},
  \href{http://adsabs.harvard.edu/abs/1991ApJ...376..223H}{\color{cyan}376},
  223

\bibitem[{{Haworth} \& {Owen}(2020)}]{2020MNRAS.492.5030H}
{Haworth}, T.~J., \& {Owen}, J.~E. 2020,
  \href{http://dx.doi.org/10.1093/mnras/staa151}{\color{magenta}\mnras},
  \href{https://ui.adsabs.harvard.edu/abs/2020MNRAS.492.5030H}{\color{cyan}492},
  5030

\bibitem[{{Heays} {et~al.}(2017){Heays}, {Bosman}, \& {van
  Dishoeck}}]{2017A&A...602A.105H}
{Heays}, A.~N., {Bosman}, A.~D., \& {van Dishoeck}, E.~F. 2017,
  \href{http://dx.doi.org/10.1051/0004-6361/201628742}{\color{magenta}\aap},
  \href{https://ui.adsabs.harvard.edu/abs/2017A&A...602A.105H}{\color{cyan}602},
  A105

\bibitem[{{Heinemann} {et~al.}(2006){Heinemann}, {Dobler}, {Nordlund}, \&
  {Brandenburg}}]{2006A&A...448..731H}
{Heinemann}, T., {Dobler}, W., {Nordlund}, {\r{A}}., \& {Brandenburg}, A. 2006,
  \href{http://dx.doi.org/10.1051/0004-6361:20053120}{\color{magenta}\aap},
  \href{https://ui.adsabs.harvard.edu/abs/2006A&A...448..731H}{\color{cyan}448},
  731

\bibitem[{{Hollenbach} \& {McKee}(1989)}]{1989ApJ...342..306H}
{Hollenbach}, D., \& {McKee}, C.~F. 1989,
  \href{http://dx.doi.org/10.1086/167595}{\color{magenta}\apj},
  \href{http://adsabs.harvard.edu/abs/1989ApJ...342..306H}{\color{cyan}342},
  306

\bibitem[{{Hollenbach} {et~al.}(1991){Hollenbach}, {Takahashi}, \&
  {Tielens}}]{1991ApJ...377..192H}
{Hollenbach}, D.~J., {Takahashi}, T., \& {Tielens}, A.~G.~G.~M. 1991,
  \href{http://dx.doi.org/10.1086/170347}{\color{magenta}\apj},
  \href{http://adsabs.harvard.edu/abs/1991ApJ...377..192H}{\color{cyan}377},
  192

\bibitem[{{Hollenbach} \& {Tielens}(1997)}]{1997ARA&A..35..179H}
{Hollenbach}, D.~J., \& {Tielens}, A.~G.~G.~M. 1997,
  \href{http://dx.doi.org/10.1146/annurev.astro.35.1.179}{\color{magenta}\araa},
  \href{http://adsabs.harvard.edu/abs/1997ARA%26A..35..179H}{\color{cyan}35},
  179

\bibitem[{{Hollenbach} \& {Tielens}(1999)}]{1999RvMP...71..173H}
---. 1999,
  \href{http://dx.doi.org/10.1103/RevModPhys.71.173}{\color{magenta}Reviews of
  Modern Physics},
  \href{http://adsabs.harvard.edu/abs/1999RvMP...71..173H}{\color{cyan}71}, 173

\bibitem[{{Hunter}(2007)}]{2007CSE.....9...90H}
{Hunter}, J.~D. 2007,
  \href{http://dx.doi.org/10.1109/MCSE.2007.55}{\color{magenta}Computing in
  Science and Engineering},
  \href{https://ui.adsabs.harvard.edu/abs/2007CSE.....9...90H}{\color{cyan}9},
  90

\bibitem[{{Igea} \& {Glassgold}(1999)}]{1999ApJ...518..848I}
{Igea}, J., \& {Glassgold}, A.~E. 1999,
  \href{http://dx.doi.org/10.1086/307302}{\color{magenta}\apj},
  \href{http://adsabs.harvard.edu/abs/1999ApJ...518..848I}{\color{cyan}518},
  848

\bibitem[{{Ilgner} \& {Nelson}(2006)}]{2006A&A...445..205I}
{Ilgner}, M., \& {Nelson}, R.~P. 2006,
  \href{http://dx.doi.org/10.1051/0004-6361:20053678}{\color{magenta}\aap},
  \href{http://adsabs.harvard.edu/abs/2006A%26A...445..205I}{\color{cyan}445},
  205

\bibitem[{{Jiang} {et~al.}(2012){Jiang}, {Stone}, \&
  {Davis}}]{2012ApJS..199...14J}
{Jiang}, Y.-F., {Stone}, J.~M., \& {Davis}, S.~W. 2012,
  \href{http://dx.doi.org/10.1088/0067-0049/199/1/14}{\color{magenta}\apjs},
  \href{http://adsabs.harvard.edu/abs/2012ApJS..199...14J}{\color{cyan}199}, 14

\bibitem[{{Jonkheid} {et~al.}(2004){Jonkheid}, {Faas}, {van Zadelhoff}, \& {van
  Dishoeck}}]{2004A&A...428..511J}
{Jonkheid}, B., {Faas}, F.~G.~A., {van Zadelhoff}, G.-J., \& {van Dishoeck},
  E.~F. 2004,
  \href{http://dx.doi.org/10.1051/0004-6361:20048013}{\color{magenta}\aap},
  \href{http://adsabs.harvard.edu/abs/2004A%26A...428..511J}{\color{cyan}428},
  511

\bibitem[{{Kama} {et~al.}(2016){Kama}, {Bruderer}, {Carney}, {Hogerheijde},
  {van Dishoeck}, {Fedele}, {Baryshev}, {Boland }, {G{\"u}sten}, {Aikutalp},
  {Choi}, {Endo}, {Frieswijk}, {Karska}, {Klaassen}, {Koumpia}, {Kristensen},
  {Leurini}, {Nagy}, {Perez Beaupuits}, {Risacher}, {van der Marel}, {van
  Kempen}, {van Weeren}, {Wyrowski}, \& {Y{\i}ld{\i}z}}]{2016A&A...588A.108K}
{Kama}, M., {Bruderer}, S., {Carney}, M., {et~al.} 2016,
  \href{http://dx.doi.org/10.1051/0004-6361/201526791}{\color{magenta}\aap},
  \href{https://ui.adsabs.harvard.edu/abs/2016A&A...588A.108K}{\color{cyan}588},
  A108

\bibitem[{{Keto} \& {Caselli}(2008)}]{2008ApJ...683..238K}
{Keto}, E., \& {Caselli}, P. 2008,
  \href{http://dx.doi.org/10.1086/589147}{\color{magenta}\apj},
  \href{http://adsabs.harvard.edu/abs/2008ApJ...683..238K}{\color{cyan}683},
  238

\bibitem[{{King} {et~al.}(2007){King}, {Pringle}, \&
  {Livio}}]{2007MNRAS.376.1740K}
{King}, A.~R., {Pringle}, J.~E., \& {Livio}, M. 2007,
  \href{http://dx.doi.org/10.1111/j.1365-2966.2007.11556.x}{\color{magenta}\mnras},
  \href{http://adsabs.harvard.edu/abs/2007MNRAS.376.1740K}{\color{cyan}376},
  1740

\bibitem[{{Kley}(1989)}]{1989A&A...208...98K}
{Kley}, W. 1989, \aap,
  \href{http://adsabs.harvard.edu/abs/1989A%26A...208...98K}{\color{cyan}208},
  98

\bibitem[{{K{\"o}nigl} \& {Salmeron}(2011)}]{2011ppcd.book..283K}
{K{\"o}nigl}, A., \& {Salmeron}, R. 2011, {The Effects of Large-Scale Magnetic
  Fields on Disk Formation and Evolution} (Chicago, IL: University of Chicago
  Press), 283--352

\bibitem[{{Krapp} {et~al.}(2018){Krapp}, {Gressel}, {Ben{\'{\i}}tez-Llambay},
  {Downes}, {Mohandas}, \& {Pessah}}]{2018ApJ...865..105K}
{Krapp}, L., {Gressel}, O., {Ben{\'{\i}}tez-Llambay}, P., {et~al.} 2018,
  \href{http://dx.doi.org/10.3847/1538-4357/aadcf0}{\color{magenta}\apj},
  \href{http://adsabs.harvard.edu/abs/2018ApJ...865..105K}{\color{cyan}865},
  105

\bibitem[{{Krasnopolsky} {et~al.}(1999){Krasnopolsky}, {Li}, \&
  {Blandford}}]{1999ApJ...526..631K}
{Krasnopolsky}, R., {Li}, Z.-Y., \& {Blandford}, R. 1999,
  \href{http://dx.doi.org/10.1086/308023}{\color{magenta}\apj},
  \href{https://ui.adsabs.harvard.edu/abs/1999ApJ...526..631K}{\color{cyan}526},
  631

\bibitem[{{Kuiper} {et~al.}(2010){Kuiper}, {Klahr}, {Dullemond}, {Kley}, \&
  {Henning}}]{2010A&A...511A..81K}
{Kuiper}, R., {Klahr}, H., {Dullemond}, C., {Kley}, W., \& {Henning}, T. 2010,
  \href{http://dx.doi.org/10.1051/0004-6361/200912355}{\color{magenta}\aap},
  \href{https://ui.adsabs.harvard.edu/abs/2010A&A...511A..81K}{\color{cyan}511},
  A81

\bibitem[{{Kuiper} \& {Klessen}(2013)}]{2013A&A...555A...7K}
{Kuiper}, R., \& {Klessen}, R.~S. 2013,
  \href{http://dx.doi.org/10.1051/0004-6361/201321404}{\color{magenta}\aap},
  \href{http://adsabs.harvard.edu/abs/2013A%26A...555A...7K}{\color{cyan}555},
  A7

\bibitem[{{Kunz} \& {Lesur}(2013)}]{2013MNRAS.434.2295K}
{Kunz}, M.~W., \& {Lesur}, G. 2013,
  \href{http://dx.doi.org/10.1093/mnras/stt1171}{\color{magenta}\mnras},
  \href{http://adsabs.harvard.edu/abs/2013MNRAS.434.2295K}{\color{cyan}434},
  2295

\bibitem[{{Landry} {et~al.}(2013){Landry}, {Dodson-Robinson}, {Turner}, \&
  {Abram}}]{2013ApJ...771...80L}
{Landry}, R., {Dodson-Robinson}, S.~E., {Turner}, N.~J., \& {Abram}, G. 2013,
  \href{http://dx.doi.org/10.1088/0004-637X/771/2/80}{\color{magenta}\apj},
  \href{http://adsabs.harvard.edu/abs/2013ApJ...771...80L}{\color{cyan}771}, 80

\bibitem[{{Latter} \& {Papaloizou}(2018)}]{2018MNRAS.474.3110L}
{Latter}, H.~N., \& {Papaloizou}, J. 2018,
  \href{http://dx.doi.org/10.1093/mnras/stx3031}{\color{magenta}\mnras},
  \href{https://ui.adsabs.harvard.edu/abs/2018MNRAS.474.3110L}{\color{cyan}474},
  3110

\bibitem[{{Le Teuff} {et~al.}(2000){Le Teuff}, {Millar}, \&
  {Markwick}}]{2000A&AS..146..157L}
{Le Teuff}, Y.~H., {Millar}, T.~J., \& {Markwick}, A.~J. 2000,
  \href{http://dx.doi.org/10.1051/aas:2000265}{\color{magenta}\aaps},
  \href{http://adsabs.harvard.edu/abs/2000A%26AS..146..157L}{\color{cyan}146},
  157

\bibitem[{{Lee} {et~al.}(1996){Lee}, {Herbst}, {Pineau des Forets}, {Roueff},
  \& {Le Bourlot}}]{1996A&A...311..690L}
{Lee}, H.-H., {Herbst}, E., {Pineau des Forets}, G., {Roueff}, E., \& {Le
  Bourlot}, J. 1996, \aap,
  \href{http://adsabs.harvard.edu/abs/1996A%26A...311..690L}{\color{cyan}311},
  690

\bibitem[{{Lesur} {et~al.}(2014){Lesur}, {Kunz}, \&
  {Fromang}}]{2014A&A...566A..56L}
{Lesur}, G., {Kunz}, M.~W., \& {Fromang}, S. 2014,
  \href{http://dx.doi.org/10.1051/0004-6361/201423660}{\color{magenta}\aap},
  \href{http://adsabs.harvard.edu/abs/2014A%26A...566A..56L}{\color{cyan}566},
  A56

\bibitem[{{Leung} \& {Ogilvie}(2019)}]{2019MNRAS.tmp.1545L}
{Leung}, P. K.~C., \& {Ogilvie}, G.~I. 2019,
  \href{http://dx.doi.org/10.1093/mnras/stz1620}{\color{magenta}\mnras}, 1545

\bibitem[{{Levermore} \& {Pomraning}(1981)}]{1981ApJ...248..321L}
{Levermore}, C.~D., \& {Pomraning}, G.~C. 1981,
  \href{http://dx.doi.org/10.1086/159157}{\color{magenta}\apj},
  \href{http://adsabs.harvard.edu/abs/1981ApJ...248..321L}{\color{cyan}248},
  321

\bibitem[{{Li} {et~al.}(2014){Li}, {Banerjee}, {Pudritz}, {J{\o}rgensen},
  {Shang}, {Krasnopolsky}, \& {Maury}}]{2014prpl.conf..173L}
{Li}, Z.~Y., {Banerjee}, R., {Pudritz}, R.~E., {et~al.} 2014, in Protostars and
  Planets VI, ed. Henrik Beuther et al., (Tucson, AZ: Univ. of Arizona Press),
  173

\bibitem[{{Lin} \& {Youdin}(2015)}]{2015ApJ...811...17L}
{Lin}, M.-K., \& {Youdin}, A.~N. 2015,
  \href{http://dx.doi.org/10.1088/0004-637X/811/1/17}{\color{magenta}\apj},
  \href{https://ui.adsabs.harvard.edu/abs/2015ApJ...811...17L}{\color{cyan}811},
  17

\bibitem[{{Lovelace} {et~al.}(1995){Lovelace}, {Romanova}, \&
  {Bisnovatyi-Kogan}}]{1995MNRAS.275..244L}
{Lovelace}, R.~V.~E., {Romanova}, M.~M., \& {Bisnovatyi-Kogan}, G.~S. 1995,
  \href{http://dx.doi.org/10.1093/mnras/275.2.244}{\color{magenta}\mnras},
  \href{https://ui.adsabs.harvard.edu/abs/1995MNRAS.275..244L}{\color{cyan}275},
  244

\bibitem[{{Mamajek}(2009)}]{2009AIPC.1158....3M}
{Mamajek}, E.~E. 2009, in American Institute of Physics Conference Series, Vol.
  1158, American Institute of Physics Conference Series, ed. T.~{Usuda},
  M.~{Tamura}, \& M.~{Ishii}, 3--10

\bibitem[{{McElroy} {et~al.}(2013){McElroy}, {Walsh}, {Markwick}, {Cordiner},
  {Smith}, \& {Millar}}]{2013A&A...550A..36M}
{McElroy}, D., {Walsh}, C., {Markwick}, A.~J., {et~al.} 2013,
  \href{http://dx.doi.org/10.1051/0004-6361/201220465}{\color{magenta}\aap},
  \href{http://adsabs.harvard.edu/abs/2013A%26A...550A..36M}{\color{cyan}550},
  A36

\bibitem[{{McKee} \& {Ostriker}(2007)}]{2007ARA&A..45..565M}
{McKee}, C.~F., \& {Ostriker}, E.~C. 2007,
  \href{http://dx.doi.org/10.1146/annurev.astro.45.051806.110602}{\color{magenta}\araa},
  \href{http://adsabs.harvard.edu/abs/2007ARA%26A..45..565M}{\color{cyan}45},
  565

\bibitem[{{McMullin} {et~al.}(2007){McMullin}, {Waters}, {Schiebel}, {Young},
  \& {Golap}}]{2007ASPC..376..127M}
{McMullin}, J.~P., {Waters}, B., {Schiebel}, D., {Young}, W., \& {Golap}, K.
  2007, in Astronomical Society of the Pacific Conference Series, Vol. 376,
  Astronomical Data Analysis Software and Systems XVI, ed. R.~A. {Shaw},
  F.~{Hill}, \& D.~J. {Bell}, 127

\bibitem[{{McNally} {et~al.}(2018){McNally}, {Nelson}, \&
  {Paardekooper}}]{2018MNRAS.477.4596M}
{McNally}, C.~P., {Nelson}, R.~P., \& {Paardekooper}, S.-J. 2018,
  \href{http://dx.doi.org/10.1093/mnras/sty905}{\color{magenta}\mnras},
  \href{http://adsabs.harvard.edu/abs/2018MNRAS.477.4596M}{\color{cyan}477},
  4596

\bibitem[{{McNally} {et~al.}(2017){McNally}, {Nelson}, {Paardekooper},
  {Gressel}, \& {Lyra}}]{2017MNRAS.472.1565M}
{McNally}, C.~P., {Nelson}, R.~P., {Paardekooper}, S.-J., {Gressel}, O., \&
  {Lyra}, W. 2017,
  \href{http://dx.doi.org/10.1093/mnras/stx2136}{\color{magenta}\mnras},
  \href{http://adsabs.harvard.edu/abs/2017MNRAS.472.1565M}{\color{cyan}472},
  1565

\bibitem[{{Meyer} {et~al.}(2012){Meyer}, {Balsara}, \&
  {Aslam}}]{2012MNRAS.422.2102M}
{Meyer}, C.~D., {Balsara}, D.~S., \& {Aslam}, T.~D. 2012,
  \href{http://dx.doi.org/10.1111/j.1365-2966.2012.20744.x}{\color{magenta}\mnras},
  \href{http://adsabs.harvard.edu/abs/2012MNRAS.422.2102M}{\color{cyan}422},
  2102

\bibitem[{{Miyoshi} \& {Kusano}(2005)}]{2005JCoPh.208..315M}
{Miyoshi}, T., \& {Kusano}, K. 2005,
  \href{http://dx.doi.org/10.1016/j.jcp.2005.02.017}{\color{magenta}\rm JCoPh},
  \href{http://adsabs.harvard.edu/abs/2005JCoPh.208..315M}{\color{cyan}208},
  315

\bibitem[{{Mohanty} {et~al.}(2013){Mohanty}, {Ercolano}, \&
  {Turner}}]{2013ApJ...764...65M}
{Mohanty}, S., {Ercolano}, B., \& {Turner}, N.~J. 2013,
  \href{http://dx.doi.org/10.1088/0004-637X/764/1/65}{\color{magenta}\apj},
  \href{http://adsabs.harvard.edu/abs/2013ApJ...764...65M}{\color{cyan}764}, 65

\bibitem[{{Najita} \& {{\'A}d{\'a}m\-kovics}(2017)}]{2017ApJ...847....6N}
{Najita}, J.~R., \& {{\'A}d{\'a}m\-kovics}, M. 2017,
  \href{http://dx.doi.org/10.3847/1538-4357/aa8632}{\color{magenta}\apj},
  \href{http://adsabs.harvard.edu/abs/2017ApJ...847....6N}{\color{cyan}847}, 6

\bibitem[{{Nakatani} {et~al.}(2018){Nakatani}, {Hosokawa}, {Yoshida}, {Nomura},
  \& {Kuiper}}]{2018ApJ...857...57N}
{Nakatani}, R., {Hosokawa}, T., {Yoshida}, N., {Nomura}, H., \& {Kuiper}, R.
  2018, \href{http://dx.doi.org/10.3847/1538-4357/aab70b}{\color{magenta}\apj},
  \href{https://ui.adsabs.harvard.edu/abs/2018ApJ...857...57N}{\color{cyan}857},
  57

\bibitem[{{Nelson} \& {Gressel}(2010)}]{2010MNRAS.409..639N}
{Nelson}, R.~P., \& {Gressel}, O. 2010,
  \href{http://dx.doi.org/10.1111/j.1365-2966.2010.17327.x}{\color{magenta}\mnras},
  \href{http://adsabs.harvard.edu/abs/2010MNRAS.409..639N}{\color{cyan}409},
  639

\bibitem[{{Nelson} {et~al.}(2013){Nelson}, {Gressel}, \&
  {Umurhan}}]{2013MNRAS.435.2610N}
{Nelson}, R.~P., {Gressel}, O., \& {Umurhan}, O.~M. 2013,
  \href{http://dx.doi.org/10.1093/mnras/stt1475}{\color{magenta}\mnras},
  \href{http://adsabs.harvard.edu/abs/2013MNRAS.435.2610N}{\color{cyan}435},
  2610

\bibitem[{{Nolan} {et~al.}(2017){Nolan}, {Salmeron}, {Federrath}, {Bicknell},
  \& {Sutherland}}]{2017MNRAS.471.1488N}
{Nolan}, C.~A., {Salmeron}, R., {Federrath}, C., {Bicknell}, G.~V., \&
  {Sutherland}, R.~S. 2017,
  \href{http://dx.doi.org/10.1093/mnras/stx1642}{\color{magenta}\mnras},
  \href{https://ui.adsabs.harvard.edu/abs/2017MNRAS.471.1488N}{\color{cyan}471},
  1488

\bibitem[{{Nomura} \& {Millar}(2005)}]{2005A&A...438..923N}
{Nomura}, H., \& {Millar}, T.~J. 2005,
  \href{http://dx.doi.org/10.1051/0004-6361:20052809}{\color{magenta}\aap},
  \href{http://adsabs.harvard.edu/abs/2005A%26A...438..923N}{\color{cyan}438},
  923

\bibitem[{{Olson} \& {Kunasz}(1987)}]{1987JQSRT..38..325O}
{Olson}, G.~L., \& {Kunasz}, P.~B. 1987,
  \href{http://dx.doi.org/10.1016/0022-4073(87)90027-6}{\color{magenta}JQSRT},
  \href{https://ui.adsabs.harvard.edu/abs/1987JQSRT..38..325O}{\color{cyan}38},
  325

\bibitem[{{Panoglou} {et~al.}(2012){Panoglou}, {Cabrit}, {Pineau Des
  For{\^e}ts}, {Garcia}, {Ferreira}, \& {Casse}}]{2012A&A...538A...2P}
{Panoglou}, D., {Cabrit}, S., {Pineau Des For{\^e}ts}, G., {et~al.} 2012,
  \href{http://dx.doi.org/10.1051/0004-6361/200912861}{\color{magenta}\aap},
  \href{http://adsabs.harvard.edu/abs/2012A%26A...538A...2P}{\color{cyan}538},
  A2

\bibitem[{{Perez-Becker} \& {Chiang}(2011)}]{2011ApJ...735....8P}
{Perez-Becker}, D., \& {Chiang}, E. 2011,
  \href{http://dx.doi.org/10.1088/0004-637X/735/1/8}{\color{magenta}\apj},
  \href{http://adsabs.harvard.edu/abs/2011ApJ...735....8P}{\color{cyan}735}, 8

\bibitem[{{Picogna} {et~al.}(2019){Picogna}, {Ercolano}, {Owen}, \&
  {Weber}}]{2019MNRAS.487..691P}
{Picogna}, G., {Ercolano}, B., {Owen}, J.~E., \& {Weber}, M.~L. 2019,
  \href{http://dx.doi.org/10.1093/mnras/stz1166}{\color{magenta}\mnras},
  \href{https://ui.adsabs.harvard.edu/abs/2019MNRAS.487..691P}{\color{cyan}487},
  691

\bibitem[{{Porth} \& {Fendt}(2010)}]{2010ApJ...709.1100P}
{Porth}, O., \& {Fendt}, C. 2010,
  \href{http://dx.doi.org/10.1088/0004-637X/709/2/1100}{\color{magenta}\apj},
  \href{https://ui.adsabs.harvard.edu/abs/2010ApJ...709.1100P}{\color{cyan}709},
  1100

\bibitem[{{Pudritz} {et~al.}(2007){Pudritz}, {Ouyed}, {Fendt}, \&
  {Brandenburg}}]{2007prpl.conf..277P}
{Pudritz}, R.~E., {Ouyed}, R., {Fendt}, C., \& {Brandenburg}, A. 2007, in
  Protostars and Planets V, 277

\bibitem[{{Ramsey} {et~al.}(2012){Ramsey}, {Clarke}, \&
  {Men'shchikov}}]{2012ApJS..199...13R}
{Ramsey}, J.~P., {Clarke}, D.~A., \& {Men'shchikov}, A.~B. 2012,
  \href{http://dx.doi.org/10.1088/0067-0049/199/1/13}{\color{magenta}\apjs},
  \href{https://ui.adsabs.harvard.edu/abs/2012ApJS..199...13R}{\color{cyan}199},
  13

\bibitem[{{Ramsey} \& {Dullemond}(2015)}]{2015A&A...574A..81R}
{Ramsey}, J.~P., \& {Dullemond}, C.~P. 2015,
  \href{http://dx.doi.org/10.1051/0004-6361/201424954}{\color{magenta}\aap},
  \href{http://adsabs.harvard.edu/abs/2015A%26A...574A..81R}{\color{cyan}574},
  A81

\bibitem[{{Richling} \& {Yorke}(2000)}]{2000ApJ...539..258R}
{Richling}, S., \& {Yorke}, H.~W. 2000,
  \href{http://dx.doi.org/10.1086/309198}{\color{magenta}\apj},
  \href{https://ui.adsabs.harvard.edu/abs/2000ApJ...539..258R}{\color{cyan}539},
  258

\bibitem[{{Rodenkirch} {et~al.}(2019){Rodenkirch}, {Klahr}, {Fendt}, \&
  {Dullemond}}]{2019arXiv191104510R}
{Rodenkirch}, P.~J., {Klahr}, H., {Fendt}, C., \& {Dullemond}, C.~P. 2019,
  arXiv e-prints, arXiv:1911.04510

\bibitem[{{Rodgers-Lee} {et~al.}(2016){Rodgers-Lee}, {Ray}, \&
  {Downes}}]{2016MNRAS.463..134R}
{Rodgers-Lee}, D., {Ray}, T.~P., \& {Downes}, T.~P. 2016,
  \href{http://dx.doi.org/10.1093/mnras/stw1980}{\color{magenta}\mnras},
  \href{http://adsabs.harvard.edu/abs/2016MNRAS.463..134R}{\color{cyan}463},
  134

\bibitem[{{R{\"o}llig} {et~al.}(2006){R{\"o}llig}, {Ossenkopf}, {Jeyakumar},
  {Stutzki}, \& {Sternberg}}]{2006A&A...451..917R}
{R{\"o}llig}, M., {Ossenkopf}, V., {Jeyakumar}, S., {Stutzki}, J., \&
  {Sternberg}, A. 2006,
  \href{http://dx.doi.org/10.1051/0004-6361:20053845}{\color{magenta}\aap},
  \href{http://adsabs.harvard.edu/abs/2006A%26A...451..917R}{\color{cyan}451},
  917

\bibitem[{{R{\"o}llig} {et~al.}(2007){R{\"o}llig}, {Abel}, {Bell}, {Bensch},
  {Black}, {Ferland}, {Jonkheid}, {Kamp}, {Kaufman}, {Le Bourlot}, {Le Petit},
  {Meijerink}, {Morata}, {Ossenkopf}, {Roueff}, {Shaw}, {Spaans}, {Sternberg},
  {Stutzki}, {Thi}, {van Dishoeck}, {van Hoof}, {Viti}, \&
  {Wolfire}}]{2007A&A...467..187R}
{R{\"o}llig}, M., {Abel}, N.~P., {Bell}, T., {et~al.} 2007,
  \href{http://dx.doi.org/10.1051/0004-6361:20065918}{\color{magenta}\aap},
  \href{http://adsabs.harvard.edu/abs/2007A%26A...467..187R}{\color{cyan}467},
  187

\bibitem[{{Sch{\"o}ier} {et~al.}(2005){Sch{\"o}ier}, {van der Tak}, {van
  Dishoeck}, \& {Black}}]{2005A&A...432..369S}
{Sch{\"o}ier}, F.~L., {van der Tak}, F.~F.~S., {van Dishoeck}, E.~F., \&
  {Black}, J.~H. 2005,
  \href{http://dx.doi.org/10.1051/0004-6361:20041729}{\color{magenta}\aap},
  \href{http://adsabs.harvard.edu/abs/2005A%26A...432..369S}{\color{cyan}432},
  369

\bibitem[{{Shadmehri} \& {Ghoreyshi}(2019)}]{2019MNRAS.tmp.1959S}
{Shadmehri}, M., \& {Ghoreyshi}, S.~M. 2019,
  \href{http://dx.doi.org/10.1093/mnras/stz2025}{\color{magenta}\mnras}, 1959

\bibitem[{{Simon} {et~al.}(2015){Simon}, {Lesur}, {Kunz}, \&
  {Armitage}}]{2015MNRAS.454.1117S}
{Simon}, J.~B., {Lesur}, G., {Kunz}, M.~W., \& {Armitage}, P.~J. 2015,
  \href{http://dx.doi.org/10.1093/mnras/stv2070}{\color{magenta}\mnras},
  \href{http://adsabs.harvard.edu/abs/2015MNRAS.454.1117S}{\color{cyan}454},
  1117

\bibitem[{{Simon} {et~al.}(2016){Simon}, {Pascucci}, {Edwards}, {Feng},
  {Gorti}, {Hollenbach}, {Rigliaco}, \& {Keane}}]{2016ApJ...831..169S}
{Simon}, M.~N., {Pascucci}, I., {Edwards}, S., {et~al.} 2016,
  \href{http://dx.doi.org/10.3847/0004-637X/831/2/169}{\color{magenta}\apj},
  \href{https://ui.adsabs.harvard.edu/abs/2016ApJ...831..169S}{\color{cyan}831},
  169

\bibitem[{{Skinner} \& {Ostriker}(2010)}]{2010ApJS..188..290S}
{Skinner}, M.~A., \& {Ostriker}, E.~C. 2010,
  \href{http://dx.doi.org/10.1088/0067-0049/188/1/290}{\color{magenta}\apjs},
  \href{http://adsabs.harvard.edu/abs/2010ApJS..188..290S}{\color{cyan}188},
  290

\bibitem[{{Skinner} \& {Ostriker}(2013)}]{2013ApJS..206...21S}
---. 2013,
  \href{http://dx.doi.org/10.1088/0067-0049/206/2/21}{\color{magenta}\apjs},
  \href{http://adsabs.harvard.edu/abs/2013ApJS..206...21S}{\color{cyan}206}, 21

\bibitem[{{Spruit}(1996)}]{1996ASIC..477..249S}
{Spruit}, H.~C. 1996, in NATO Advanced Science Institutes (ASI) Series C, ed.
  R.~A.~M.~J. {Wijers}, M.~B. {Davies}, \& C.~A. {Tout}, Vol. 477, 249--286

\bibitem[{{Sternberg} \& {Dalgarno}(1989)}]{1989ApJ...338..197S}
{Sternberg}, A., \& {Dalgarno}, A. 1989,
  \href{http://dx.doi.org/10.1086/167193}{\color{magenta}\apj},
  \href{http://adsabs.harvard.edu/abs/1989ApJ...338..197S}{\color{cyan}338},
  197

\bibitem[{{Stone} {et~al.}(1992){Stone}, {Mihalas}, \&
  {Norman}}]{1992ApJS...80..819S}
{Stone}, J.~M., {Mihalas}, D., \& {Norman}, M.~L. 1992,
  \href{http://dx.doi.org/10.1086/191682}{\color{magenta}\apjs},
  \href{http://adsabs.harvard.edu/abs/1992ApJS...80..819S}{\color{cyan}80}, 819

\bibitem[{{Suriano} {et~al.}(2017){Suriano}, {Li}, {Krasnopolsky}, \&
  {Shang}}]{2017MNRAS.468.3850S}
{Suriano}, S.~S., {Li}, Z.-Y., {Krasnopolsky}, R., \& {Shang}, H. 2017,
  \href{http://dx.doi.org/10.1093/mnras/stx735}{\color{magenta}\mnras},
  \href{https://ui.adsabs.harvard.edu/abs/2017MNRAS.468.3850S}{\color{cyan}468},
  3850

\bibitem[{{Suriano} {et~al.}(2018){Suriano}, {Li}, {Krasnopolsky}, \&
  {Shang}}]{2018MNRAS.477.1239S}
---. 2018,
  \href{http://dx.doi.org/10.1093/mnras/sty717}{\color{magenta}\mnras},
  \href{http://adsabs.harvard.edu/abs/2018MNRAS.477.1239S}{\color{cyan}477},
  1239

\bibitem[{{Suriano} {et~al.}(2019){Suriano}, {Li}, {Krasnopolsky}, {Suzuki}, \&
  {Shang}}]{2019MNRAS.484..107S}
{Suriano}, S.~S., {Li}, Z.-Y., {Krasnopolsky}, R., {Suzuki}, T.~K., \& {Shang},
  H. 2019,
  \href{http://dx.doi.org/10.1093/mnras/sty3502}{\color{magenta}\mnras},
  \href{https://ui.adsabs.harvard.edu/abs/2019MNRAS.484..107S}{\color{cyan}484},
  107

\bibitem[{{Suzuki} \& {Inutsuka}(2009)}]{2009ApJ...691L..49S}
{Suzuki}, T.~K., \& {Inutsuka}, S.-i. 2009,
  \href{http://dx.doi.org/10.1088/0004-637X/691/1/L49}{\color{magenta}\apjl},
  \href{http://adsabs.harvard.edu/abs/2009ApJ...691L..49S}{\color{cyan}691},
  L49

\bibitem[{{Suzuki} {et~al.}(2016){Suzuki}, {Ogihara}, {Morbidelli}, {Crida}, \&
  {Guillot}}]{2016A&A...596A..74S}
{Suzuki}, T.~K., {Ogihara}, M., {Morbidelli}, A., {Crida}, A., \& {Guillot}, T.
  2016,
  \href{http://dx.doi.org/10.1051/0004-6361/201628955}{\color{magenta}\aap},
  \href{http://adsabs.harvard.edu/abs/2016A%26A...596A..74S}{\color{cyan}596},
  A74

\bibitem[{{Tielens} \& {Hollenbach}(1985)}]{1985ApJ...291..722T}
{Tielens}, A.~G.~G.~M., \& {Hollenbach}, D. 1985,
  \href{http://dx.doi.org/10.1086/163111}{\color{magenta}\apj},
  \href{http://adsabs.harvard.edu/abs/1985ApJ...291..722T}{\color{cyan}291},
  722

\bibitem[{{Tsukagoshi} {et~al.}(2015){Tsukagoshi}, {Momose}, {Saito},
  {Kitamura}, {Shimajiri}, \& {Kawabe}}]{2015ApJ...802L...7T}
{Tsukagoshi}, T., {Momose}, M., {Saito}, M., {et~al.} 2015,
  \href{http://dx.doi.org/10.1088/2041-8205/802/1/L7}{\color{magenta}\apjl},
  \href{https://ui.adsabs.harvard.edu/abs/2015ApJ...802L...7T}{\color{cyan}802},
  L7

\bibitem[{{Tsukamoto} {et~al.}(2017){Tsukamoto}, {Okuzumi}, {Iwasaki},
  {Machida}, \& {Inutsuka}}]{2017PASJ...69...95T}
{Tsukamoto}, Y., {Okuzumi}, S., {Iwasaki}, K., {Machida}, M.~N., \& {Inutsuka},
  S.-i. 2017,
  \href{http://dx.doi.org/10.1093/pasj/psx113}{\color{magenta}\pasj},
  \href{https://ui.adsabs.harvard.edu/abs/2017PASJ...69...95T}{\color{cyan}69},
  95

\bibitem[{{Turner} \& {Drake}(2009)}]{2009ApJ...703.2152T}
{Turner}, N.~J., \& {Drake}, J.~F. 2009,
  \href{http://dx.doi.org/10.1088/0004-637X/703/2/2152}{\color{magenta}\apj},
  \href{http://adsabs.harvard.edu/abs/2009ApJ...703.2152T}{\color{cyan}703},
  2152

\bibitem[{{Turner} {et~al.}(2014){Turner}, {Fromang}, {Gammie}, {Klahr},
  {Lesur}, {Wardle}, \& {Bai}}]{2014prpl.conf..411T}
{Turner}, N.~J., {Fromang}, S., {Gammie}, C., {et~al.} 2014, in Protostars and
  Planets VI, ed. Henrik Beuther et al., (Tucson, AZ: Univ. of Arizona Press),
  411

\bibitem[{{Turner} {et~al.}(2007){Turner}, {Sano}, \&
  {Dziourkevitch}}]{2007ApJ...659..729T}
{Turner}, N.~J., {Sano}, T., \& {Dziourkevitch}, N. 2007,
  \href{http://dx.doi.org/10.1086/512007}{\color{magenta}\apj},
  \href{https://ui.adsabs.harvard.edu/abs/2007ApJ...659..729T}{\color{cyan}659},
  729

\bibitem[{{Turner} \& {Stone}(2001)}]{2001ApJS..135...95T}
{Turner}, N.~J., \& {Stone}, J.~M. 2001,
  \href{http://dx.doi.org/10.1086/321779}{\color{magenta}\apjs},
  \href{https://ui.adsabs.harvard.edu/abs/2001ApJS..135...95T}{\color{cyan}135},
  95

\bibitem[{{Ustyugova} {et~al.}(1999){Ustyugova}, {Koldoba}, {Romanova},
  {Chechetkin}, \& {Lovelace}}]{1999ApJ...516..221U}
{Ustyugova}, G.~V., {Koldoba}, A.~V., {Romanova}, M.~M., {Chechetkin}, V.~M.,
  \& {Lovelace}, R.~V.~E. 1999,
  \href{http://dx.doi.org/10.1086/307093}{\color{magenta}\apj},
  \href{https://ui.adsabs.harvard.edu/abs/1999ApJ...516..221U}{\color{cyan}516},
  221

\bibitem[{{van der Tak} {et~al.}(2007){van der Tak}, {Black}, {Sch{\"o}ier},
  {Jansen}, \& {van Dishoeck}}]{2007A&A...468..627V}
{van der Tak}, F.~F.~S., {Black}, J.~H., {Sch{\"o}ier}, F.~L., {Jansen}, D.~J.,
  \& {van Dishoeck}, E.~F. 2007,
  \href{http://dx.doi.org/10.1051/0004-6361:20066820}{\color{magenta}\aap},
  \href{http://adsabs.harvard.edu/abs/2007A%26A...468..627V}{\color{cyan}468},
  627

\bibitem[{{Vanajakshi} {et~al.}(1989){Vanajakshi}, {Thompson}, \&
  {Black}}]{1989JCoPh..84..343V}
{Vanajakshi}, T.~C., {Thompson}, K.~W., \& {Black}, D.~C. 1989,
  \href{http://dx.doi.org/10.1016/0021-9991(89)90237-4}{\color{magenta}\rm
  JCoPh},
  \href{https://ui.adsabs.harvard.edu/abs/1989JCoPh..84..343V}{\color{cyan}84},
  343

\bibitem[{{Vlemmings} {et~al.}(2019){Vlemmings}, {Lankhaar}, {Cazzoletti},
  {Ceccobello}, {Dall'Olio}, {van Dishoeck}, {Facchini}, {Humphreys},
  {Persson}, {Testi}, \& {Williams}}]{2019A&A...624L...7V}
{Vlemmings}, W.~H.~T., {Lankhaar}, B., {Cazzoletti}, P., {et~al.} 2019,
  \href{http://dx.doi.org/10.1051/0004-6361/201935459}{\color{magenta}\aap},
  \href{https://ui.adsabs.harvard.edu/abs/2019A&A...624L...7V}{\color{cyan}624},
  L7

\bibitem[{{Wang} {et~al.}(2019){Wang}, {Bai}, \&
  {Goodman}}]{2019ApJ...874...90W}
{Wang}, L., {Bai}, X.-N., \& {Goodman}, J. 2019,
  \href{http://dx.doi.org/10.3847/1538-4357/ab06fd}{\color{magenta}\apj},
  \href{http://adsabs.harvard.edu/abs/2019ApJ...874...90W}{\color{cyan}874}, 90

\bibitem[{{Wang} \& {Goodman}(2017)}]{2017ApJ...847...11W}
{Wang}, L., \& {Goodman}, J. 2017,
  \href{http://dx.doi.org/10.3847/1538-4357/aa8726}{\color{magenta}\apj},
  \href{https://ui.adsabs.harvard.edu/abs/2017ApJ...847...11W}{\color{cyan}847},
  11

\bibitem[{{Wardle} \& {K{\"o}nigl}(1993)}]{1993ApJ...410..218W}
{Wardle}, M., \& {K{\"o}nigl}, A. 1993,
  \href{http://dx.doi.org/10.1086/172739}{\color{magenta}\apj},
  \href{http://adsabs.harvard.edu/abs/1993ApJ...410..218W}{\color{cyan}410},
  218

\bibitem[{{Williams} \& {Cieza}(2011)}]{2011ARA&A..49...67W}
{Williams}, J.~P., \& {Cieza}, L.~A. 2011,
  \href{http://dx.doi.org/10.1146/annurev-astro-081710-102548}{\color{magenta}\araa},
  \href{http://adsabs.harvard.edu/abs/2011ARA\%26A..49...67W}{\color{cyan}49},
  67

\bibitem[{{Wilson} \& {Rood}(1994)}]{1994ARA&A..32..191W}
{Wilson}, T.~L., \& {Rood}, R. 1994,
  \href{http://dx.doi.org/10.1146/annurev.aa.32.090194.001203}{\color{magenta}\araa},
  \href{http://adsabs.harvard.edu/abs/1994ARA%26A..32..191W}{\color{cyan}32},
  191

\bibitem[{{Woitke} {et~al.}(2009){Woitke}, {Kamp}, \&
  {Thi}}]{2009A&A...501..383W}
{Woitke}, P., {Kamp}, I., \& {Thi}, W.-F. 2009,
  \href{http://dx.doi.org/10.1051/0004-6361/200911821}{\color{magenta}\aap},
  \href{http://adsabs.harvard.edu/abs/2009A%26A...501..383W}{\color{cyan}501},
  383

\bibitem[{{Woodall} {et~al.}(2007){Woodall}, {Ag{\'u}ndez}, {Markwick-Kemper},
  \& {Millar}}]{2007A&A...466.1197W}
{Woodall}, J., {Ag{\'u}ndez}, M., {Markwick-Kemper}, A.~J., \& {Millar}, T.~J.
  2007,
  \href{http://dx.doi.org/10.1051/0004-6361:20064981}{\color{magenta}\aap},
  \href{http://adsabs.harvard.edu/abs/2007A%26A...466.1197W}{\color{cyan}466},
  1197

\bibitem[{{Zanni} {et~al.}(2007){Zanni}, {Ferrari}, {Rosner}, {Bodo}, \&
  {Massaglia}}]{2007A&A...469..811Z}
{Zanni}, C., {Ferrari}, A., {Rosner}, R., {Bodo}, G., \& {Massaglia}, S. 2007,
  \href{http://dx.doi.org/10.1051/0004-6361:20066400}{\color{magenta}\aap},
  \href{https://ui.adsabs.harvard.edu/abs/2007A&A...469..811Z}{\color{cyan}469},
  811

\bibitem[{{Zhang} {et~al.}(2019){Zhang}, {Arce}, {Mardones}, {Cabrit},
  {Dunham}, {Garay}, {Noriega-Crespo}, {Offner}, {Raga}, \&
  {Corder}}]{2019ApJ...883....1Z}
{Zhang}, Y., {Arce}, H.~G., {Mardones}, D., {et~al.} 2019,
  \href{http://dx.doi.org/10.3847/1538-4357/ab3850}{\color{magenta}\apj},
  \href{https://ui.adsabs.harvard.edu/abs/2019ApJ...883....1Z}{\color{cyan}883},
  1

\bibitem[{{Ziegler}(2004)}]{2004JCoPh.196..393Z}
{Ziegler}, U. 2004,
  \href{http://dx.doi.org/10.1016/j.jcp.2003.11.003}{\color{magenta}\rm JCoPh},
  \href{http://adsabs.harvard.edu/abs/2004JCoPh.196..393Z}{\color{cyan}196},
  393

\bibitem[{{Ziegler}(2011)}]{2011JCoPh.230.1035Z}
---. 2011, \rm JCoPh, 230, 1035

\bibitem[{{Ziegler}(2016)}]{2016A&A...586A..82Z}
---. 2016,
  \href{http://dx.doi.org/10.1051/0004-6361/201527262}{\color{magenta}\aap},
  \href{http://adsabs.harvard.edu/abs/2016A%26A...586A..82Z}{\color{cyan}586},
  A82

\end{thebibliography}


\end{document}